\documentclass[aps, preprint, superscriptaddress,preprintnumbers,floatfix,nofootinbib,amsmath,amssymb]{revtex4}

\usepackage{url}
\usepackage{hyperref}
\usepackage{cleveref}

\usepackage{graphicx}
\usepackage[T1]{fontenc} 
\usepackage{epsfig}
\usepackage{amstext,amssymb}
\usepackage{amsmath}
\usepackage{amsfonts}

\usepackage{xspace}
\usepackage{subfigure}
\usepackage{color}
\usepackage{bm}

\newcommand{\beqa}{\begin{eqnarray}}
\newcommand{\eeqa}{\end{eqnarray}}
\newcommand{\beq}{\begin{equation}}
\newcommand{\eeq}{\end{equation}}

\newcommand{\nn}{\nonumber}
\newcommand{\bmt}{\begin{pmatrix}}
\newcommand{\emt}{\end{pmatrix}}
\usepackage[toc,page]{appendix}
\usepackage{comment}
\newcommand{\be}{\begin{equation}}
\newcommand{\ee}{\end{equation}}
\newcommand{\bea}{\begin{eqnarray}}
\newcommand{\eea}{\end{eqnarray}}

\begin{document}
\title{Exploring the role of new physics in $b \to u \tau \bar \nu$ decays}

\author{Aishwarya Bhatta}
\email{aish.bhatta@gmail.com}
\affiliation{School of Physics, University of Hyderabad, Hyderabad - 500046,  India}
\author{Atasi Ray}
\email{atasiray92@gmail.com}
\affiliation{School of Physics, University of Hyderabad, Hyderabad - 500046,  India}

\author{Rukmani Mohanta}
\email{rmsp@uohyd.ac.in}
\affiliation{School of Physics, University of Hyderabad, Hyderabad - 500046,  India}
%%%%%%%%%%%%%%%%%%%%%%%%%%%%%%%%%%%%%%%%%%%%%%%%%%%%%%%%%%%%
\begin{abstract}
%%%%%%%%%%%%%%%%%%%%%%%%%%%%%%%%%%%%%%%%%%%%%%%%%%%%%%%%%
The recent  measurements on $R_D$, $R_{D^*}$ and $R_{J/\psi}$ by three pioneering experiments, BaBar, Belle and LHCb, indicate that the notion of  lepton flavour universality is violated in the weak  charged-current processes, mediated through $b \to c \ell \bar \nu_\ell$ transitions. These intriguing results, which delineate a tension with their standard model  predictions  at  the level of $(2-3)\sigma$ have triggered many new physics propositions in recent times, and are generally attributed to the possible implication of new physics in $ b \to c \tau \bar \nu$ transition.  This, in turn, opens up another avenue, i.e., $ b \to u \tau \bar \nu$ processes, to look for new physics. Since these processes are doubly Cabibbo suppressed, the impact of new physics could be significant enough, leading to sizeable effects in some of the observables. In this work, we investigate in detail the role of new physics in $B \to (\pi,\rho,\omega)\tau \bar \nu$ and $B_s \to (K,K^*) \tau \bar \nu$ processes considering a model independent approach. 
In particular, we focus on the standard observables like branching fraction, lepton flavour non-universality (LNU) parameter,   forward-backward asymmetry and polarization asymmetries. We find significant deviations in some of these observables, which can be explored by the currently running experiments LHCb and Belle-II.   We also briefly comment on the impact of scalar leptoquark $R_2(3,2,7/6)$  and vector leptoquark $U_1(3,1,2/3)$ on these decay modes.

\end{abstract}
%\pacs{13.30.-a,14.20.Mr, 14.80.Sv}
\maketitle
%%%%%%%%%%%%%%%%%%%%%%%%%%%%%%%%%%%%%%%%%%%%%%%%%%%%%%%%%%%%%%%%%%%%%%%%%%%%%%%%
\section{Introduction}

Looking for physics beyond Standard Model (SM) is one of the the prime objectives of present day particle physics research. With no direct evidence of any kind of new physics (NP) signal at the LHC, much attention has been paid in recent times towards the various observed anomalies, which may be considered as smoking-gun signals of  NP and require thorough and careful investigation. In this context, semileptonic $B$ decays, both  the charged-current  $b \to c \ell \bar \nu_\ell$ as well as neutral-current  $b \to s \ell^+ \ell^-$ mediated transitions play a crucial role in probing the nature of physics beyond the SM.

 In the last few years, several  enthralling anomalies at the level of $(2-4)\sigma$ have been observed by the $B$-physics experiments, i.e.,   Belle \cite{Huschle:2015rga, Hirose:2016wfn,Abdesselam:2019wac,Abdesselam:2019dgh, Abdesselam:2019lab}, Babar \cite{Lees:2012xj, Lees:2013uzd} and LHCb \cite{Aaij:2013aln, Aaij:2013qta, Aaij:2014pli, Aaij:2014ora,  Aaij:2015yra, Aaij:2015esa, Aaij:2017vbb,  Aaij:2017uff, Aaij:2017tyk, Aaij:2019wad},  in the form of lepton flavour universality (LFU) violation in semileptonic $B$ decays associated with charged current  and neutral current   transitions. These discrepancies could be interpreted as hints of lepton flavour universality violation, which can't be accommodated in the SM and hence, suggest the necessity of NP contributions. 
 In the charged-current sector these observables are characterized by the ratio of  branching fractions $R_{D^{(*)}}\equiv {\rm Br}(B \to D^{(*)} \tau \bar \nu)/{\rm Br}(B \to D^{(*)} \ell \bar \nu_\ell)$, where $\ell =(e,\mu)$ and their present world average  values  $R_D^{\rm exp}=0.340 \pm 0.027 \pm 0.013$ and  $R_{D^*}^{\rm exp}=0.295 \pm 0.011 \pm 0.008$ from  Heavy Flavour Averaging Group (HFLAV)~\cite{Amhis:2019ckw}, have $3.1\sigma$ deviation (considering their correlation of $-0.38$) from their corresponding SM values. Analogous observable in the decay of  $B_c$ meson, symbolized by $R_{J/\psi}=0.71 \pm 0.17\pm 0.18$  \cite{Aaij:2017tyk} also exhibits $1.7\sigma$ discrepancy with its SM prediction.
Motivated by these results, a legion of studies have been performed from different points of view, e.g.,  revaluation of form factors in the SM predictions, studies to accommodate $R_{D^{(*)}}$ anomalies in a model-independent way as well as incorporating various  NP scenarios  and making use of  other observables to probe the NP effects, (see for example a representative list \cite{Iguro:2020cpg,Blanke:2018yud,Sahoo:2019hbu, Murgui:2019czp,Alok:2019uqc, Iguro:2018vqb,Robinson:2018gza,Sahoo:2016pet}
and references therein).

The dearth of evidence of similar deviations in semileptonic or leptonic  decays of  $K$ and $\pi$ mesons, or in electroweak precision observables, supports the idea in which the potential NP contribution responsible for LFU violation  is coupled  only to the  third generation fermions. Thus, for resolving the $R_{D^{(*)}}$ anomalies, it is generally presumed that only $b \to c \tau \bar \nu_\tau$ decay channel is sensitive to NP.  Hence, it is natural to expect that the same class of NP might also affect the related charged current transitions mediated through  $b \to u \tau \bar \nu_\tau$. In this regard, the study of $B \to (\pi,\rho, \omega)\tau \bar \nu$  and $B_s \to (K,K^*) \tau \bar\nu$  charged current processes, involving the  quark level transitions $b \to u \tau \bar \nu_\tau$ are quite enthralling and in this work, we would like to perform a detailed  analysis of these decay modes. Rather than considering any specific NP scenario, we adopt a model-independent approach, wherein we consider all possible Lorentz invariant terms in the effective Lagrangian, describing the process. Using the available experimental data  to constrain the possible new coefficients allows us to deduce the information on the nature of NP without any prejudice.  We then scrutinize the impact of these new coefficients on the branching fraction, forward-backward asymmetry, LFU observable and lepton polarization asymmetry of these decay modes. It should  be emphasized  that as  these modes are relatively rare due to Cabbibo suppression, the impact of NP  could be significant enough leading to observable effects in some of the observables. This in turn, leads the possibility that they could be observed at LHCb or Belle II experiments. Recently, some groups have looked into  these decay modes  in the context of various new physics scenarios~\cite{Rajeev:2018txm,Colangelo:2019axi,Sahoo:2020wnk,Colangelo:2020vhu, Kumbhakar:2020okw}.

%These ratios are advantageous both theoretically and experimentally, as 
%the  hadronic uncertainties involved, specifically from the CKM factors and  form factors cancelled out to a large extent.  

 The outline of the  paper is as follows. In Sec. II we present the required theoretical framework to calculate the decay rate and other observables sensitive to NP, starting from  the general effective Lagrangian containing new Wilson coefficients. Section-III deals with the constrained parameter space of the new physics couplings. Section-IV is comprised of the effect of NP on various parameters and their sensitivity towards NP. Here we show the $q^2$ variation of different observables and compute their numerical values. In Sec. V, we briefly comment on the effect of scalar leptoquark $R_2(3,2,7/6)$  and vector leptoquark $U_1(3,1,2/3)$ on these observables.  Finally, we conclude our work in Section-VI. 
  
\section{Theoretical Framework}
In effective field theory approach, the most general effective Hamiltonian  describing the transition $b\to u \tau \bar{\nu}_\tau$ is  expressed as \cite{Sakaki:2013bfa},
\beqa
{\cal H}_{eff}=\frac{4G_F}{\sqrt{2}}V_{ub}\big[(1+V_L)  O_{V_L}+V_R
 O_{V_R}+ S_L  O_{S_L} +S_R   O_{S_R}  +T_L  O_{T_L}\big],\label{Ham}
\eeqa
where $G_F$ is the Fermi constant, $V_{ub}$ is the CKM matrix element,  $ O_i$ are the dimension-six four fermion operators and  $V_L, V_R, S_L,  S_R, T_L$ are the corresponding new Wilson coefficients, which are zero in the SM. Here, we consider the neutrinos as left-chiral. The operator ${\cal O}_{V_L}$ corresponds to the  SM  operator having the usual $(V-A)\times(V-A)$ structure, whereas the other operators $ O_{V_R,S_L,S_R,T_L}$ arise only in some new physics scenarios. The explicit form of these operators are 
\bea
&&O_{V_L}=(\bar u \gamma_\mu P_L b)(\bar \tau \gamma^\mu P_L \nu),~~~~~
O_{V_R}=(\bar u \gamma_\mu P_R b)(\bar \tau \gamma^\mu P_L \nu),~~~~~
O_{S_L}=(\bar u  P_L b)(\bar \tau  P_L \nu),\nn\\
&&
O_{S_R}=(\bar u  P_R b)(\bar \tau  P_L \nu),~~~~~~~~~~~O_{T_L}=(\bar u \sigma_{\mu \nu} P_L b)(\bar \tau \sigma^{\mu \nu} P_L \nu),
\eea
where $P_{L,R}=(1 \mp \gamma_5)/2$ represent  the chiral projection operators. 

Including all  new physics operators of the effective Hamiltonian (\ref{Ham}), the differential decay distribution for the   $\bar B \to  P \tau \bar \nu$ processes (where $P$ denotes a psedoscalar meson), can be represented in terms of helicity amplitudes  \cite{Sakaki:2013bfa}
\bea
\frac{d\Gamma(\bar B \to  P \tau \bar \nu)}{dq^2} &=& {G_F^2 |V_{ub}|^2 \over 192\pi^3 m_B^3} q^2 \sqrt{\lambda_P(q^2)} \left( 1 - {m_\tau^2 \over q^2} \right)^2  \nn \\   & \times& \Bigg \lbrace \Big | 1 + V_L + V_R \Big |^2 \left[ \left( 1 + {m_\tau^2 \over 2q^2} \right) H_{V,0}^{s~2} + {3 \over 2}{m_\tau^2 \over q^2}  H_{V,t}^{s~2} \right] \nn \\ & +& {3 \over 2} \left |S_L + S_R \right |^2 \, H_S^{s~2} + 8 \left |T_L \right |^2 \left(1+ \frac{2m_\tau^2}{q^2} \right) H_T^{s~2} \nn \\ & +&3{\rm Re}\left[ ( 1 + V_L + V_R ) (S_L^* + S_R^* ) \right] {m_\tau \over \sqrt{q^2}} \, H_S^s H_{V,t}^s  \nn \\ & -&12{\rm Re}\left[ \left( 1 + V_L + V_R \right) T_L^* \right] \frac{m_\tau}{\sqrt{q^2}} H_T^s H_{V,0}^s  \Bigg \rbrace,  \label{br-exp}
\eea
where $q^2$ is the momentum transfer squared, $m_B (m_P)$ and $ m_\tau$  represent the masses of  $B (P)$ meson and $\tau$ lepton  respectively. $\lambda_P \equiv\lambda (m_B^2, m_P^2, q^2)=((m_B-m_P)^2-q^2)((m_B+m_P)^2-q^2)$, is the triangle function. $H_{V(0,t),S,T}^s$ are the  helicity amplitudes, related to the hadronic form factors $(f_{+,0,T})$  describing $B \to P$ transitions are  expressed as
\bea
&&H_{V,0}^s(q^2)= \sqrt{  \frac{\lambda_P(q^2)}{q^2}}f_+(q^2),~~~~~~~H_{V,t}^s(q^2)= \frac{m_B^2-m_P^2}{\sqrt{q^2}}f_0(q^2),\nn\\
&&H_S^s(q^2)\simeq \frac{m_B^2-m_P^2}{m_b-m_u}f_0(q^2),~~~~~~~
H_T^s(q^2)=- \frac{\sqrt{\lambda_P(q^2)}}{m_B+m_P}f_T(q^2)\;.\label{P-form}
\eea
Similarly, the differential decay distribution for $ B \to  V \tau  \bar\nu$ processes, where $V$ represents a vector meson,     in terms of the helicity amplitudes ($H_{i,\pm},~ H_{i,0}, H_{V,t}$, where ($i=V,T$)   is expressed as \cite{Sakaki:2013bfa}
\bea
 {d\Gamma( \bar B \to  V \tau \bar \nu) \over dq^2} &=& {G_F^2 |V_{ub}|^2 \over 192\pi^3 m_B^3} q^2 \sqrt{\lambda_{V} (q^2)} \left( 1 - {m_\tau^2 \over q^2} \right)^2 \nn  \\ &\times & \bigg \{  \left( \left|1 + V_L \right|^2 + \left| V_R\right|^2 \right)  \left[ \left( 1 + {m_\tau^2 \over 2q^2} \right) \left( H_{V, +}^2 + H_{V,-}^2 + H_{V,0}^2 \right) + {3 \over 2}{m_\tau^2 \over q^2} \, H_{V,t}^2 \right]  \nn \\ &-&  2{\rm Re}\left[\left(1+ V_L \right) V_R^* \right] \left[ \left( 1 + {m_\tau^2 \over 2q^2} \right) \left( H_{V,0}^2 + 2 H_{V,+} H_{V,-} \right) + {3 \over 2}{m_\tau^2 \over q^2} \, H_{V,t}^2 \right] \nn \\ &+&   {3 \over 2} |S_R - S_L|^2 \, H_S^2 + 8 |T_L|^2 \left(1+\frac{2m_\tau^2}{q^2} \right) \left(H_{T, +}^2 + H_{T, -}^2 + H_{T, 0}^2 \right) \nn \\ &+&   3{\rm Re}\left [ \left ( 1 + V_L - V_R \right) \left (S_R^* - S_L^* \right) \right ] {m_\tau\over \sqrt{q^2}} \, H_S H_{V,t} \nn \\ &-& 12{\rm Re}\left[ \left(1+V_L \right)T_L^* \right] \frac{m_\tau}{\sqrt{q^2}} \left(H_{T,0}H_{V,0}+H_{T,+}H_{V, +} - H_{T,-} H_{V, -} \right) \nn \\ &+& 
 12{\rm Re}\left[ V_R T_L^* \right ] \frac{m_\tau}{\sqrt{q^2}}  \left(H_{T,0}H_{V,0}+H_{T,+}H_{V, -} - H_{T,-} H_{V, +} \right) \bigg \}, 
\eea
where  $\lambda_{V}= ((m_B-m_V)^2-q^2)((m_B+m_V)^2-q^2)$. The relations between the helicity amplitudes  and the $B \to V$ form factors are depicted as
\bea
&&H_{V,\pm}(q^2)=(m_B+m_V) A_1(q^2) \mp \frac{\sqrt{\lambda_V(q^2)}}{m_B+m_V}V(q^2),\nn\\
&&H_{V,0}(q^2)= \frac{m_B+m_V}{2 m_V \sqrt{q^2}}\Big[-(m_B^2-m_V^2-q^2)A_1(q^2)+ \frac{\lambda_V(q^2)}{(m_B +m_V)^2}A_2(q^2)  \Big], \nn\\
&&H_{V,t}(q^2)= - \sqrt{\frac{\lambda_V(q^2)}{q^2}}A_0(q^2),~~~~H_S(q^2) \simeq - \frac{\sqrt{\lambda_V(q^2)}}{m_b+m_u}A_0(q^2),\nn\\
&& H_{T,\pm}(q^2)= \frac{1}{\sqrt{q^2}}\Big[ \pm(m_B^2-m_V^2)T_2(q^2) +\sqrt{\lambda_V(q^2)}T_1(q^2) \Big],\nn\\
&&H_{T,0}(q^2)=\frac{1}{2 m_V}\Big[-(m_B^2+3m_V^2-q^2)T_2(q^2) + \frac{\lambda_V(q^2)}{m_B^2-m_V^2}T_3(q^2) \Big ].
\eea
%The $q^2$ dependence of the form factors will be discussed for later for each processes separately.
 
In addition to branching fraction, other observables, which are sensitive to new physics are presented below:
\begin{itemize}
\item  Lepton flavour universality violating parameter:
\bea
R_{P,V}^{\tau/\ell}(q^2)= \frac{d\Gamma(B \to (P,V) \tau \bar \nu)/dq^2}{d\Gamma(B \to (P,V) \ell \bar \nu)/dq^2},~~~~~(\ell=e,\mu)
\eea
\item Forward-backward asymmetry of final $\tau$ lepton:
\bea
A_{\rm FB}(q^2)=\left ( \int_{-1}^0 d \cos \theta \frac{d^2 \Gamma}{d q^2 d \cos \theta}- \int_0^1 d \cos \theta \frac{d^2 \Gamma}{d q^2 d \cos \theta}\right )\Big {/}\frac{d \Gamma}{d q^2}\equiv\frac{ b_\theta(q^2)}{d \Gamma/d q^2},
\eea
where $\theta$  represents the angle between  $ \tau$ lepton and $B$ meson three-momenta,  in the rest frame of $\tau \bar \nu$. The expressions for $b_\theta(q^2)$ for $B \to (P,V) \tau \bar \nu$ processes are given as
\bea
b_\theta^P(q^2) &=& \frac{G_F^2 |V_{ub}|^2}{128 \pi^3 m_B^3}q^2 \sqrt{\lambda_P(q^2)}\left ( 1-\frac{m_\tau^2}{q^2} \right )^2 \Big\{|1+V_L+V_R|^2 \frac{m_\tau^2}{q^2} H_{V,0}^s H_{V,t}^s \nn\\
&+&{\rm Re}[ (1+V_L+V_R)(S_L^* +S_R^*)]\frac{m_\tau}{\sqrt{q^2}}H_S^s H_{V,0}^s -4 {\rm Re}[(S_L+S_R)T_L^* ]H_T^s H_S^s\nn\\
&-&4 {\rm Re} [(1+V_L+V_R)T_L^*] \frac{m_\tau}{\sqrt{q^2}}H_T^s H_{V,t}^s
\Big \},\\
b_\theta^V(q^2) &=& \frac{G_F^2 |V_{ub}|^2}{128 \pi^3 m_B^3}q^2 \sqrt{\lambda_V(q^2)}\left ( 1-\frac{m_\tau^2}{q^2} \right )^2 \Big\{
\frac{1}{2}(|1+V_L|^2 -|V_R|^2)(H_{V,+}^2-H_{V,-}^2)\nn\\
&+&|1+V_L-V_R|^2\frac{m_\tau^2}{q^2}H_{V,0} H_{V,t}+8 |T_L|^2 \frac{m_\tau^2}{q^2}(H_{T,+}^2-H_{T,-}^2) -4 {\rm Re}[(S_R-S_L)T_L^*]H_{T,0}H_S\nn\\
&+&{\rm Re}[(1+V_L-V_R)(S_R^*-S_L^*)]\frac{m_\tau}{\sqrt{q^2}} H_S H_{V,0}\nn\\
&-&4 {\rm Re}[(1+V_L)T_L^*] \frac{m_\tau}{\sqrt{q^2}}(H_{T,0}H_{V,t}+H_{T,+}H_{V,+}+H_{T,-}H_{V,-})\nn\\
&+&4{\rm Re}[V_R T_L^*]\frac{m_\tau}{\sqrt{q^2}}(H_{T,0}H_{V,t}+H_{T,+}H_{V,-}+H_{T,-}H_{V,+})
\Big\}.
\eea

\item  Tau polarization asymmetry:
\bea\nonumber
P_\tau(q^2)=\frac{d\Gamma(\lambda_\tau=1/2)/dq^2-d\Gamma(\lambda_\tau=-1/2)/dq^2}{d\Gamma(\lambda_\tau=1/2)/dq^2+d\Gamma(\lambda_\tau=-1/2)/dq^2}\;,
\eea
where $d\Gamma(\lambda_\tau=\pm1/2)/dq^2$ are the differential decay rates of $B \to (P,V)$ processes with the tau polarization,  $\lambda_\tau = \pm 1/2$.   
\item Longitudinal polarization of final $V$ meson:
\bea
F_L^V(q^2)=\frac{d\Gamma(\lambda_V=0)/dq^2}{d\Gamma/dq^2}\;,
\eea
where $d\Gamma(\lambda_V=0)/dq^2$ is the $B \to V$ differential  decay
rate with the polarization of the vector meson, $\lambda_V=0$.
The expressions for  $d\Gamma(\lambda_\tau=\pm 1/2)/dq^2$ and
$d\Gamma(\lambda_V=0)/dq^2$ are provided in the Appendix.
\end{itemize}
%%%%%%%%%%%%%%%%%%%%%%%%%%%%%%%%%%%%%
%{\color{red} We now briefly discuss about the possible beyond the SM scenarios, without going into any detail, which could potentially contribute to these new couplings.  
%\begin{itemize}
%\item The vector coupling $V_L$ can generally arise in various leptoquark scenarios \cite{Dorsner:2016wpm}, e.g., the vector leptoquark $U_1(3,1,2/3)$
%\end{itemize}}
 \section{Constraints on new physics coefficients}
Though there are no appreciable discrepancies observed in the observables associated with $b \to u \tau \bar \nu$ transitions, but there exist few measurements which show some tension with their SM predictions by more than one sigma.  One such confrontation  is  observed in the leptonic decay channel $B^- \to \tau^- \bar \nu_\tau$ where the measured branching fraction ${\rm Br}(B^- \to \tau \bar \nu)=(1.09 \pm 0.24)\times 10^{-4}$  \cite{Tanabashi:2018oca} shows a slight disagreement with its SM prediction ${\rm Br}(B^- \to \tau \bar \nu)|^{\rm SM}=(8.48 \pm 0.28)\times 10^{-5}$ \cite{Sahoo:2017bdx}.
Another discrepancy  is observed in the ratio of branching fractions ($R_\pi^\ell$), which is defined as
\bea
R_\pi^\ell= \frac{\tau_{B^0}}{\tau_{B^-}}\frac{{\rm Br}(B^- \to \tau \bar \nu)}{{\rm Br}(B^0 \to \pi^+ \ell^- \bar \nu_{\ell})},~~~~~(\ell=e, \mu)
\eea
where $\tau_{B^0}~(\tau_{B^-})$ represents the lifetime of $B^0~(B^-)$ meson. Using the measured values of these observables from \cite{Tanabashi:2018oca}, one can obtain
\bea
R_\pi^\ell |^{\rm Expt}=0.699 \pm 0.156,
\eea
which depicts nearly $1\sigma$ deviation from its SM prediction  $R_\pi^\ell |^{\rm SM}=0.583 \pm 0.055$.
The SM predicted branching ratio of the semileptonic decay ${\rm Br}(B^0 \to \pi^+ \tau ^- \bar \nu)|^{\rm SM}=(9.40 \pm 0.75) \times 10^{-5}$, is also considerably lower than its existing experimental upper limit ${\rm Br}(B^0 \to \pi^+ \tau ^- \bar \nu)< 2.5 \times 10^{-4}$ \cite{Tanabashi:2018oca}.

Considering the above observables, we have performed a  $\chi^2$-fit in \cite{Ray:2019gkv}   to constrain the new physics Wilson coefficients. Since there is no update in the values of these observables, we will use same constrained values of the  new coefficients, in this analysis. For completeness, the best-fit and $1\sigma$ allowed values of these coefficients are presented in Table I.  Since the observables, ${\rm Br}(B^- \to \tau^-\bar \nu_\tau)$ and $R_\pi^\ell$ are not sensitive to the tensor current, reliable constraint on tensor  coupling  would not  be possible to obtain, and hence,  we are not considering the effect of tensor contribution  in the analysis. 
 
\begin{table}[htb]
\centering
\label{Tab:con}
\begin{tabular}{|c|c|c|}
\hline
~New coefficients~&~Best-fit~&~$1\sigma$ range~~\\
\hline
\hline
~~$({\rm Re}[V_L], {\rm Im}[V_L]) $~~&~$(-0.915,1.108)$~&~$([-1.45,-0.65],~[1.02,1.19])$ ~\\
~$({\rm Re}[V_R], {\rm Im}[V_R]) $~&~$(-0.116,0)$~&~$([-0.205,-0.025],~[-0.41,0.41])$ ~\\
~$({\rm Re}[S_L], {\rm Im}[S_L]) $~&~$(-0.024,0)$~&~$([-0.042,-0.004],~[-0.092,0.092])$ ~\\
~$({\rm Re}[S_R], {\rm Im}[S_R]) $~&~$(-0.439,0.005)$~&~~$([-0.457,-0.421],~[-0.092,0.092])$ ~\\
\hline
\end{tabular}
\caption{Best-fit values and the corresponding $1\sigma$ ranges of new  coefficients associated with $b\to u \tau \bar{\nu}_\tau$ transition are taken from \cite{Ray:2019gkv}.}
\end{table}
\section{Results and Discussions}
Using the obtained  fit results on the new coefficients from Ref.  \cite{Ray:2019gkv}, we now proceed to investigate the impact of NP  on various observables of $B \to (P,V)\tau^- \bar \nu_\tau$ processes. For simplicity we will consider the effect of one NP operator at a time, and  discuss each decay process  individually in the following subsections. 

\subsection{$B^0 \to \pi^+ \tau^- \bar \nu_\tau$ decay process}
In order to analyze the decay distribution as well as other observables, we need to know the values of the hadronic form factors  in Eq. (\ref{P-form}), which describe the $B \to P$ transitions and are defined as 
\bea
&&\langle P(p_P) |\bar u \gamma^\mu b |\bar B(p_B) \rangle= f_+(q^2)\Big [(p_B+p_P)^\mu- \frac{m_B^2-m_P^2}{q^2} q^\mu \Big]+f_0(q^2) \frac{m_B^2-m_P^2}{q^2}q^\mu\;,\nn\\
&&\langle P(p_P)|\bar u b|\bar B(p_B) \rangle= (m_B+m_P)f_S(q^2)\;.
%&&\langle P(p_P) |\bar u i\sigma^{\mu \nu} b |\bar B(p_B) \rangle=\frac{2}{m_B+m_P}f_T(q^2)(p_B^\mu p_P^\nu-p_B^\nu p_P^\mu ),
\eea
%where $f_{i=+,0,S,T}(q^2)$ are the form factors.

For $B \to \pi$ transition, we use the BCL parametrization \cite{Bourrely:2008za}, which are given as  
\bea
f_+(q^2)= \frac{1}{(1-q^2/m_{B^*}^2)}\sum_{n=0}^{N-1} b_n^+\Big[ z^n - (-1)^{n-N} \frac{n}{N}~z^N\Big]\;,~~~~
f_0(q^2)= \sum_{n=0}^{N-1} b_n^0 z^n\;,
\eea
where $m_{B^*}=5.325$ is the $B^*$ meson mass and $b_n^{+,0}$ are the expansion coefficients. The expansion parameter is defined as
\bea
z\equiv z(q^2) = \frac{\sqrt{t_+-q^2}-\sqrt{t_+-t_0}}{\sqrt{t_+-q^2}+\sqrt{t_+-t_0}}\;,
\eea
where $t_+ =(m_B+m_\pi)^2$ and $t_0=(m_B+m_\pi)(\sqrt{m_B}-\sqrt{m_\pi})^2$. The expansion coefficients extracted from the combined fit to the experimental data of the $B \to \pi \ell \bar \nu_\ell$ $q^2$ distribution and the lattice results \cite{Lattice:2015tia, Bailey:2015nbd}:
\bea
&&b_0^+=0.419 \pm 0.013,~~b_1^+=-0.495 \pm0.054,~~b_2^+=-0.43\pm0.13,~~b_3^+=0.22 \pm 0.31,\nn\\
&&b_0^0=0.510 \pm 0.019,~~b_1^0=-1.700 \pm 0.082,~~~b_2^0=-1.53\pm0.19,~~b_3^0=4.52 \pm 0.83.
%&&b_0^T=0.393 \pm 0.017,~~b_1^T=-0.65 \pm 0.23,~~~~~b_2^T=-0.6\pm 1.5,~~~~~b_3^T=0.1 \pm 2.8.
\eea
As the lattice results are not available for the scalar form factor $f_S$, we use the equation of motion to relate it to $f_0$, i.e., $f_S(q^2)= f_0(q^2) (m_B-m_\pi)/(m_b-m_u)$.
%%%%%%%%%%%%%%%%%%%%%%%%%%%%%%%
\begin{figure}
\includegraphics[scale=0.4]{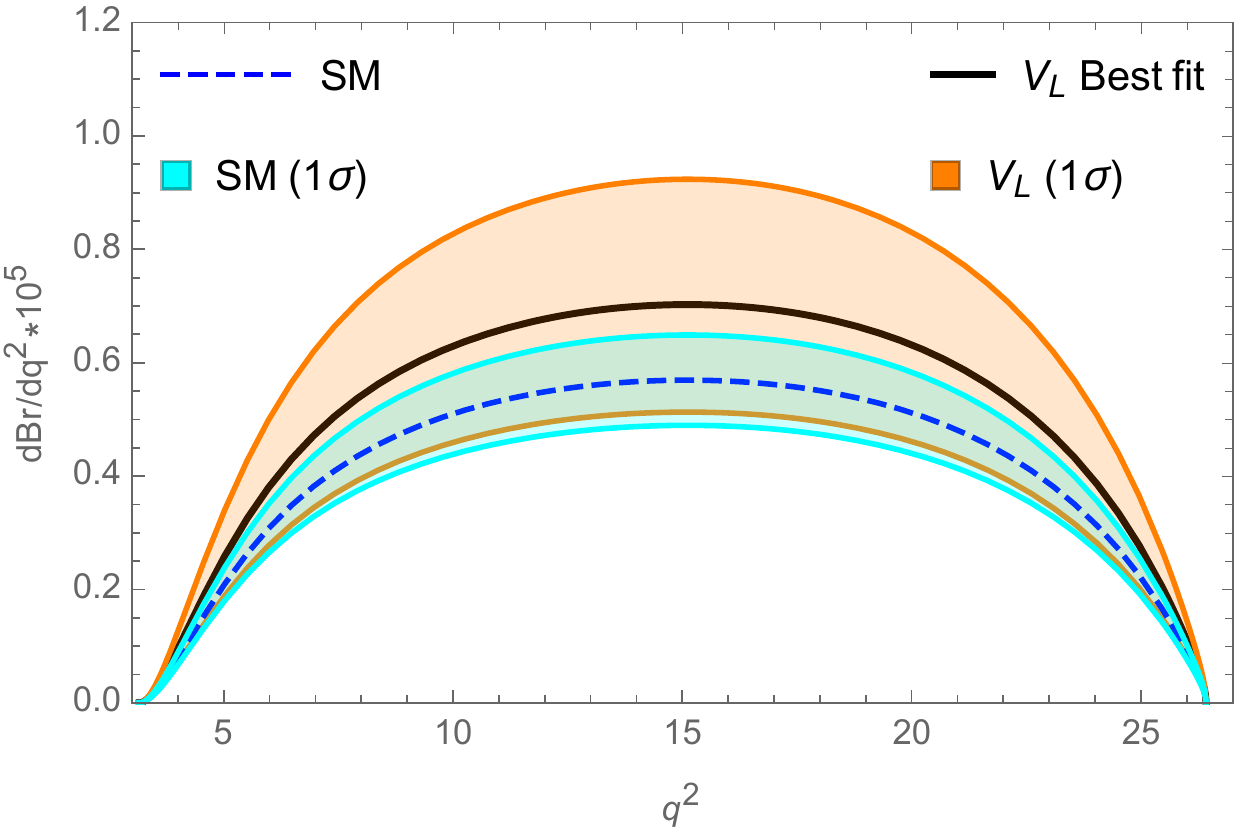}
\quad
\includegraphics[scale=0.4]{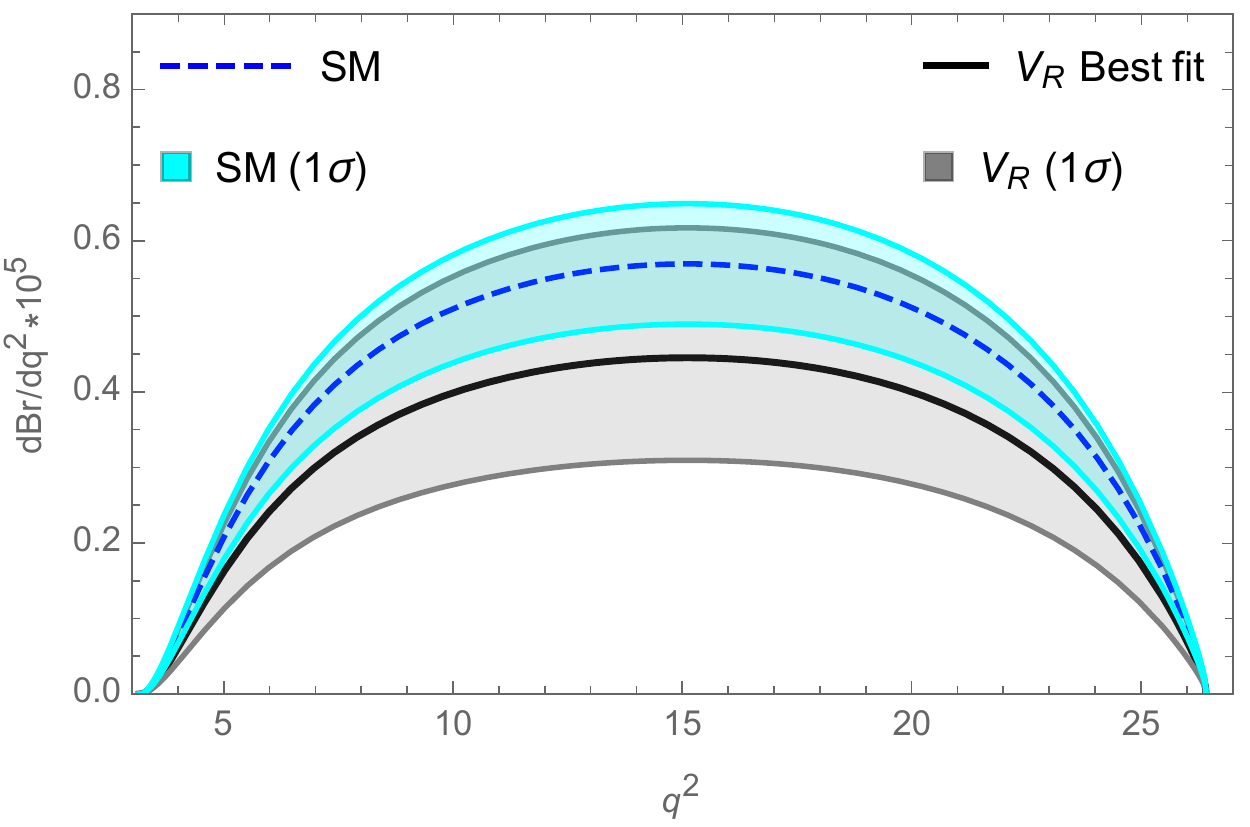}
\quad
\includegraphics[scale=0.4]{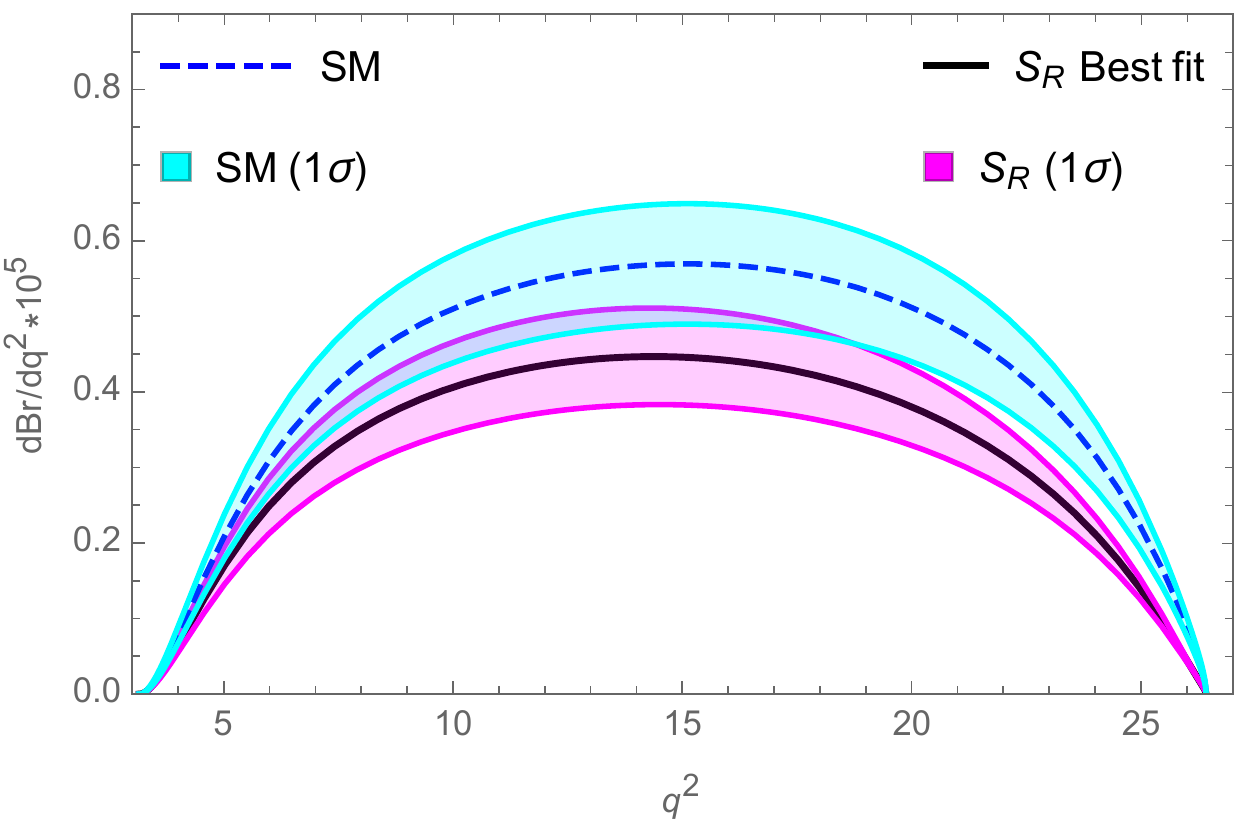}
\quad
\includegraphics[scale=0.4]{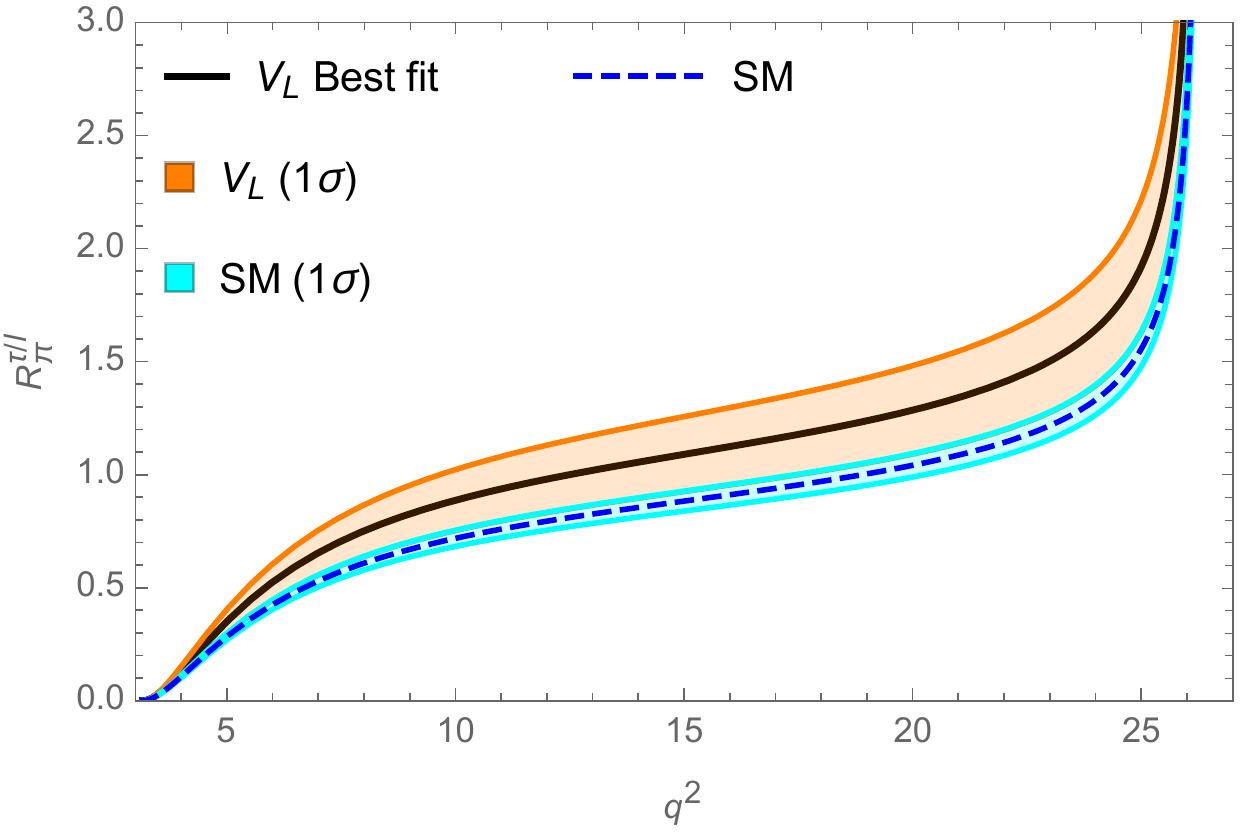}
\quad
\includegraphics[scale=0.4]{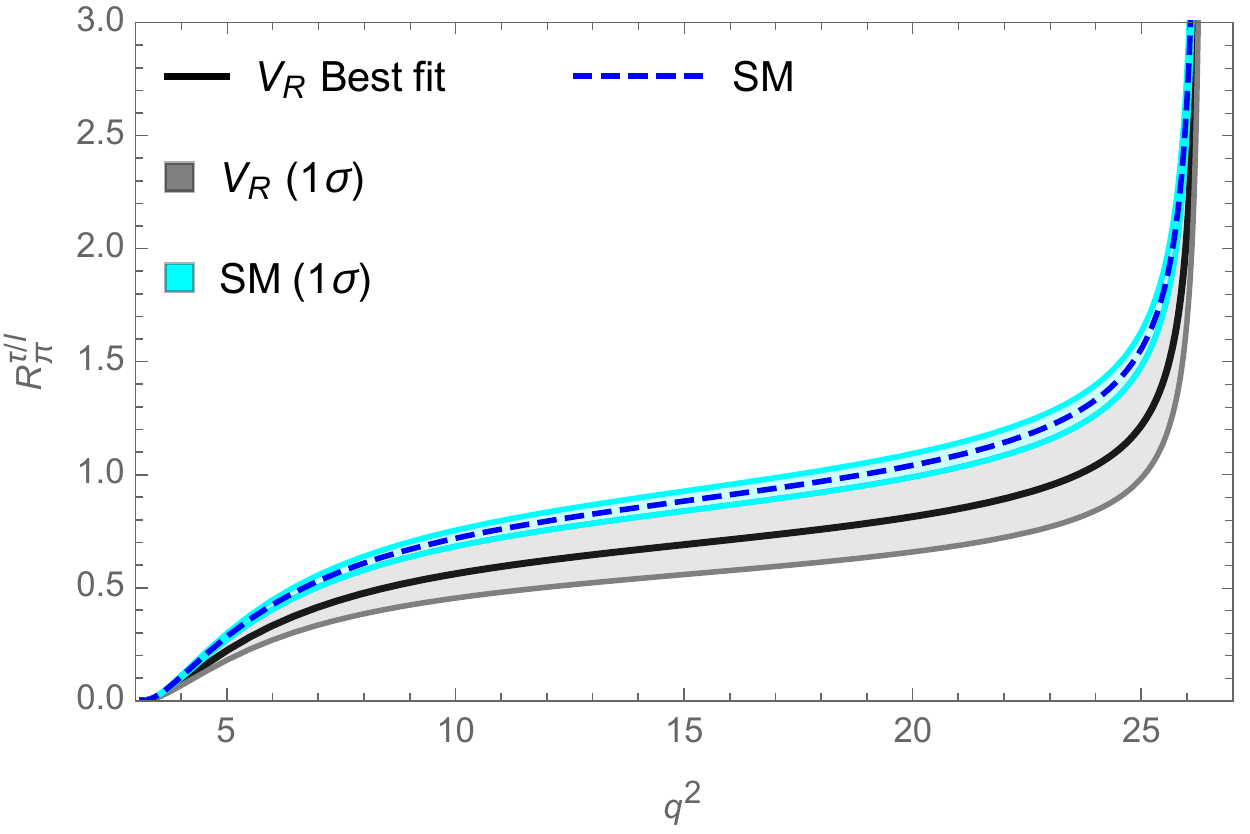}
\quad
\includegraphics[scale=0.4]{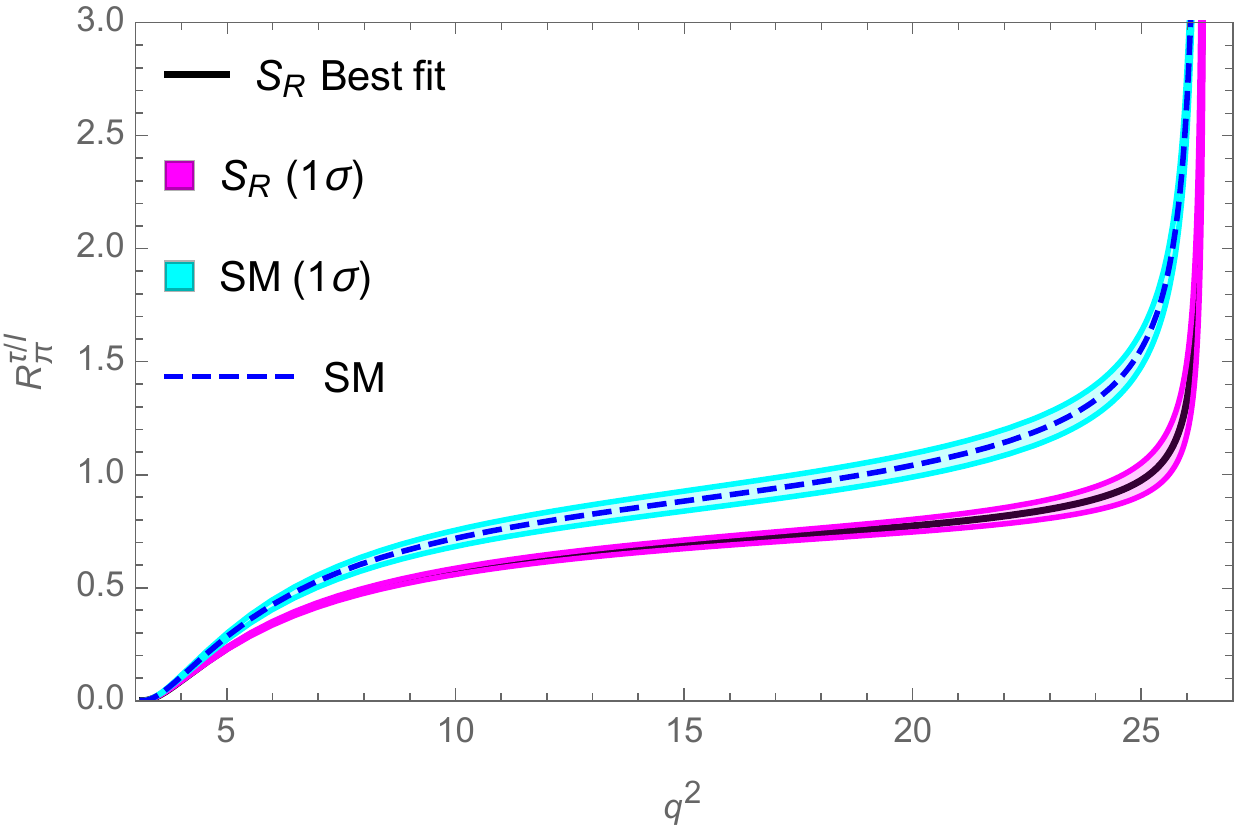}
\quad
\includegraphics[scale=0.4]{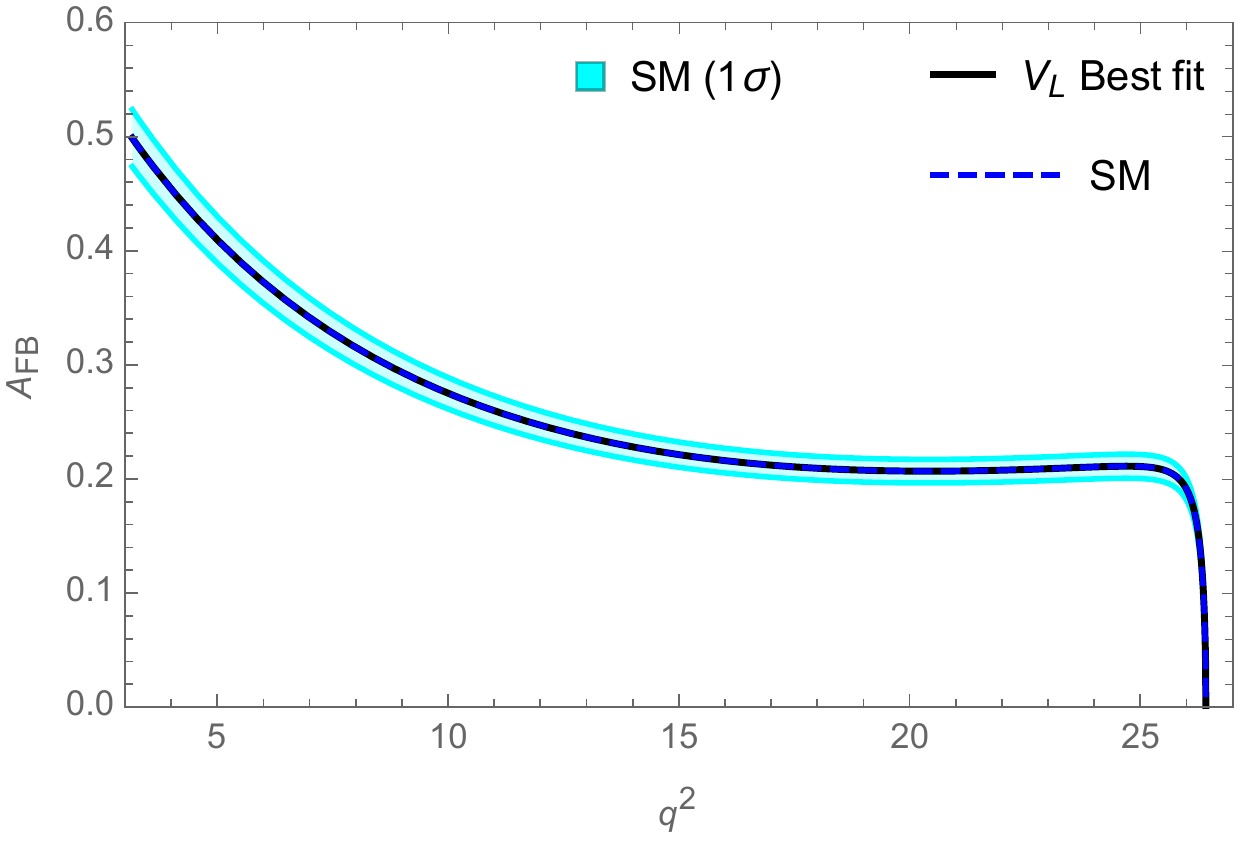}
\quad
\includegraphics[scale=0.4]{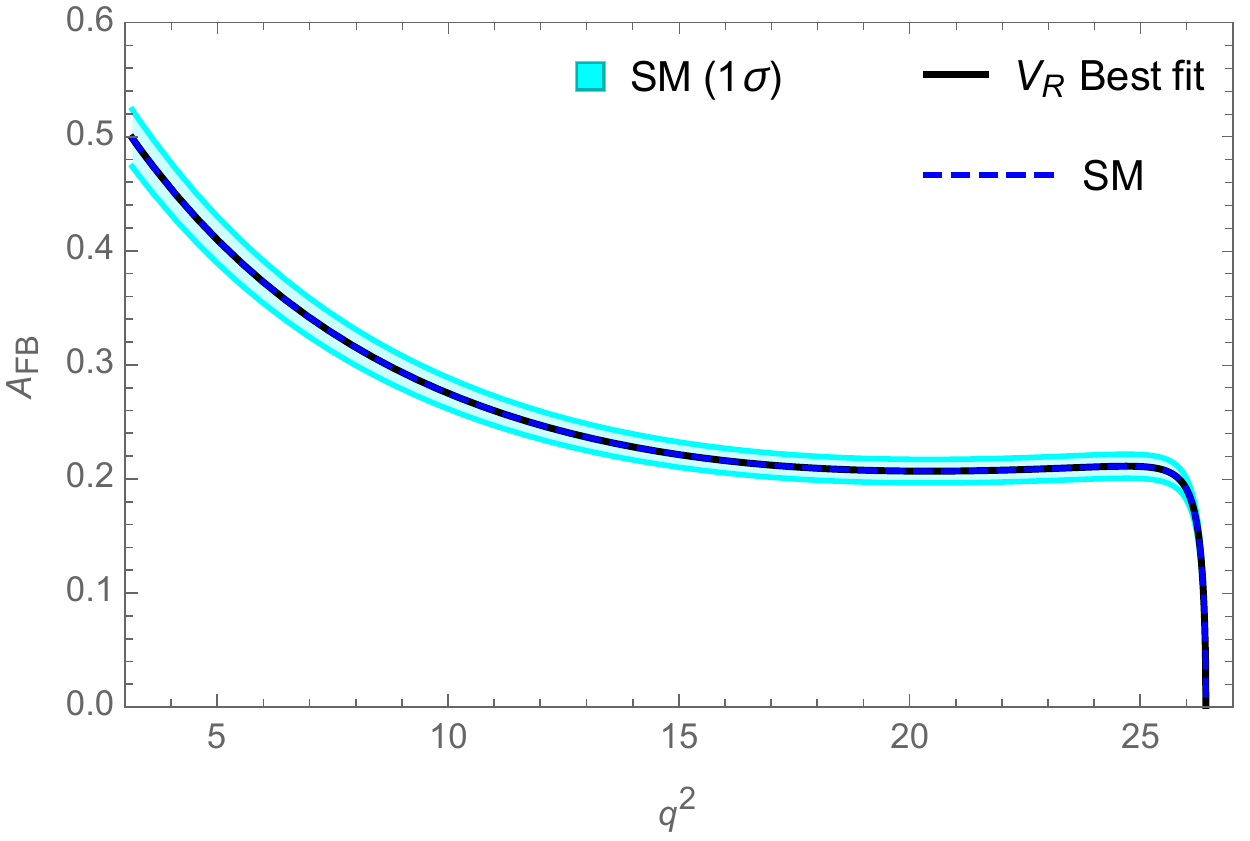}
\quad
\includegraphics[scale=0.4]{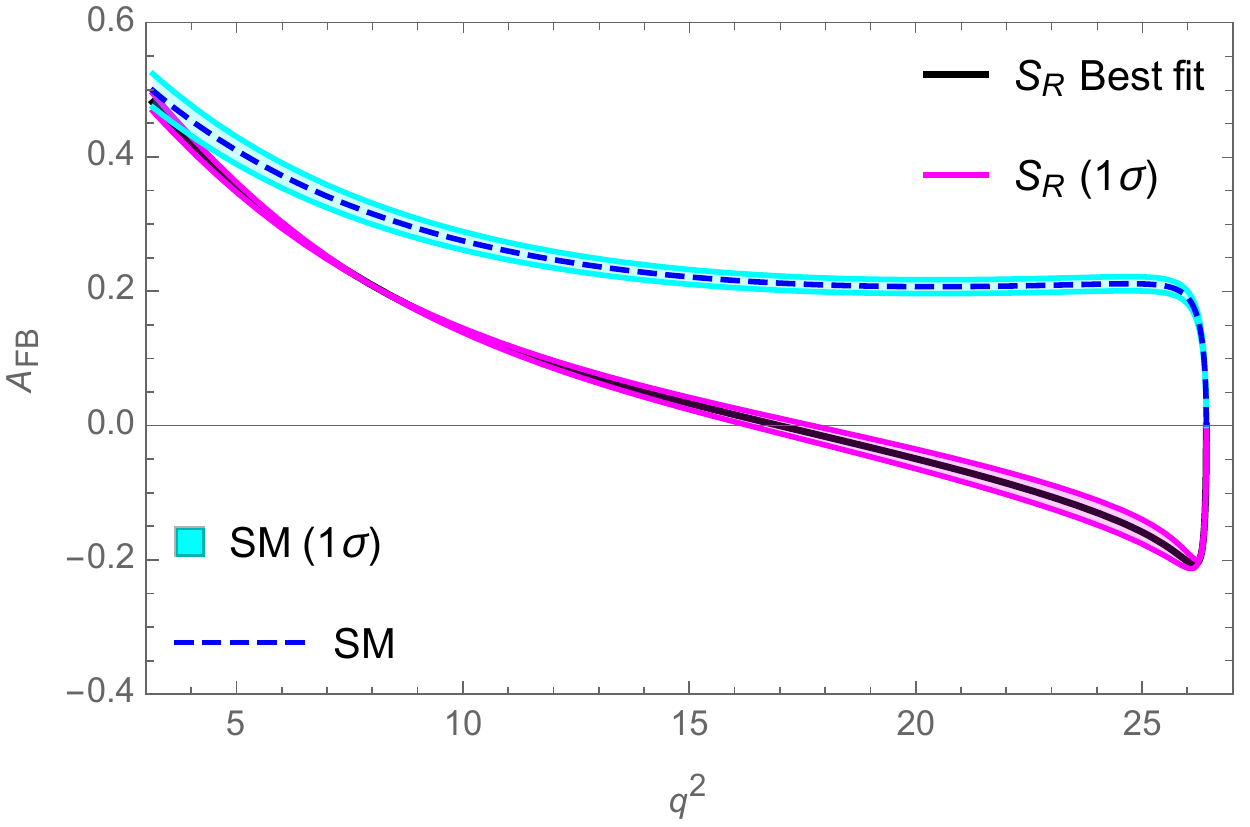}
\quad
\includegraphics[scale=0.4]{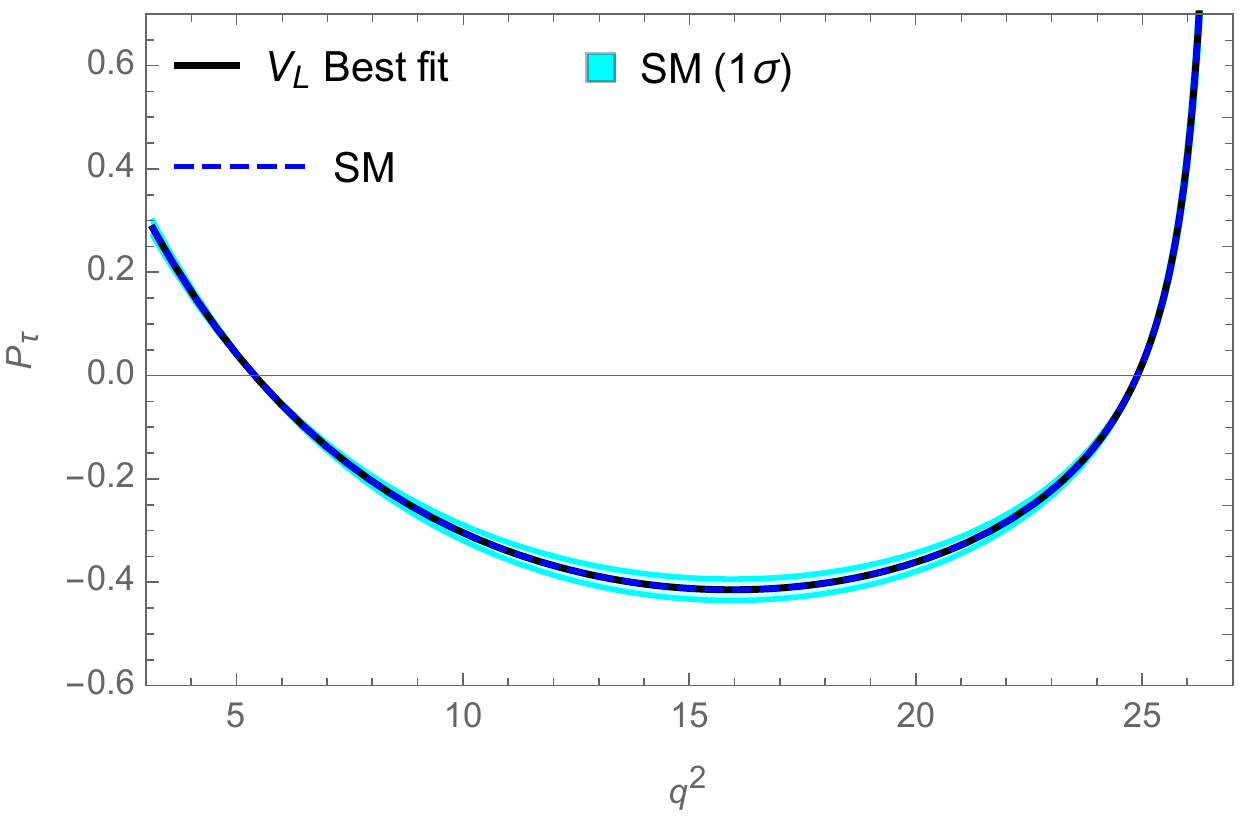}
\quad
\includegraphics[scale=0.4]{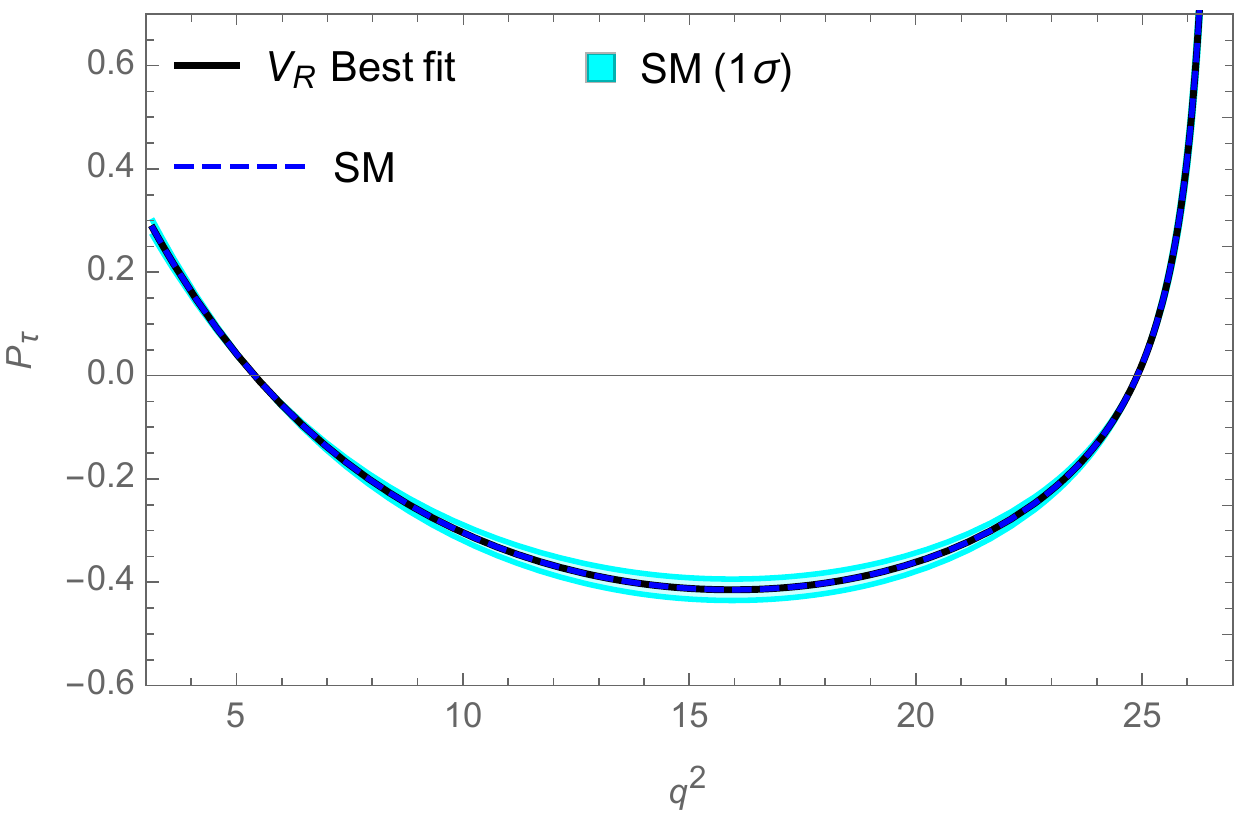}
\quad
\includegraphics[scale=0.4]{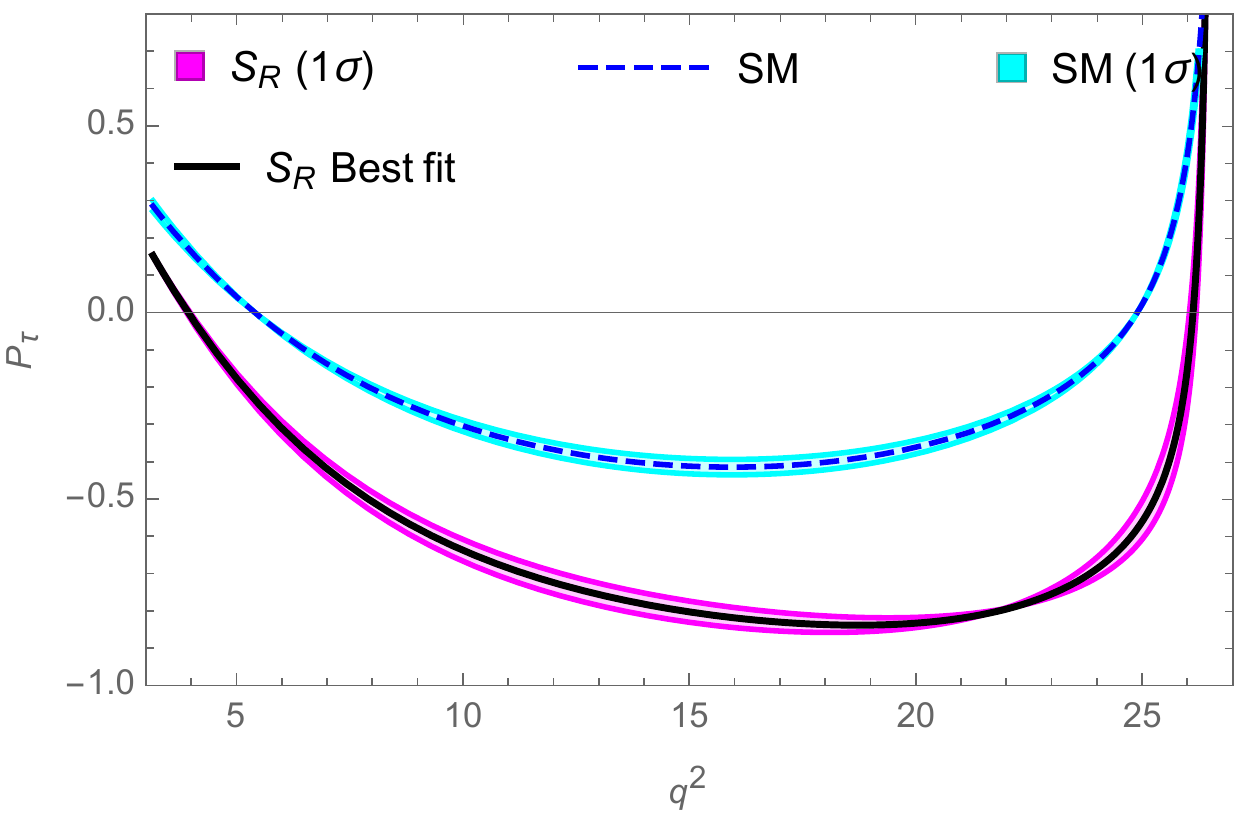}
\caption{The $q^2$ (in ${\rm GeV}^2$) variation of differential branching fraction, lepton non-universaity parameter, forward-backward asymmetry and tau-polarization asymmetry of $ \bar B^0 \to \pi^+ \tau^- \bar{\nu}$  process in the presence of  additional $V_L$ coupling (left panel), $V_R$ coupling (middle  panel) and  $S_R$  coefficient (right panel).  }
\label{B2pi}
\end{figure}

%%%%%%%%%%%%%%%%%%%%%%%%%%%%%%%%%
\begin{table}[h!]
\caption{Branching fractions of $B \to (P,V) \mu^- \bar\nu_\mu$ in the Standard Model.}
\begin{tabular}{|c|c|}
\hline
Decay Process & SM  Branching ratio \\
\hline
\hline
$~{\rm Br}(\bar B^0 \to \pi^+ \mu^- \bar \nu_\mu)$ ~&~ $ (1.533 \pm 0.215) \times 10^{-4}$~\\
\hline
${\rm Br}(\bar B^0 \to \rho^+ \mu^- \bar \nu_\mu)$ &~ $ (4.024 \pm 0.563) \times 10^{-4}$~\\
\hline
${\rm Br}( B^- \to \omega \mu^- \bar \nu_\mu)$ & ~$ (3.640 \pm 0.510) \times 10^{-4}$~\\
\hline
${\rm Br}( B_s \to K \mu^- \bar \nu_\mu)$ &~ $ (0.950 \pm 0.133) \times 10^{-4}$~\\
\hline
${\rm Br}( B_s \to K^* \mu^- \bar \nu_\mu)$ &~ $ (3.396 \pm 0.475) \times 10^{-4}$~\\
\hline
\end{tabular}
\label{Br:SM}
\end{table}
%%%%%%%%%%%%%%%%%%%%%%%%%%%%%%%%%%%%%%%%%
\begin{table}[h!]
\caption{Predicted values of branching fractions (in units of $10^{-4}$) and other observables for $B\to (P,V)\tau\bar{\nu}_\tau $ processes, both in the SM and NP scenarios.}
\begin{tabular}{|c|c|c|c|c|c|}
\hline
Observables & SM prediction & with $V_L$ NP & with $V_R$ NP & with $S_L$ NP & with $S_R$ NP \\
\hline
\hline
${\rm Br}(\bar B^0 \to \pi \tau \bar \nu)$ & $0.983 \pm 0.138$&$0.886 \to 1.596 $&$0.534 \to 1.066 $&$0.808 \to 1.116 $& $0.649 \to 0.857 $\\
$R_\pi^{\tau/\ell}$ &$0.641 \pm 0.127$&$0.578 \to 1.041$&$0.348 \to 0.695$&$0.527 \to 0.728$&$0.423 \to 0.559$\\
$A_{\rm FB}$ &$0.246 \pm 0.012$ &$0.234\to 0.258$ &$0.234\to 0.258$ &$0.237 \to 0.245$& $0.045 \to 0.067$\\
$P_{\tau}$ &$-(0.298\pm 0.015) $&$-(0.313 \to 0.283) $ &$-(0.313 \to 0.283) $ &$-(0.360 \to 0.304)$&
$-(0.698 \to 0.693)$\\
\hline
\small{${\rm Br}(\bar B^0 \to \rho \tau \bar \nu)$} & $2.142 \pm 0.300$&$1.930 \to 3.475 $&  $1.903 \to 3.191 $& $1.844 \to 2.479 $&$1.661 \to 2.192 $\\
$R_\rho^{\tau/\ell}$ &$0.532 \pm 0.105$& $0.480 \to 0.864$& $0.473 \to 0.793$& $0.458 \to 0.616$& $0.413 \to 0.545$\\
$A_{\rm FB}$ &$-(0.178 \pm 0.009)$ &$-(0.187\to 169)$ & $-(0.166 \to 0.080)$ & $-(0.177 \to 0.168) $& $-(0.287 \to 0.279)$\\
$P_{\tau}$ &$-(0.544\pm 0.027) $&$-(0.571 \to 0.517)$ & $-(0.54 \to 0.52)$ &  $-(0.542 \to 0.521)$&
 $-(0.720 \to 0.712)$\\
$F_{L}^\rho$ &$0.502\pm 0.025 $&$0.477\to 0.527  $ & $0.510 \to 0.557 $ &     $0.502 \to 0.509$&$0.445 \to 0.457$\\
\hline
%%%%%%%%%   B -> omega tau nu%%%%%%%%
\small{${\rm Br}(B^- \to \omega \tau \bar \nu)$} & $1.948 \pm 0.273$&$1.755 \to 3.161 $&  $1.731 \to 2.905 $& $1.678 \to 2.255 $&$1.506 \to 1.988 $\\

$R_\omega^{\tau/\ell}$ &$0.535 \pm 0.106$&    $0.482 \to 0.868$&    $0.475 \to 0.798$& $0.461 \to 0.619$& $0.414 \to 0.546$\\

$A_{\rm FB}$ &$-(0.119\pm 0.006)$ &$-(0.125 \to 0.113)$ & $-(0.111 \to 0.054)$ &    $-(0.119\to 0.112)$& $-(0.194 \to 0.189)$\\

$P_{\tau}$ &$-(0.538\pm 0.027) $&$-(0.565\to 0.511)$ & $-(0.534 \to 0.514)$ &  $-(0.537 \to 0.515)$&
 $-(0.719 \to 0.711)$\\
 
 $F_{L}^\omega$ &$(0.498 \pm 0.025) $&$0.473 \to 0.523$ & $0.506 \to 0.552 $ &     $0.498 \to 0.505$&$0.434 \to 0.441$\\
\hline
%%%%%%%%%%% B_s to K tau nu   %%%%%%%%

\small{${\rm Br}(B_s \to K \tau \bar \nu)$ }& $0.729 \pm 0.102 $ &$0.657 \to 1.183$  &  $0.396 \to 0.790 $  &  $0.596 \to 0.827 $&$0.456 \to 0.604 $\\

$R_K^{\tau/\ell}$ &$0.767 \pm 0.152$&     $0.692 \to 1.245$&    $0.417 \to 0.832$& $0.627\to 0.870 $ & $0.48 \to 0.636$\\

$A_{\rm FB}$ &$(0.253 \pm 0.013)$ &$0.24 \to  0.266$ & $0.24 \to  0.266 $&    $0.245 \to 0.253 $& $0.048 \to 0.071 $\\

$P_{\tau}$ &$-(0.244 \pm 0.012) $&$-(0.256 \to 0.232)$ & $-(0.256 \to 0.232)$ &  $-(0.309\to 0.250)$&
 $-(0.713 \to 0.710)$\\
 \hline
 %%%%%%% Bs to K* tau nu %%%%%%%%%%%%%%%%%
 ${\rm Br}(B_s \to K^* \tau \bar \nu)$ & $1.817 \pm 0.254 $ &$1.637 \to 2.949$  &  $1.614\to 2.708 $  &  $1.565\to 2.10 $&$1.423 \to 1.879 $\\

$R_K^{\tau/\ell}$ &$0.535 \pm 0.106$&     $0.482 \to 0.868$&    $0.475 \to 0.797$& $0.461\to 0.618$ & $0.419\to 0.553$\\

$A_{\rm FB}$ &$-(0.130\pm 0.006)$ &$-(0.136 \to 0.124)$ & $-(0.122 \to 0.064) $&    $-(0.130 \to 0.124) $& $-(0.197 \to 0.192) $\\

$P_{\tau}$ &$-(0.565 \pm 0.028) $&$-(0.593 \to 0.537)$ & $-(0.561 \to 0.543)$ &  $-(0.563 \to 0.544)$&
 $-(0.726 \to 0.719)$\\
 
 $F_{L}^{K^*}$ &$(0.481\pm 0.024) $&$0.457 \to 0.505 $ & $0.489 \to 0.534 $ &  $0.481\to 0.487$&
 $0.427 \to 0.430$\\
 \hline
\end{tabular}
\label{Results}
\end{table}
%%%%%%%%%%%%%%%%%%%%%%%%%%%%%%%%%%%%%%%%%
%%%%%%%%%%%%%%%%%%%%%%%%%%%%%%%%%%%%
Using these form factors, and the other input parameters e.g., the particle masses, lifetime of $B$ meson from \cite{Tanabashi:2018oca}, the branching fraction, $R_\pi^{\tau/\ell}$, forward-backward asymmetry and lepton polarization asymmetry parameters for $\bar B^0 \to \pi^+ \tau^- \bar \nu_\tau$ process are studied for various NP scenarios. The SM predicted branching ratio of $\bar B^0 \to \pi^+ \mu^- \bar \nu_\mu$ decay mode is   presented in Table \ref{Br:SM}. The graphical representation of our results is displayed  in Fig. \ref{B2pi}, where we have shown the $q^2$ variation of various observables in  different NP frameworks. The plots in the left panel (from top to bottom) represent the variation of differential branching fraction,  LFU violating parameter,  forward-backward asymmetry and the tau-polarization asymmetry respectively. In these plots, the blue-dashed lines correspond to SM result with central values of the input parameters, while the cyan band in the differential branching fraction plot is due to $1 \sigma$ uncertainties in the form factor, CKM matrix element and other input parameters. The black solid lines depict the contribution from  $V_L$ type NP (best-fit value), while the orange bands denote the  corresponding $1\sigma$ uncertainties. Analogously, the results for $V_R$ type NP coupling are shown in the plots of the middle panel, while the plots in the right panel are for $S_R$ coupling and  the colour-coding of these plots are provided in the plot legends. From the figures it should be noted that the branching fraction  and the $R_\pi^{\tau/\ell}$ observable deviate substantially from their SM predictions. The interesting point to be noted  from these plots is that, due to the NP contribution from $V_L$ type coupling, the values of these observables are enhanced with respect to their SM results, whereas they are reduced for $V_R$ and $S_R$ couplings. The forward-backward asymmetry and $P_\tau$ observables are insensitive to $V_L$ and $V_R$ couplings, while they differ considerably from their SM values for $S_R$  coupling. So we have not shown explicitly the $1\sigma$ uncertainties of these observables in the plots due to $V_L$ and $V_R$ couplings. 
Since  these observables behave  quite differently in various NP scenarios, their  
measurements will definitely shed light on the nature of the NP. Furthermore, as the effect of $S_L$ coupling is very nominal, the corresponding plots are not displayed explicitly. However, the integrated values of these observables in all four NP scenarios are presented in Table \ref{Results}.

%%%%%%%%%%%%%%%%%%%%%%%%%%%%%%%%%%%%%%%%%
\begin{table}
\begin{tabular}{|c|c|c|c|}
\hline
&~~$B \to \rho$~~ & ~~$B \to \omega$ ~~& ~~$B_s \to K^*$ ~~\\
\hline
$a_0^{V} $ ~~&~~ $0.33 \pm 0.03 $ ~~&~~ $0.30 \pm 0.04 $ ~~& ~~$ 0.30 \pm 0.03$\\
$a_1^{V} $ ~~&~~ $-0.86 \pm 0.18 $ ~~&~~ $-0.83 \pm 0.29 $ ~~& ~~$- 0.90 \pm 0.27$\\
$a_2^{V} $ ~~&~~ $1.80 \pm 0.97 $ ~~&~~ $1.72 \pm 1.24 $ ~~& ~~$ 2.65 \pm 1.33$\\
\hline
$a_0^{A_0} $ ~~&~~ $0.36 \pm 0.04 $ ~~&~~ $0.33 \pm 0.05 $ ~~& ~~$ 0.31 \pm 0.05$\\
$a_1^{A_0} $ ~~&~~ $-0.83 \pm 0.20 $ ~~&~~ $-0.83 \pm 0.30 $ ~~& ~~$- 0.66 \pm 0.23$\\
$a_2^{A_0} $ ~~&~~ $1.33 \pm 1.05 $ ~~&~~ $1.42 \pm 1.25 $ ~~& ~~$ 2.57 \pm 1.44$\\
\hline
$a_0^{A_1} $ ~~&~~ $0.26 \pm 0.03 $ ~~&~~ $0.24 \pm 0.03 $ ~~& ~~$ 0.23 \pm 0.03$\\
$a_1^{A_1} $ ~~&~~ $0.39 \pm 0.14 $ ~~&~~ $0.34 \pm 0.24 $ ~~& ~~$ 0.27 \pm 0.19$\\
$a_2^{A_1} $ ~~&~~ $0.16 \pm 0.41 $ ~~&~~ $0.09 \pm 0.57 $ ~~& ~~$ 0.13 \pm 0.56$\\
\hline
$a_0^{A_{12}} $ ~~&~~ $0.30 \pm 0.03 $ ~~&~~ $0.27 \pm 0.04 $ ~~& ~~$ 0.23 \pm 0.03$\\
$a_1^{A_{12}} $ ~~&~~ $0.76 \pm 0.20 $ ~~&~~ $0.66 \pm 0.26 $ ~~& ~~$ 0.60 \pm 0.21$\\
$a_2^{A_{12}} $ ~~&~~ $0.46 \pm 0.76 $ ~~&~~ $0.28 \pm 0.98 $ ~~& ~~$ 0.54 \pm 1.12$\\
\hline
\end{tabular}
\caption{Values of the various expansion coefficients ($a_k^i$) for $B\to \rho$, $B \to \omega$ and $B_s \to K^*$ processes.}
\label{Table:FF}
\end{table}
%%%%%%%%%%%%%%%%%%%%%%%%%%%%%%%%%%%%%%%%%
%%%%%%%%%%%%%%%%%%%%%%%%%%%%%%%%%%%%%%%
\subsection{$B \to (\rho,\omega) \tau \bar \nu$ decay}
The  matrix elements of the vector and scalar currents associated with $\bar{B}\to V \ell \bar{\nu}_\ell$ decay process can be expressed as,
\begin{eqnarray}
&&\langle V(k,\varepsilon)|\bar{u}\gamma _\mu b|\bar{B}|(p_B)\rangle =-i\epsilon_{\mu\nu\alpha\beta}\varepsilon^{\nu*}p_B^\alpha k^\beta \frac{2V(q^2)}{m_{B}+m_V},\nn\\
&&\langle V(k,\varepsilon )|\bar{u}\gamma_\mu \gamma_5 b|\bar{B}|(p_B)\rangle  =\varepsilon_\mu^{ *}(m_{B}+m_V)A_1(q^2)-(p_B+k)_\mu (\varepsilon^*\cdot  q)\frac{A_2(q^2)}{m_{B}+m_V}\nn\\ 
&&~~~~~~~~~~~~~~~~~\quad \quad~~~~~~~~~~~ -q_\mu(\varepsilon^*\cdot q)\frac{2m_{V}}{q^2}\left[A_3(q^2)-A_0(q^2)\right],\nn\\
&&
\langle V(k,\varepsilon )|\bar{u} \gamma_5 b|\bar{B}|(p_B)\rangle =-\frac{1}{m_b+m_u}q_\mu \langle V(k,\epsilon )|\bar{u}\gamma^\mu \gamma^5 b|\bar{B}|(p_B)\rangle \nn\\
& &~~~~~~~~~~~~~~~~~~~~~~~~~~~~~= -(\varepsilon^*\cdot q)\frac{2m_V}{m_b+m_u}A_0(q^2).\label{FFV}
\end{eqnarray}
%%%%%%%%%%%%%%%%%%%%%%%%%%%%%%%%%%%%%%%%%%
\begin{figure}
\includegraphics[scale=0.4]{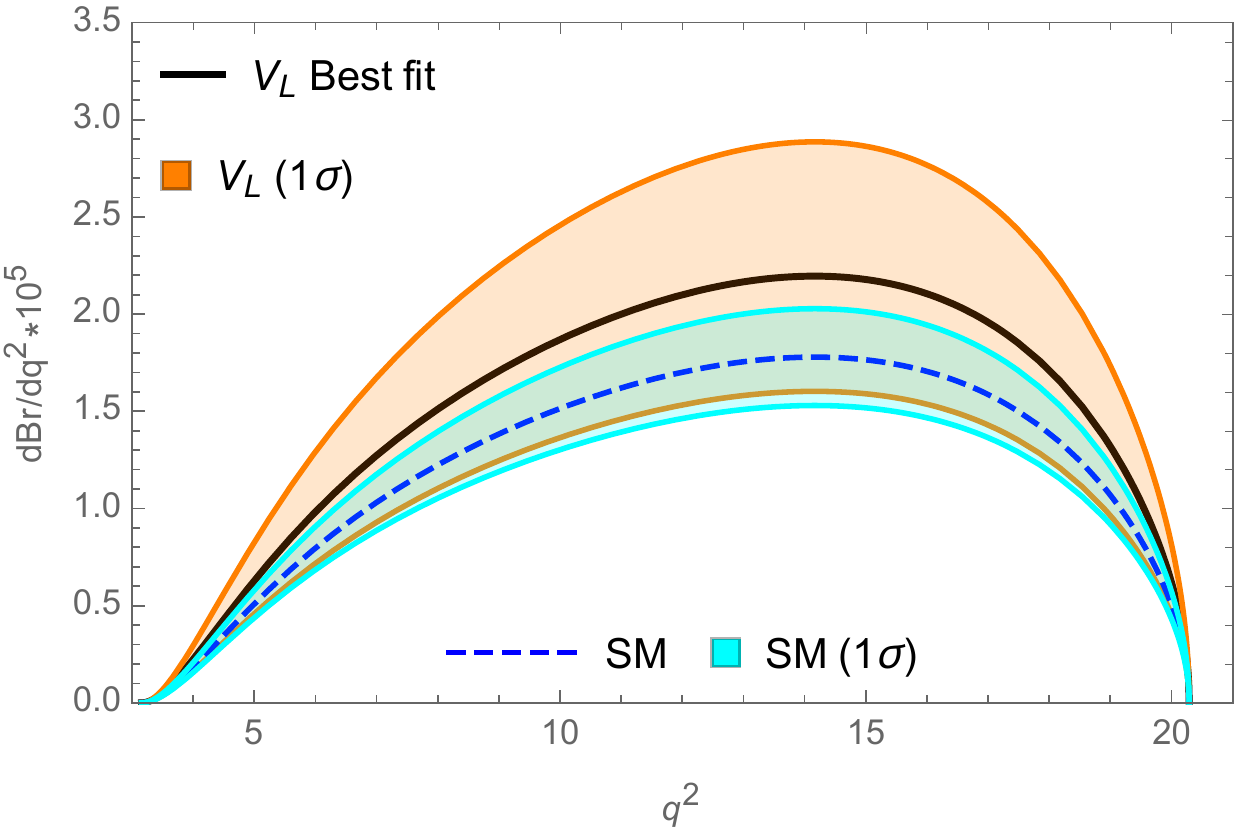}
\quad
\includegraphics[scale=0.4]{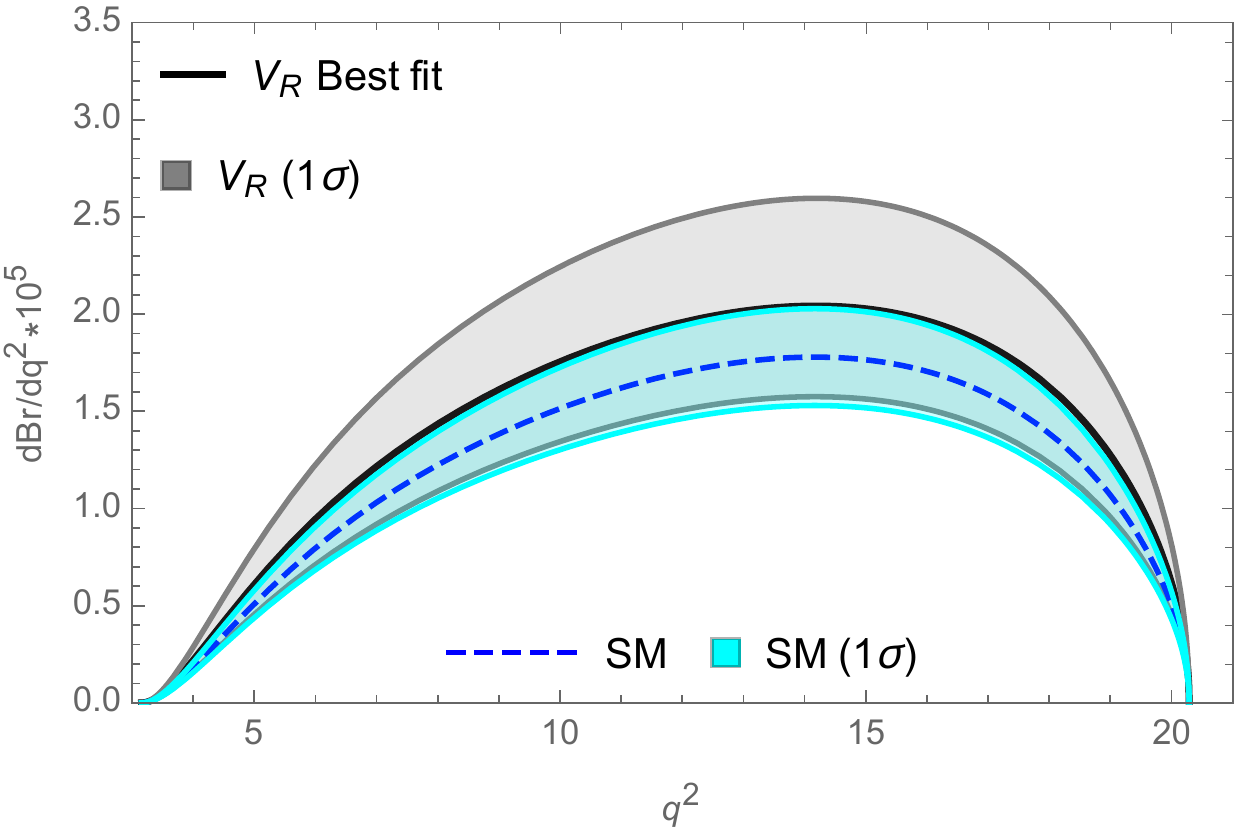}
\quad
\includegraphics[scale=0.4]{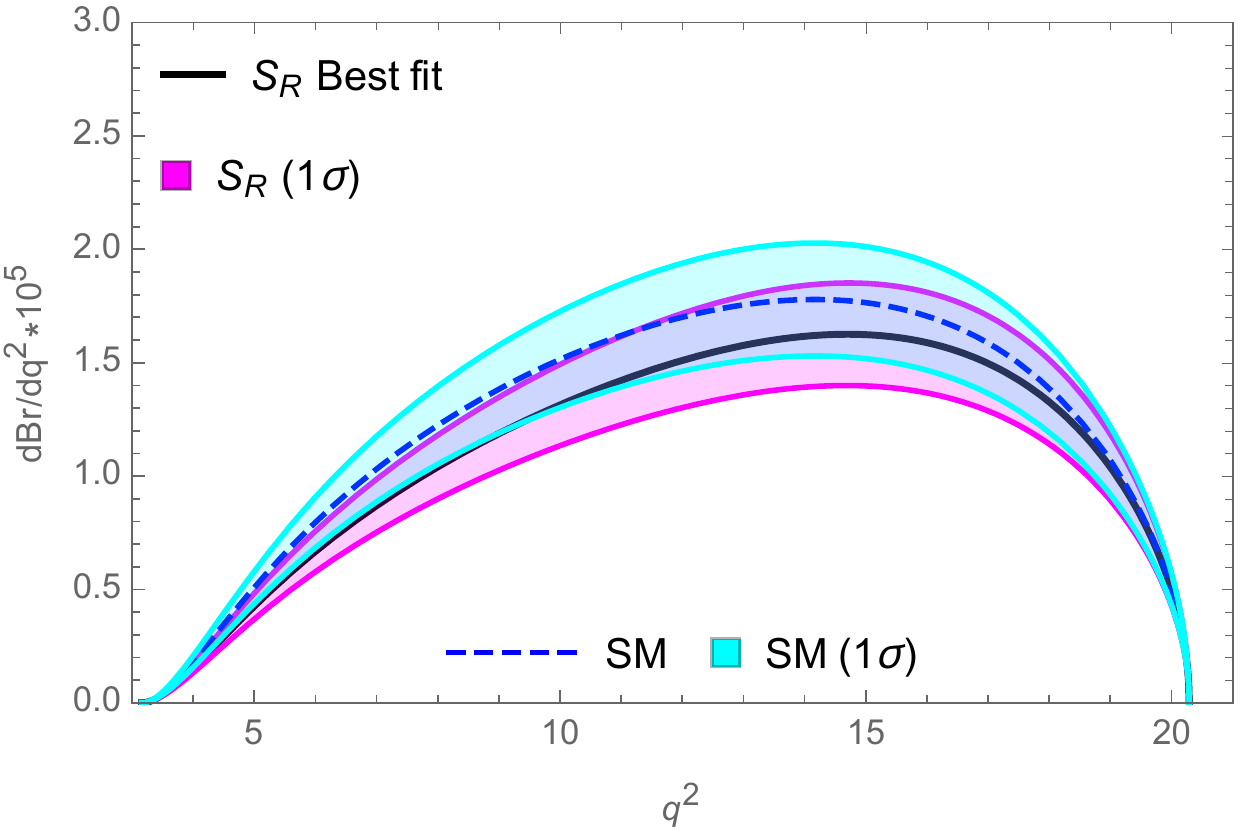}
\quad
\includegraphics[scale=0.4]{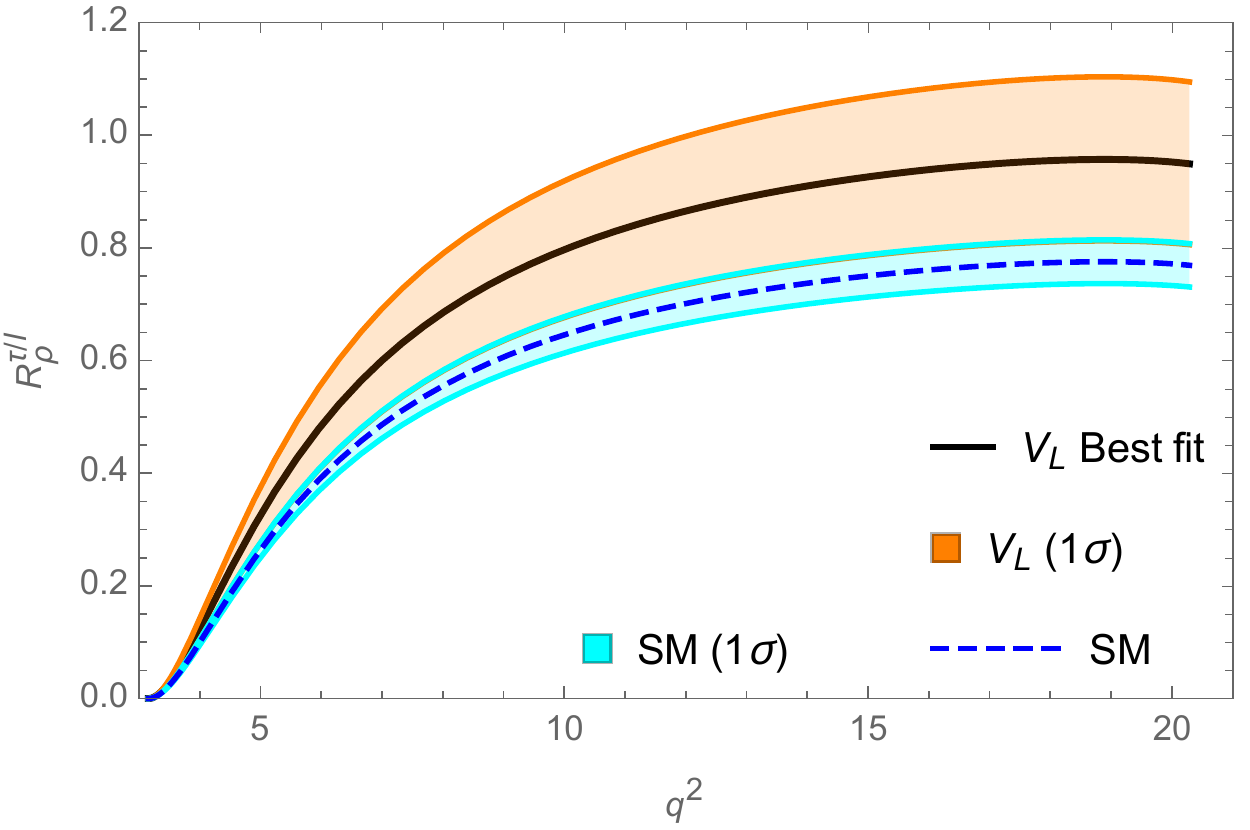}
\quad
\includegraphics[scale=0.4]{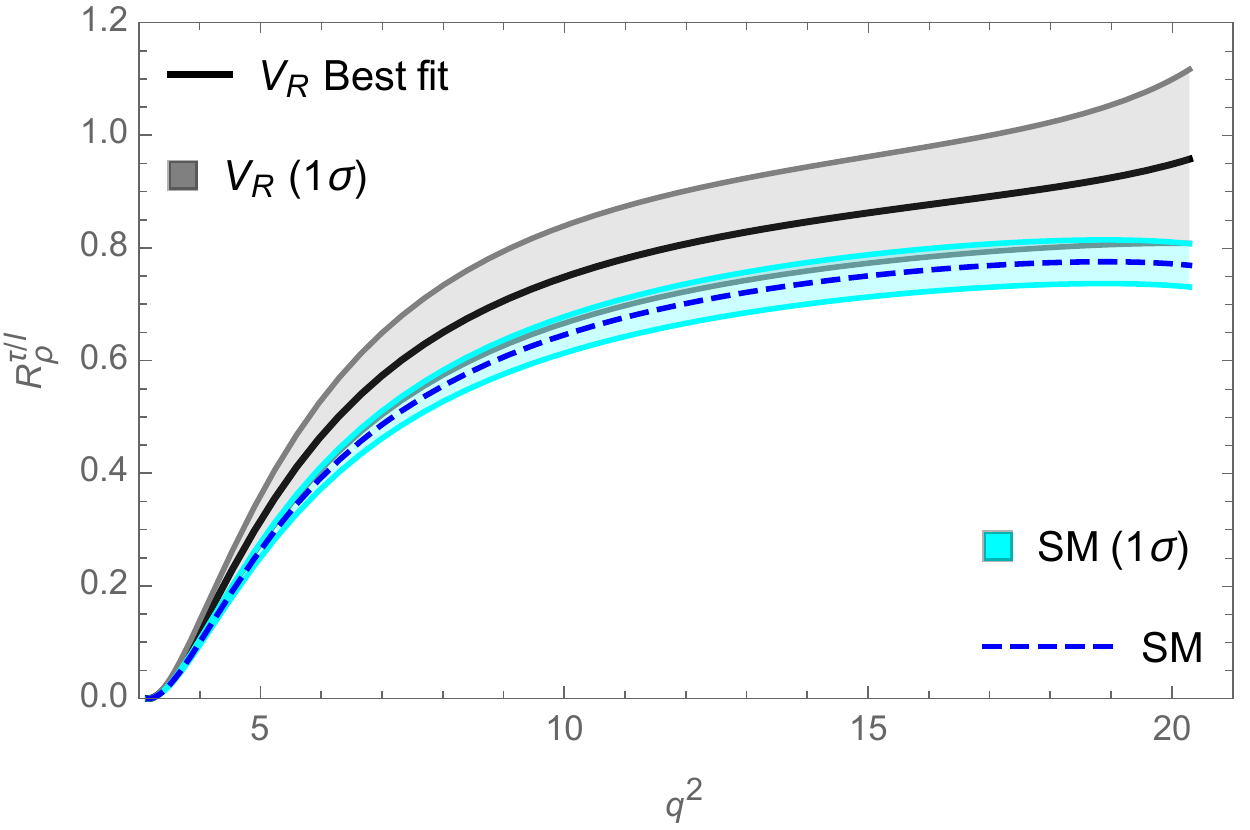}
\quad
\includegraphics[scale=0.4]{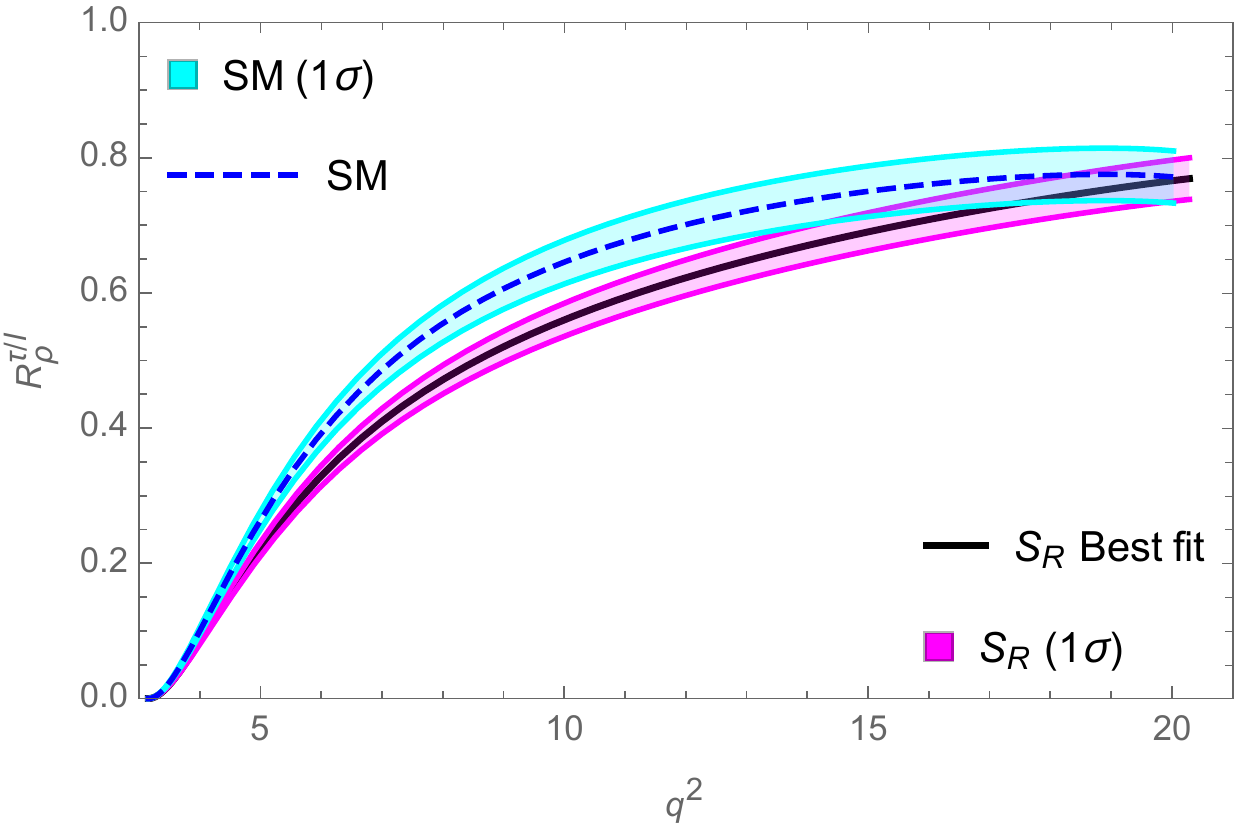}
\quad
\includegraphics[scale=0.4]{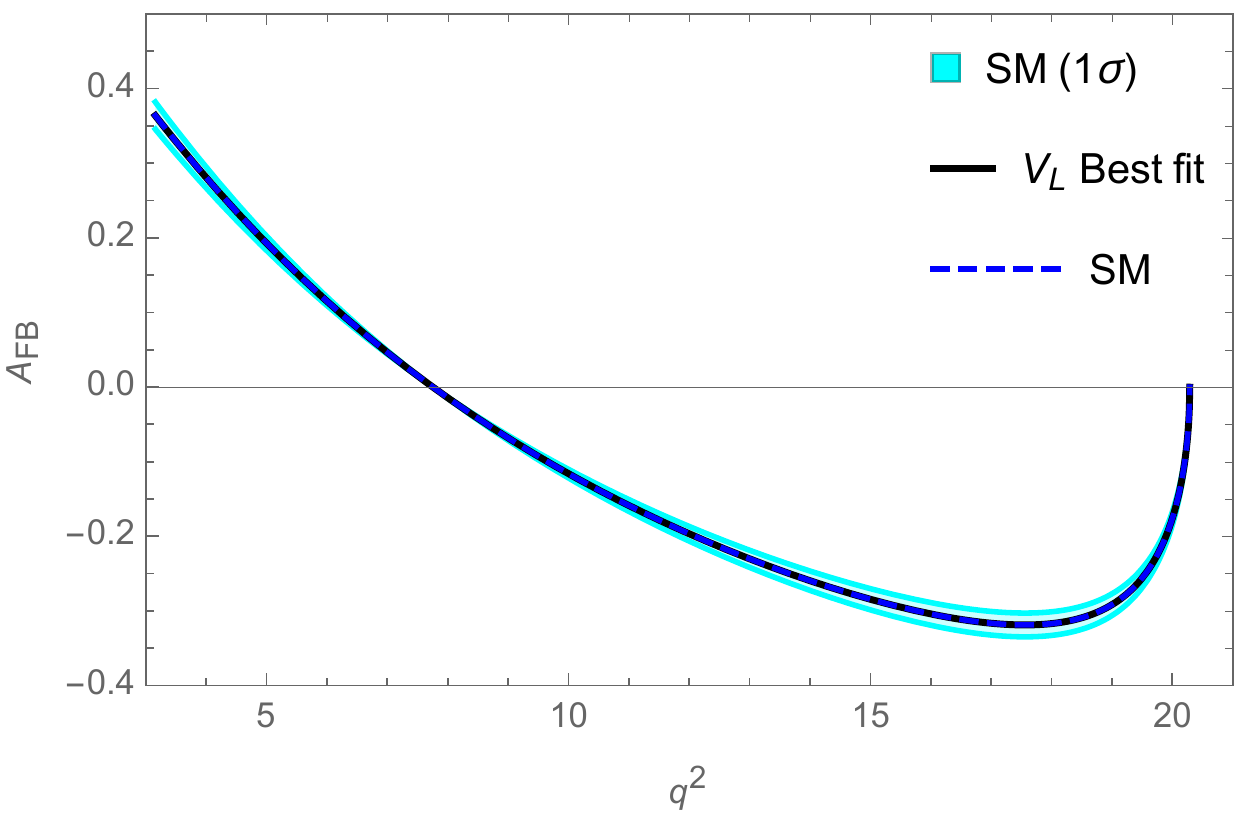}
\quad
\includegraphics[scale=0.4]{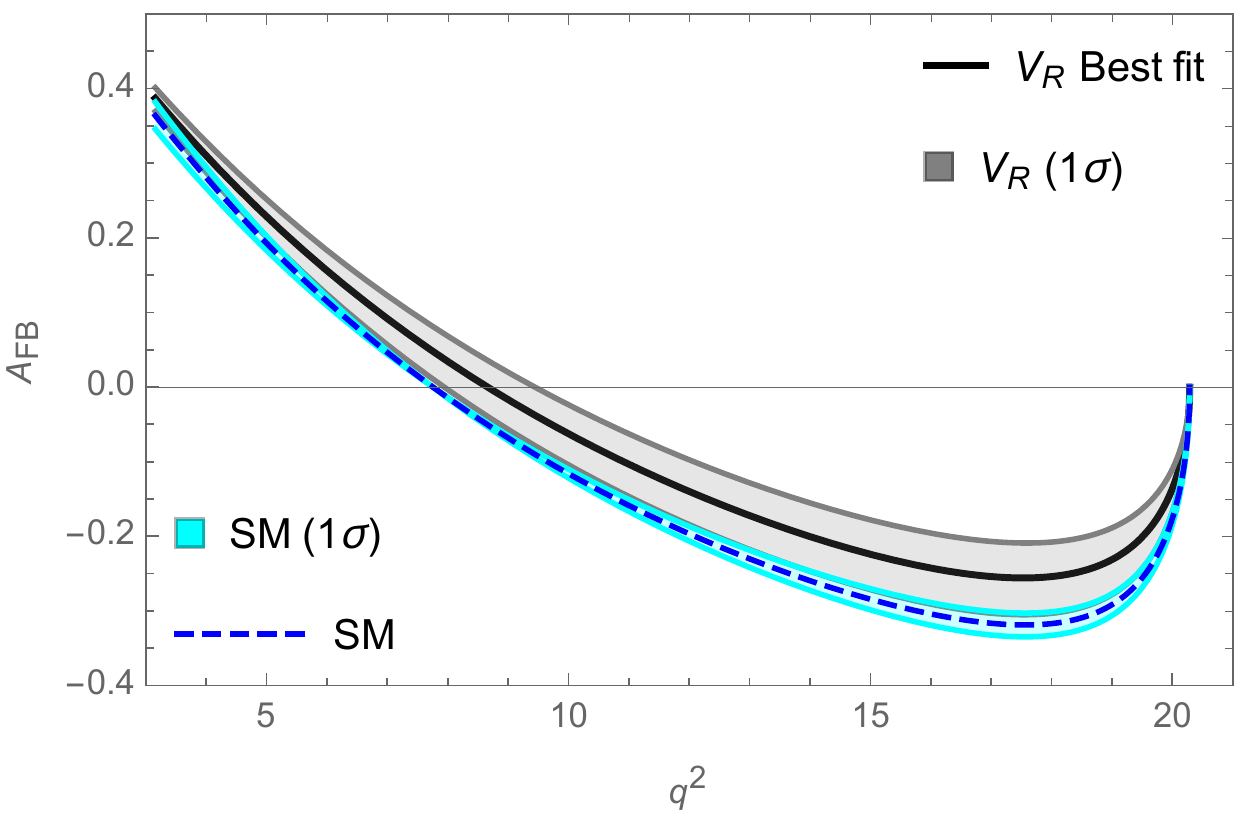}
\quad
\includegraphics[scale=0.4]{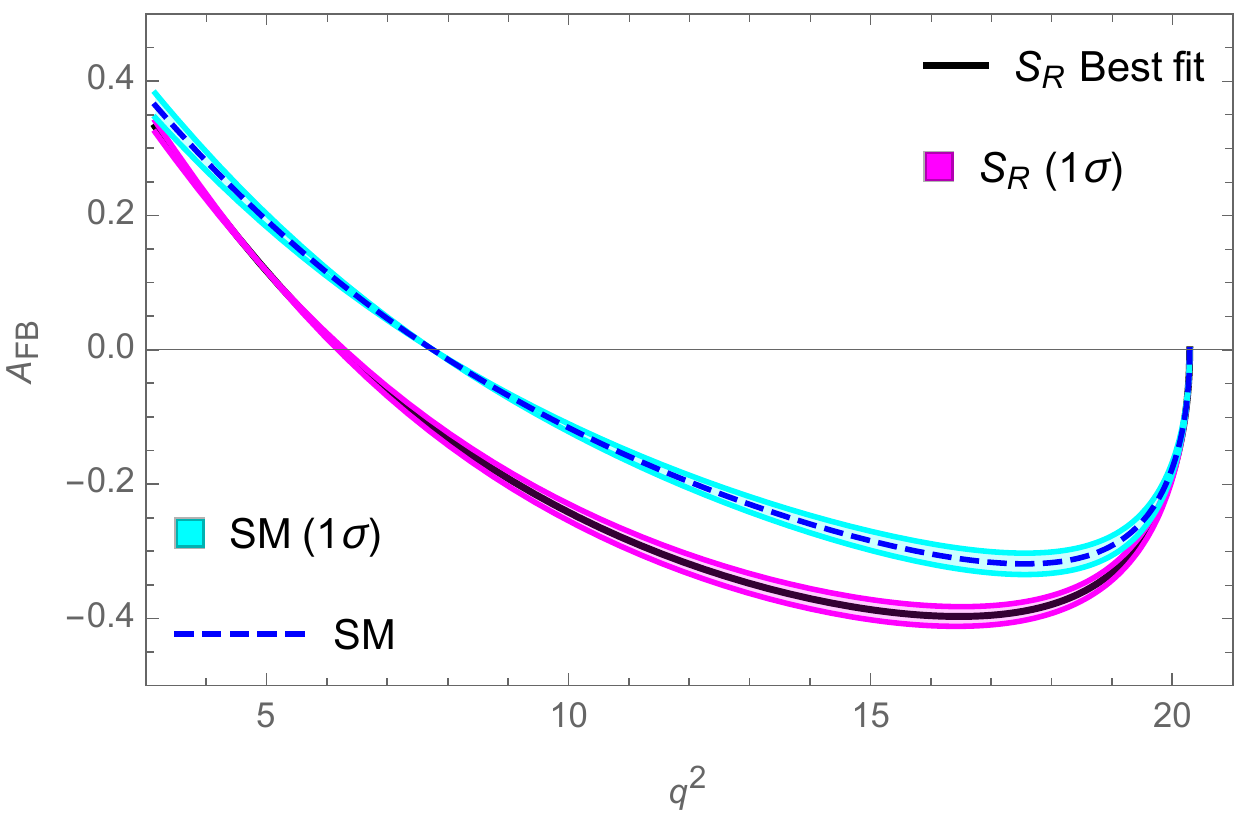}
\quad
\includegraphics[scale=0.4]{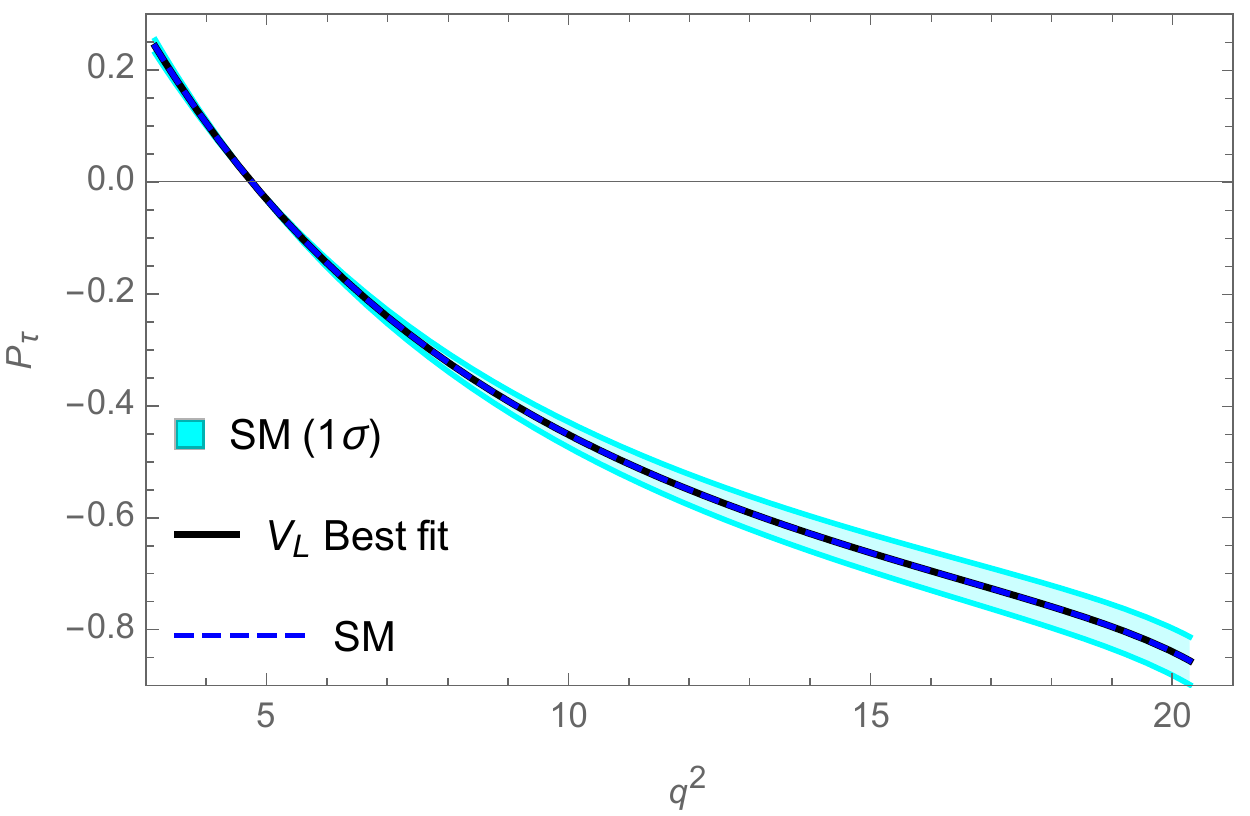}
\quad
\includegraphics[scale=0.4]{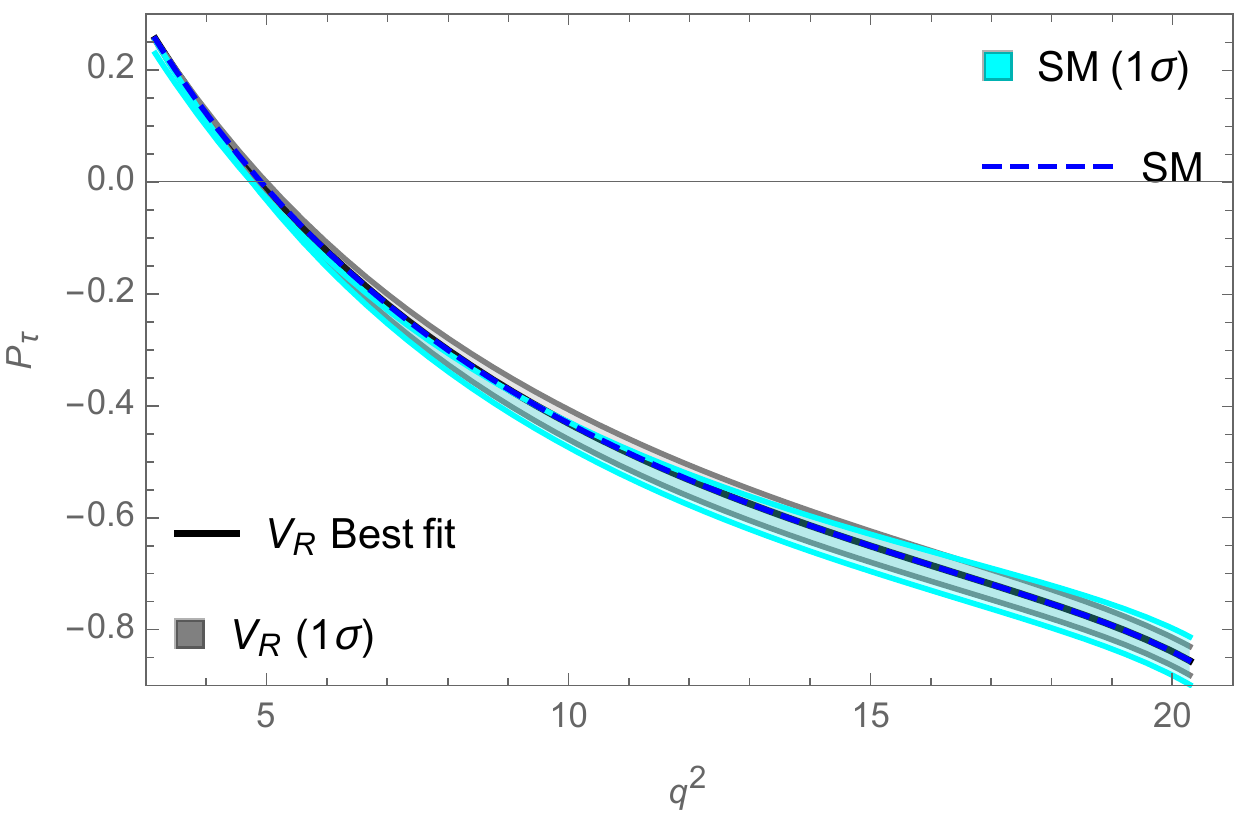}
\quad
\includegraphics[scale=0.4]{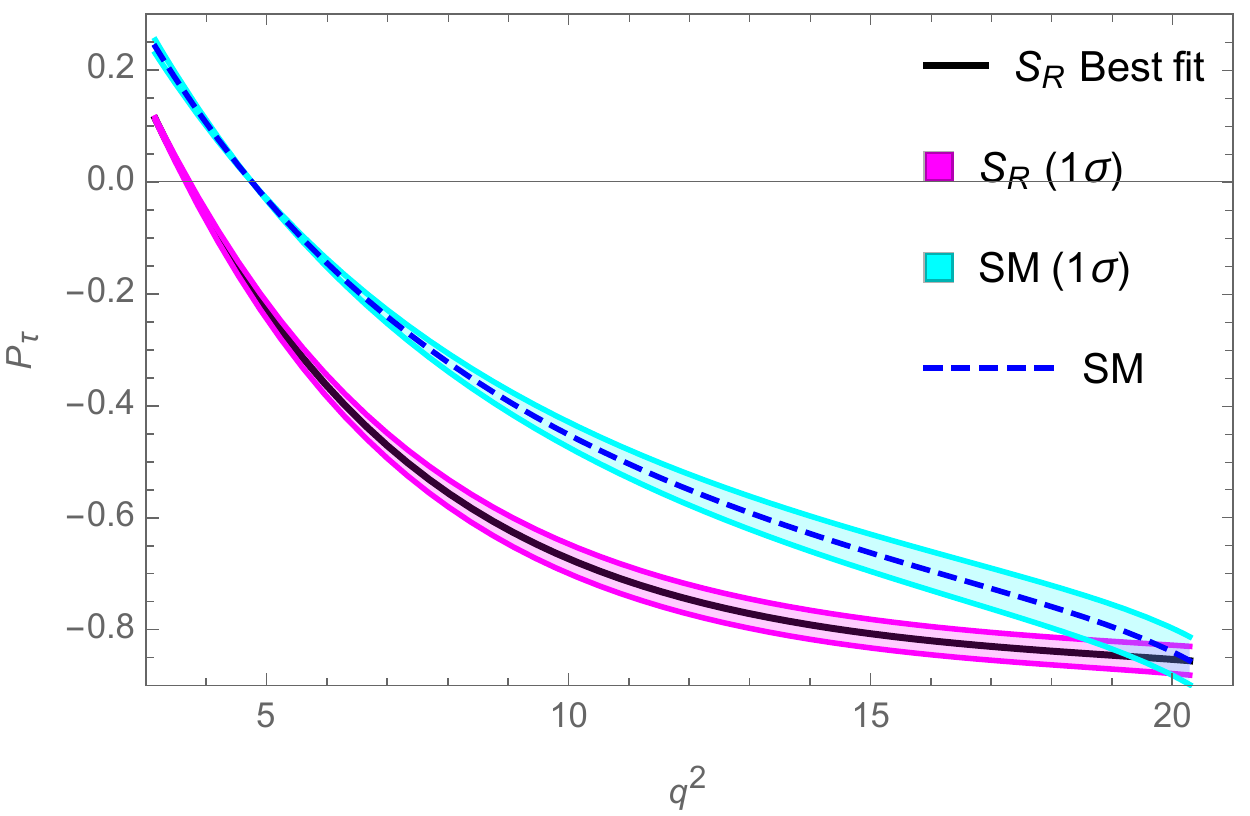}
\quad
\includegraphics[scale=0.4]{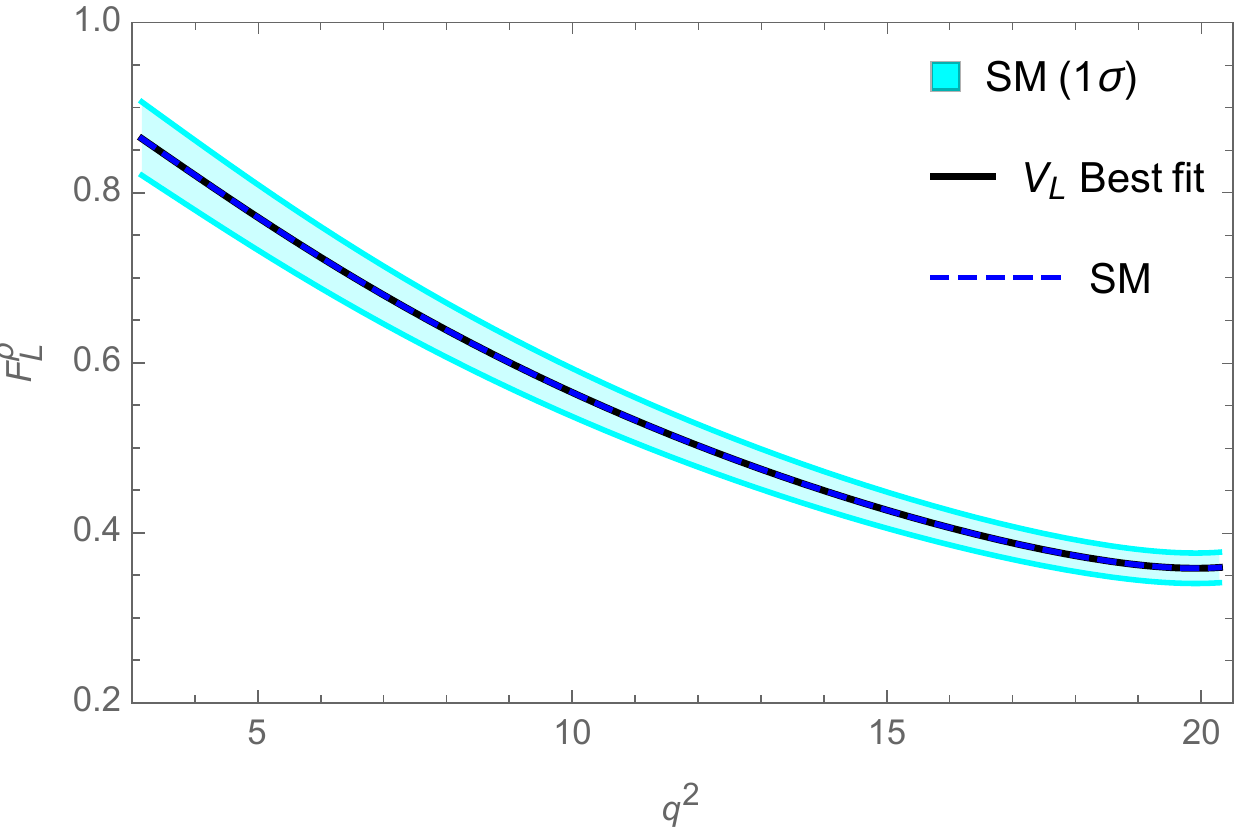}
\quad
\includegraphics[scale=0.4]{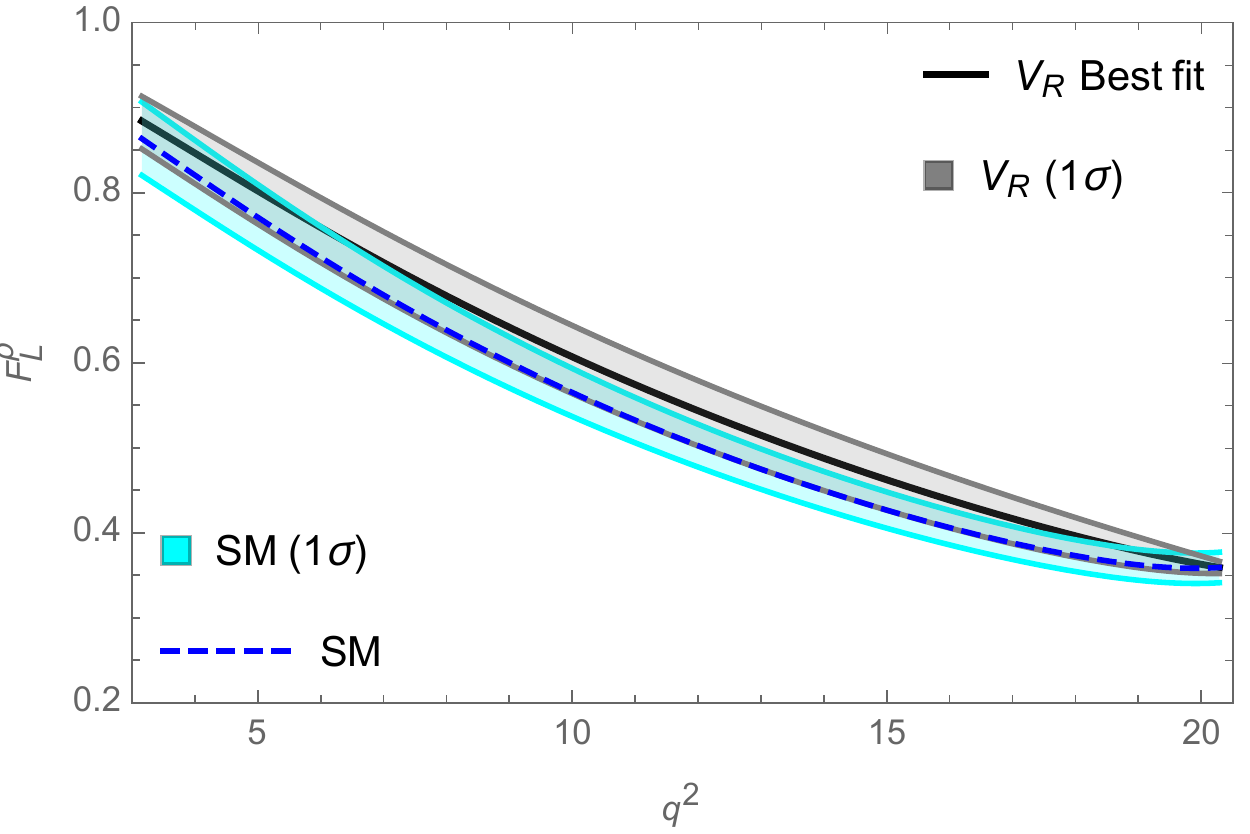}
\quad
\includegraphics[scale=0.4]{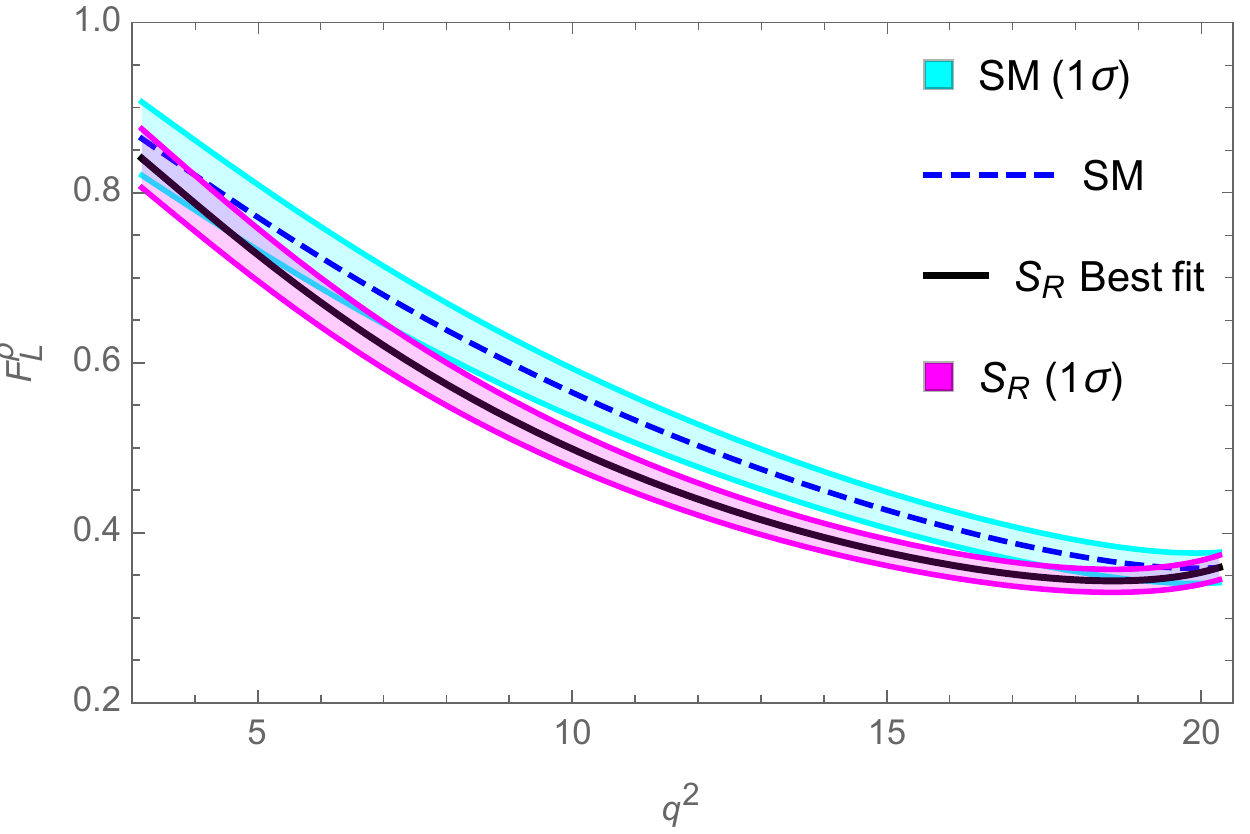}
\caption{The $q^2$ (in ${\rm GeV}^2$) variation of different observables of $\bar B^0 \to \rho \tau^- \bar \nu_\tau$ in the presence of new $V_L$ coupling (left panel), $V_R$ coupling (middle panel) and $S_R$ coupling (right panel). }
\label{B2rho}
\end{figure}
%%%%%%%%%%%%%%%%%%%%%%%%%%%%%%%%%%%%
The $q^2$ dependence of the form factors are determined by performing  a combined fit to lattice and LCSR results, which are valid for   the entire kinematic range \cite{Straub:2015ica}, and are parametrized as
\bea
F_i(q^2)= \frac{1}{(1-q^2/m_{R,i}^2)}~ \sum_{k=0} a_k^i \Big[z(q^2)-z(0)\Big]^k, \label{FFV1}
\eea
where $z(q^2)=\frac{\sqrt{t_+-q^2}-\sqrt{t_+-t_0}}{\sqrt{t_+-q^2}+\sqrt{t_++t_0}}$
with $t_\pm =( m_B\pm m_V)^2$ and  $t_0= t_+(1- \sqrt{1-t_-/t_+})$.
The form-factor  $F_i$ refers to $V(q^2)$, $A_0(q^2)$, $A_1(q^2)$ and $A_{12}(q^2)$, where $A_{12}(q^2)$ is defined as
\bea
A_{12}(q^2)= \frac{(m_B+m_V)^2(m_B^2-m_V^2-q^2)A_1(q^2) -\lambda_V(q^2) A_2(q^2)}{16 m_B m_V^2 (m_B+m_V)}.
\eea
The values of the different $a_k^i$ coefficients used in our analysis are presented in Table \ref{Table:FF}. Using these values and other input parameters from \cite{Tanabashi:2018oca}, we estimate  the branching fraction,   $R^{\tau/\ell}_{\rho}$, $A_{FB}$, $P_{\tau}$  and $F_L^\rho $  observables for 
$\bar B^0 \to \rho \tau^- \bar \nu_\tau$ process in the presence of the NP coefficients $V_L$, $V_R$, $S_L$ and $S_R$, and the  $q^2$ variation of these observables are displayed in Fig \ref{B2rho}. Since there is almost negligible  deviation of these observables  from their  SM prediction in the presence of $S_L$ coefficient, the corresponding   results are not shown in the figure. It can be noticed from the figure that the branching fraction and the LFU violating observable have significant deviation from their SM results in the presence of $V_L$, $V_R$ and $S_R$ NP scenarios, whereas  only $V_R$ and $S_R$ NP contributions  can affect $A_{\rm FB}$, $P_\tau$ and $F_L^\rho$  observables.  
The estimated average values of these observables are presented in Tab. \ref{Results} and the branching fraction of $\bar B^0 \to \rho^- \mu^+  \nu_\mu$ is furnished in Table \ref{Br:SM}.

Similarly   for $B^- \to \omega \tau^- \bar \nu_\tau$ process, use the form factors from \cite{Straub:2015ica}, we calculate the values of various observables. Since the  $q^2$ dependence of these observables have almost the similar behaviour as  $B^0 \to \rho \tau^- \bar \nu_\tau$ process, we do not provide the graphical results, however, their numerical results are presented in Table \ref{Results}. In this case also the branching fraction deviates significantly from the SM prediction with $V_L$, $V_R$ and $S_R$ type of new physics. Furthermore, $V_R$ and $S_R$ kind of new physics affect marginally the forward backward asymmetry and the longitudinal polarization of $\omega$ meson.

\subsection{$B_s \to (K,K^*) \tau \bar \nu$ decay}
We use the form factors for $B_s \to K \ell \bar \nu$ transition from lattice QCD calculation \cite{Bazavov:2019aom}, with the BCL parametrization
\bea
&&f_+(q^2)= \frac{1}{(1-q^2/m_{B^*(1^-)})}\sum_{n=1}^{N-1} b_n^+(t_0)\Big[ z^n -(-1)^{n-N}\frac{n}{N} z^N \Big]\nn\\
&&f_0(q^2)=\frac{1}{(1-q^2/m_{B^*(0^+))}}\sum_{n=1}^{N-1} b_n^0(t_0)~ z^n ,
\eea
where the factor $1/(1-q^2/m_{B^*}^2)$  take the poles into account and ensure the asymptotic scaling. The expansion parameter $z$ is defined as
\bea
z(q^2,t_0)= \frac{\sqrt{t_{\rm cut}-q^2}-\sqrt{t_{\rm cut}-t_0}}{\sqrt{t_{\rm cut}-q^2}+\sqrt{t_{\rm cut}-t_0}}
\eea
where $t_{\rm cut}$ is the particle pair production threshold   with value ${\sqrt t_{\rm cut}}=5.414$ GeV and $t_0=t_{\rm cut}-\sqrt{t_{\rm cut}(t_{\rm cut}-t_-)}$ with $t_-=(m_{B_s}-m_K)^2$. The values of the pole masses are $m_{B^*}(1^-)=5.325$ GeV and $m_{B^*}(0^+)=5.68$ GeV, and  the expansion parameters have values~\cite{Bazavov:2019aom}
\begin{eqnarray}
& & b^+_0 = 0.3623(0.0178),\quad b^+_1 = -0.9559(0.1307), \quad b^+_2 = -0.8525(0.4783), \nonumber \\
& & b^+_3 = 0.2785(0.6892), \quad b^0_0 = 0.1981(0.0101), \quad b^0_1 = -0.1661(0.1130),\nonumber \\
& & b^0_2 = -0.6430(0.4385), \quad b^0_3 = -0.3754(0.4535).
\end{eqnarray}

%\begin{figure}
%\includegraphics[scale=0.4]{BR-B2Omg-VL.pdf}
%\quad
%\includegraphics[scale=0.4]{BR-B2Omg-VR.pdf}
%\quad
%\includegraphics[scale=0.4]{BR-B2Omg-SR.pdf}
%\quad
%\includegraphics[scale=0.4]{R-B2Omg-VL.pdf}
%\quad
%\includegraphics[scale=0.4]{R-B2Omg-VR.pdf}
%\quad
%\includegraphics[scale=0.4]{R-B2Omg-SR.pdf}
%\quad
%\includegraphics[scale=0.4]{AFB-B2Omg-VL.pdf}
%\quad
%\includegraphics[scale=0.4]{AFB-B2Omg-VR.pdf}
%\quad
%\includegraphics[scale=0.4]{AFB-B2Omg-SR.pdf}
%\quad
%\includegraphics[scale=0.4]{Ptau-B2Omg-VL.pdf}
%\quad
%\includegraphics[scale=0.4]{Ptau-B2Omg-VR.pdf}
%\quad
%\includegraphics[scale=0.4]{Ptau-B2Omg-SR.pdf}
%\quad
%\includegraphics[scale=0.4]{FL-B2Omg-VL.pdf}
%\quad
%\includegraphics[scale=0.4]{FL-B2Omg-VR.pdf}
%\quad
%\includegraphics[scale=0.4]{FL-B2Omg-SR.pdf}
%
%\caption{The $q^2$ variation of differential decay rate, lepton nonuniversaity parameter and forward-backward asymmetry of $ \bar B_d^{*} \to D^+ \tau^- \bar{\nu}$ (left panel) and  $\bar{B}_d^* \to \pi^+  \tau \bar{\nu}$ (right panel) in presence of new $V_R$ coefficient. The black solid lines and the green bands are obtained by using the best-fit values and corresponding $1\sigma$ range of $V_R$ coefficient. }.
%\label{variation-VR}
%\end{figure}
%
%%%%%%%%%%%%%%%%%%%%%%%%%%%%%%%%%%%%%%%%%%
\begin{figure}
\includegraphics[scale=0.4]{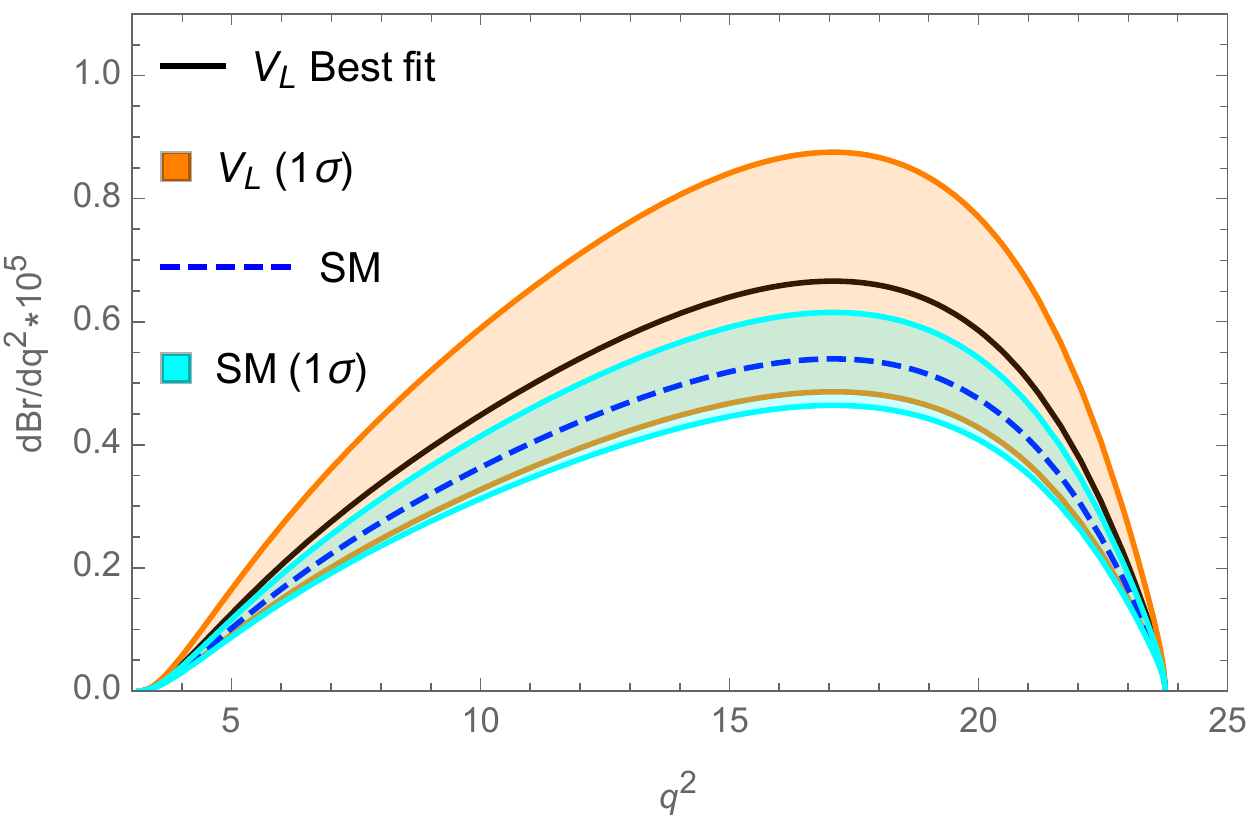}
\quad
\includegraphics[scale=0.4]{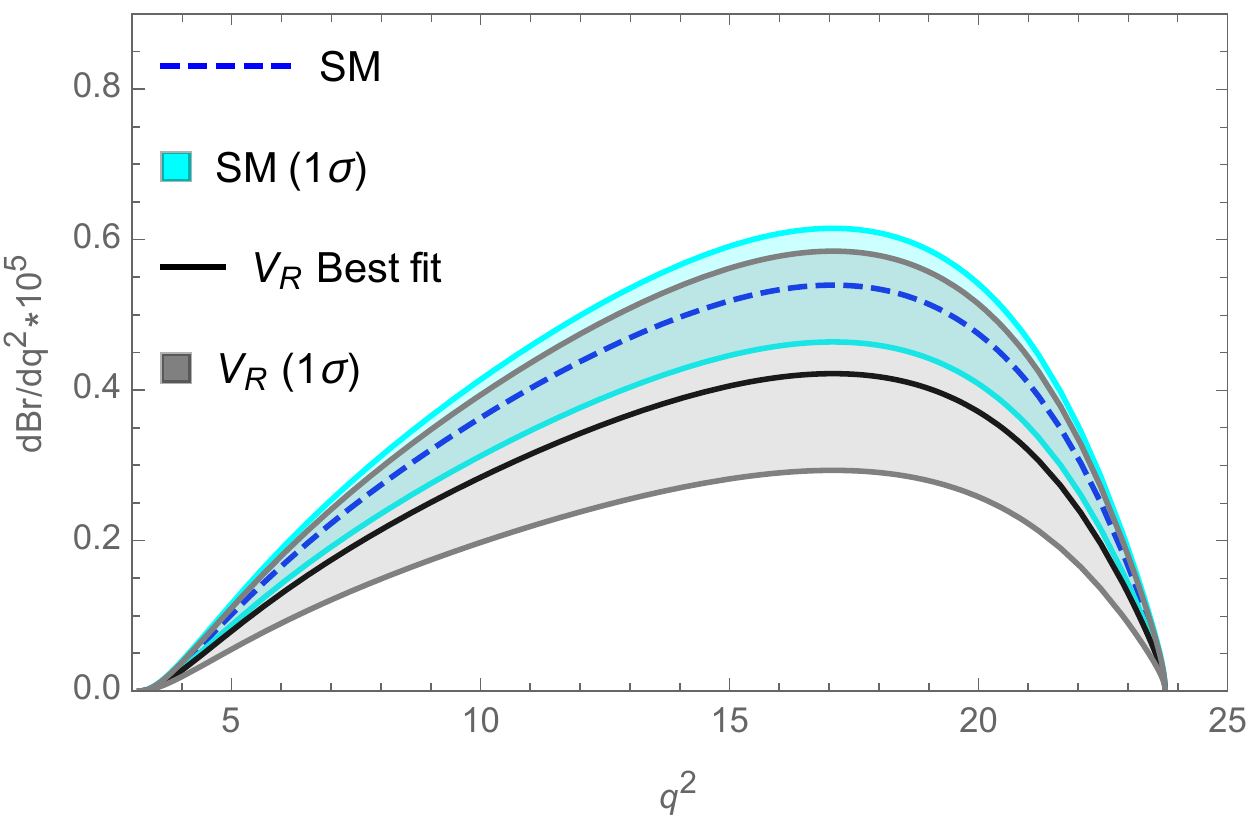}
\quad
\includegraphics[scale=0.4]{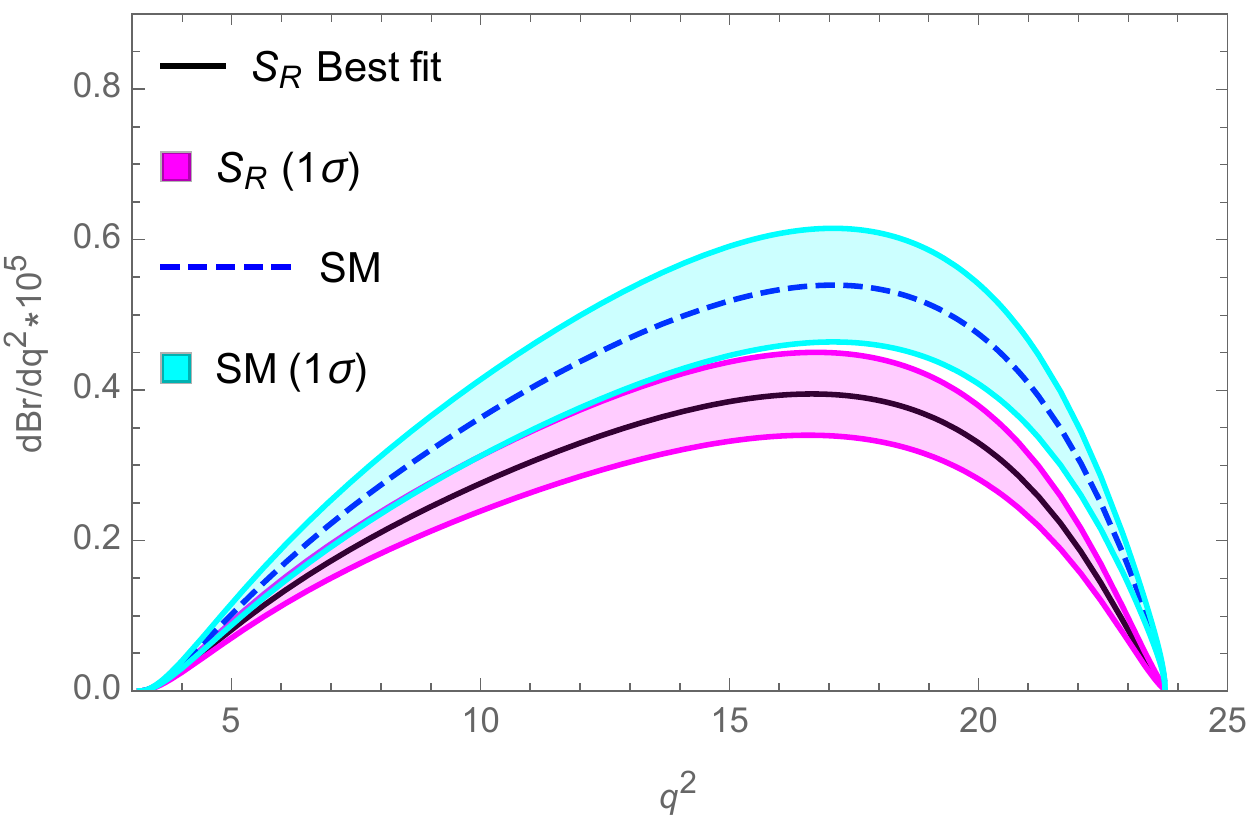}
\quad
\includegraphics[scale=0.4]{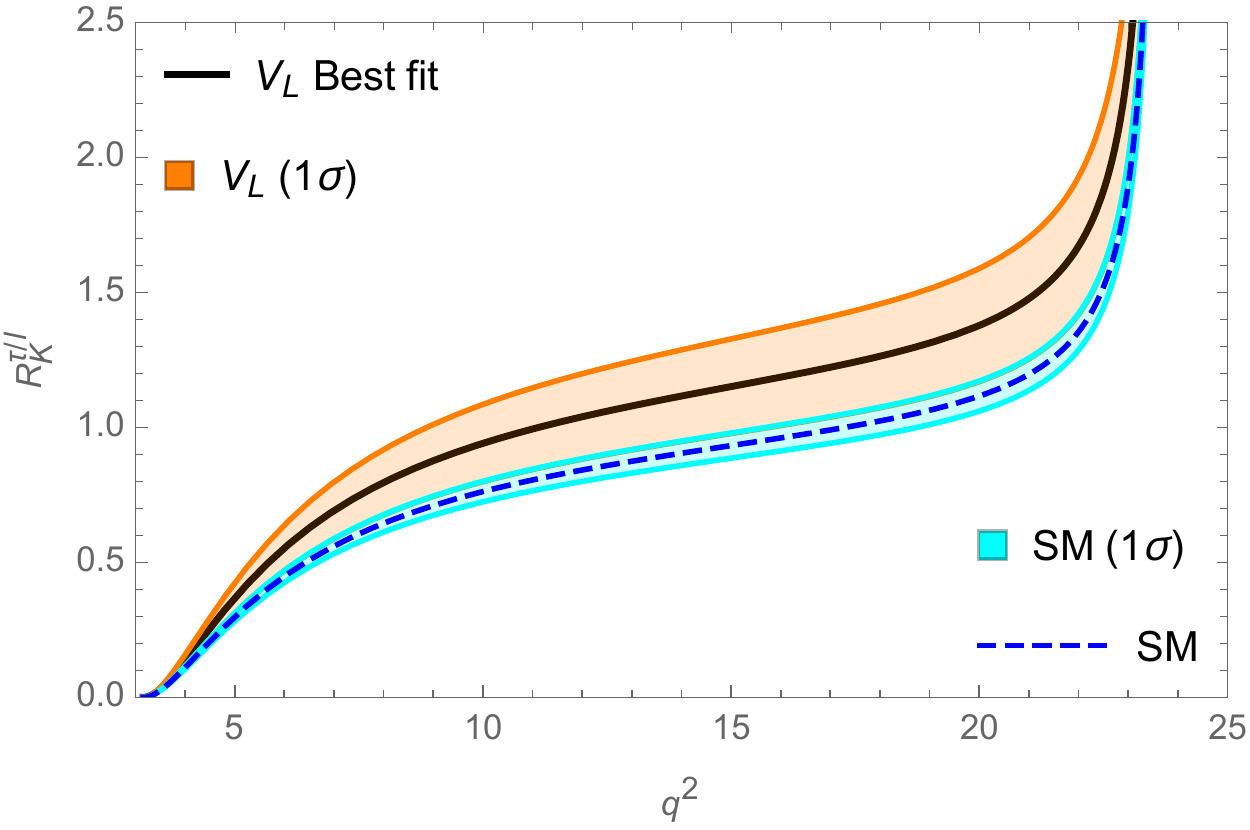}
\quad
\includegraphics[scale=0.4]{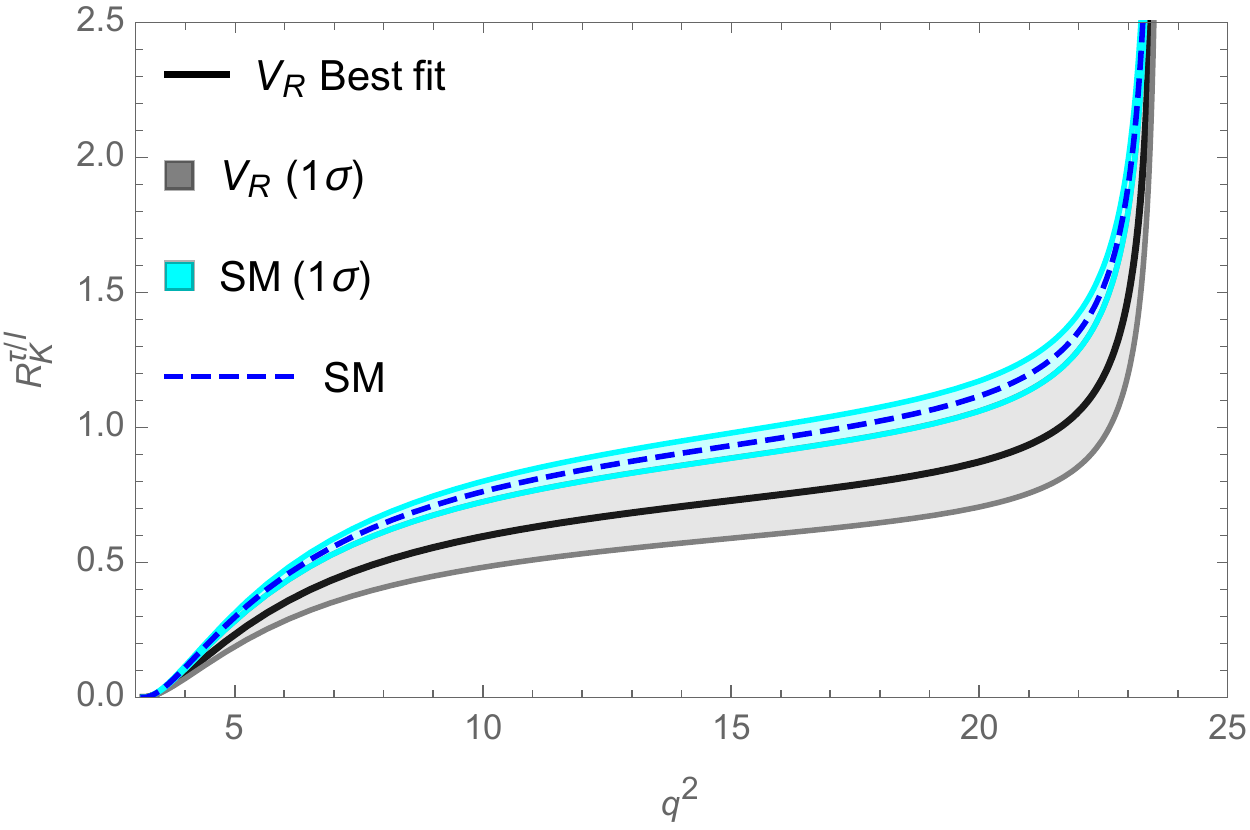}
\quad
\includegraphics[scale=0.4]{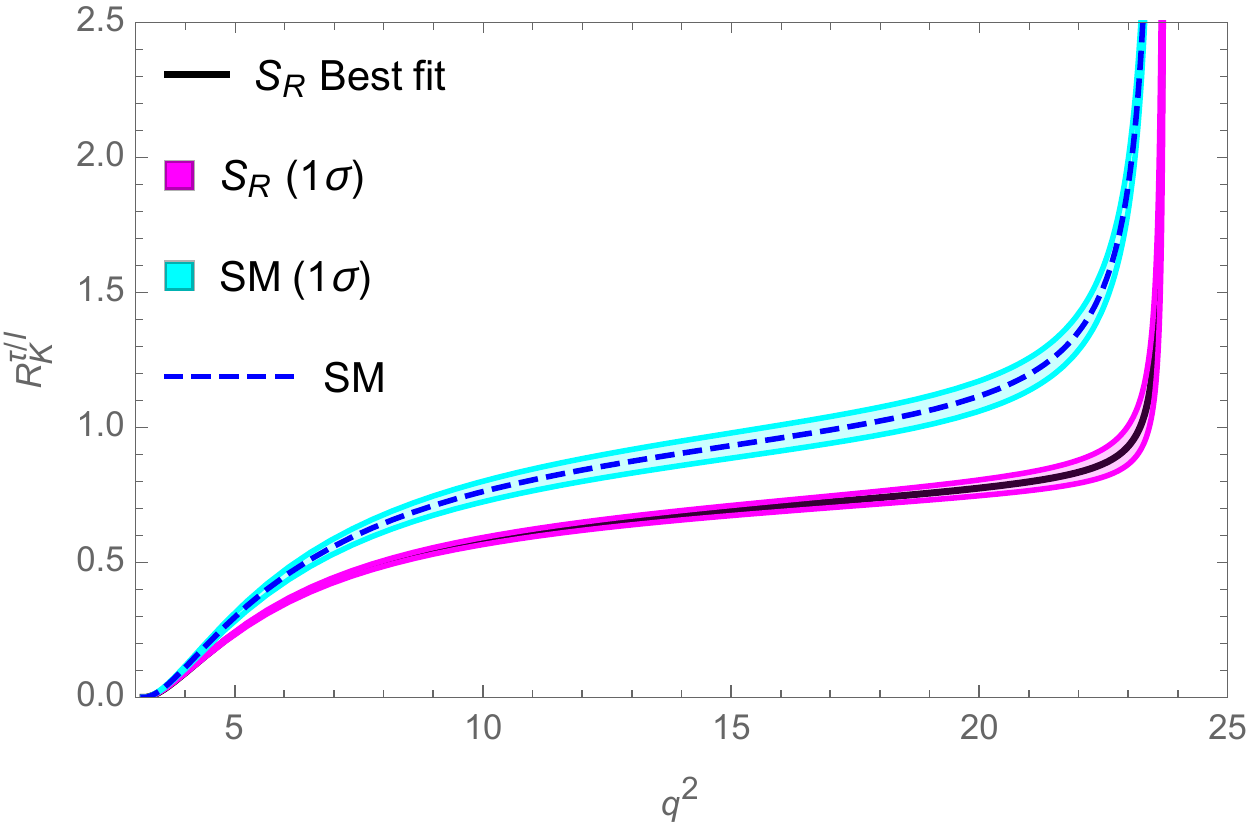}
\quad
\includegraphics[scale=0.4]{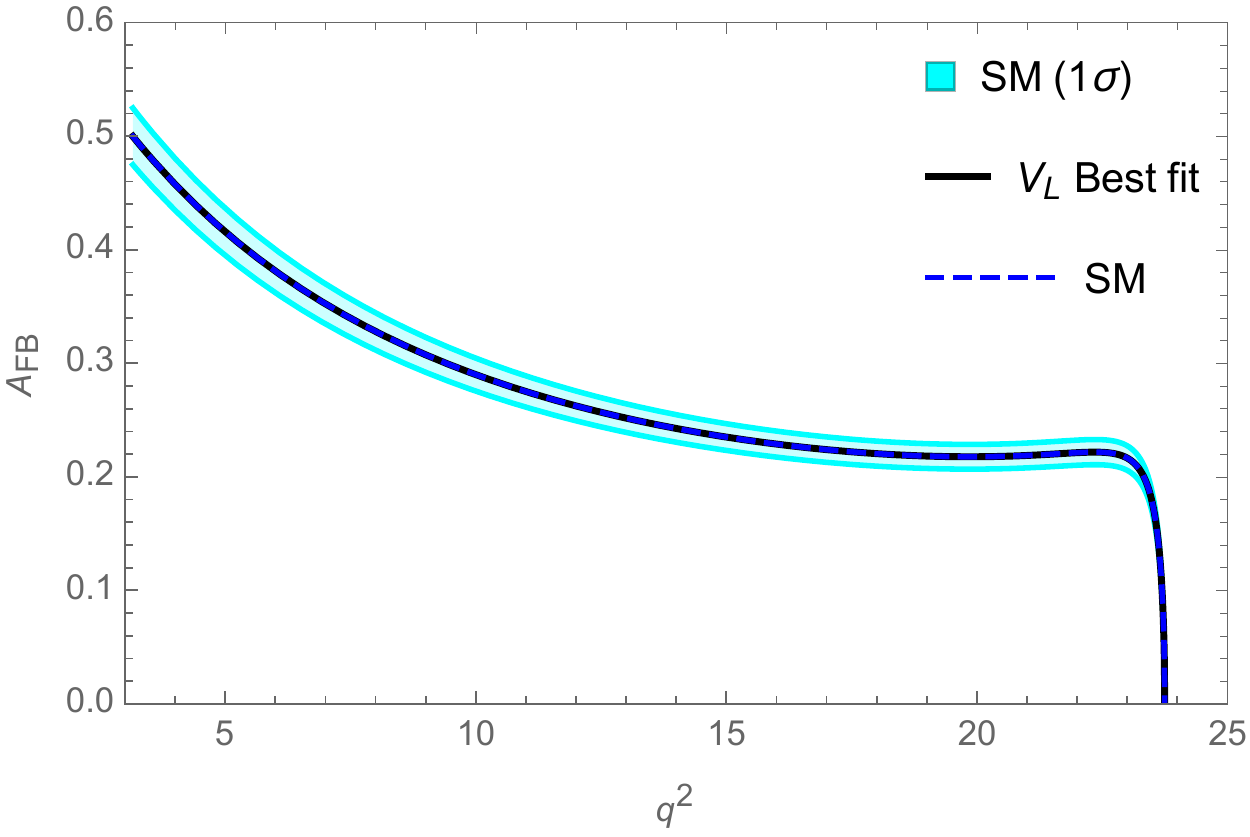}
\quad
\includegraphics[scale=0.4]{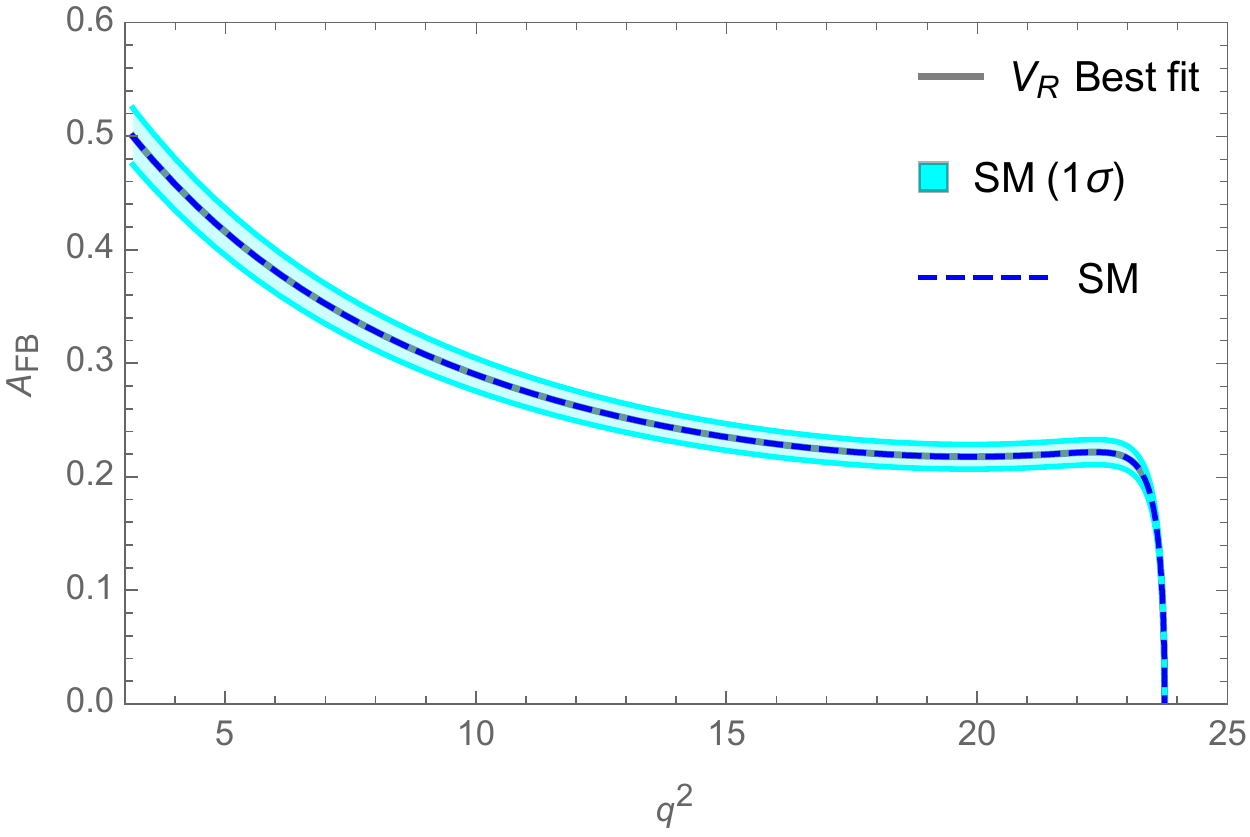}
\quad
\includegraphics[scale=0.4]{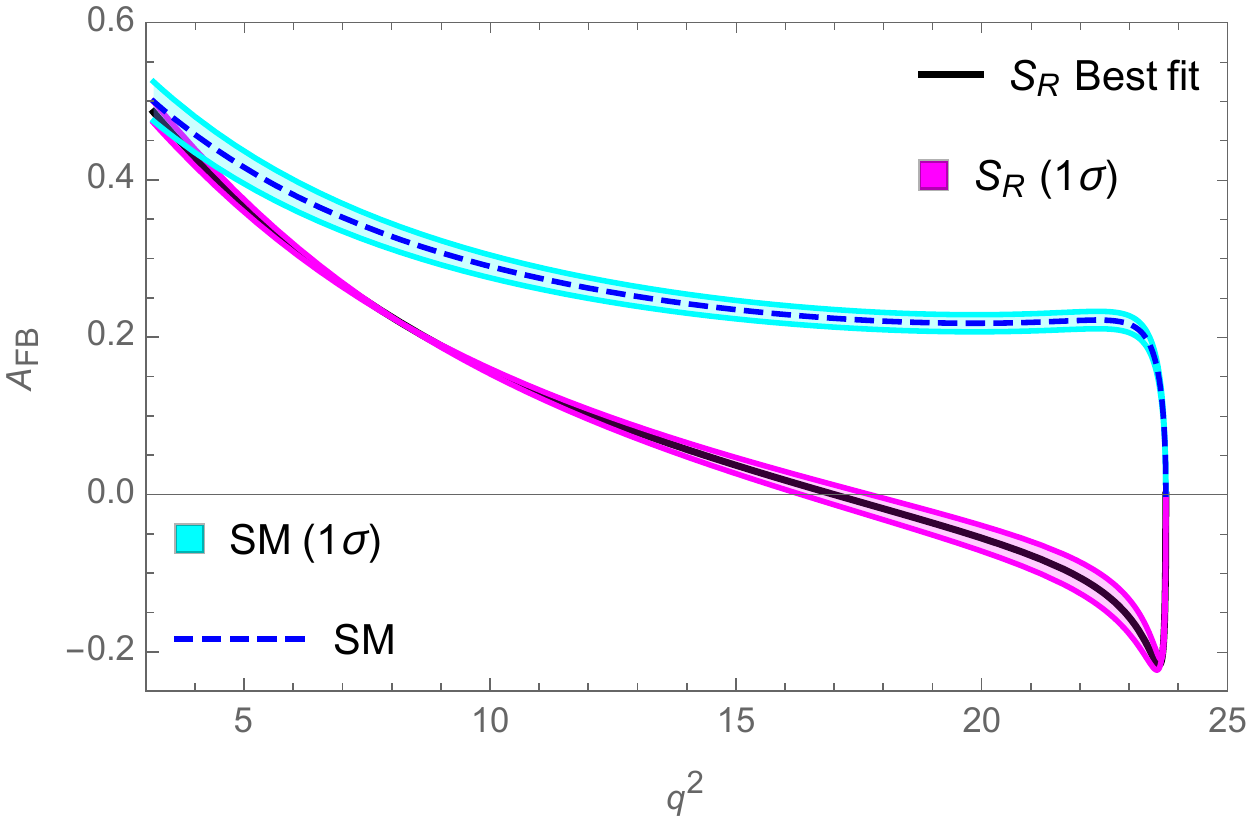}
\quad
\includegraphics[scale=0.4]{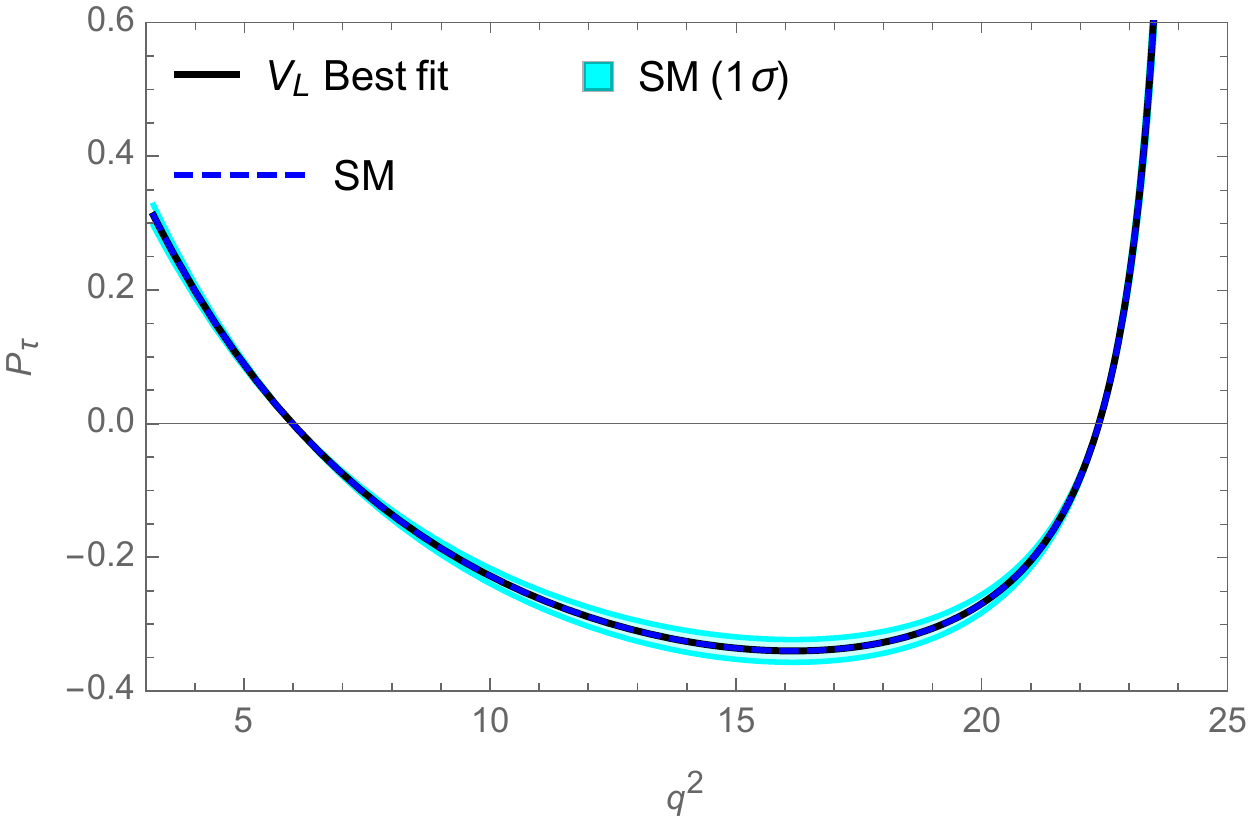}
\quad
\includegraphics[scale=0.4]{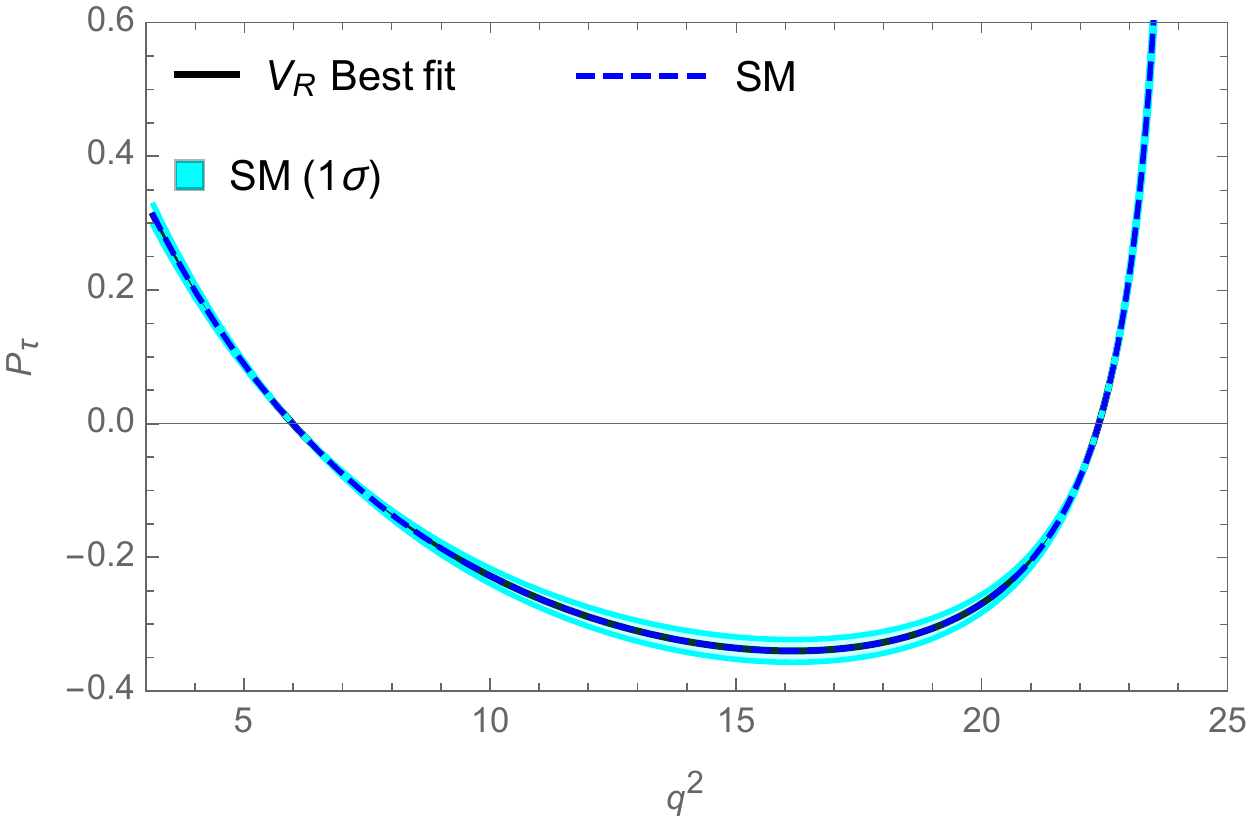}
\quad
\includegraphics[scale=0.4]{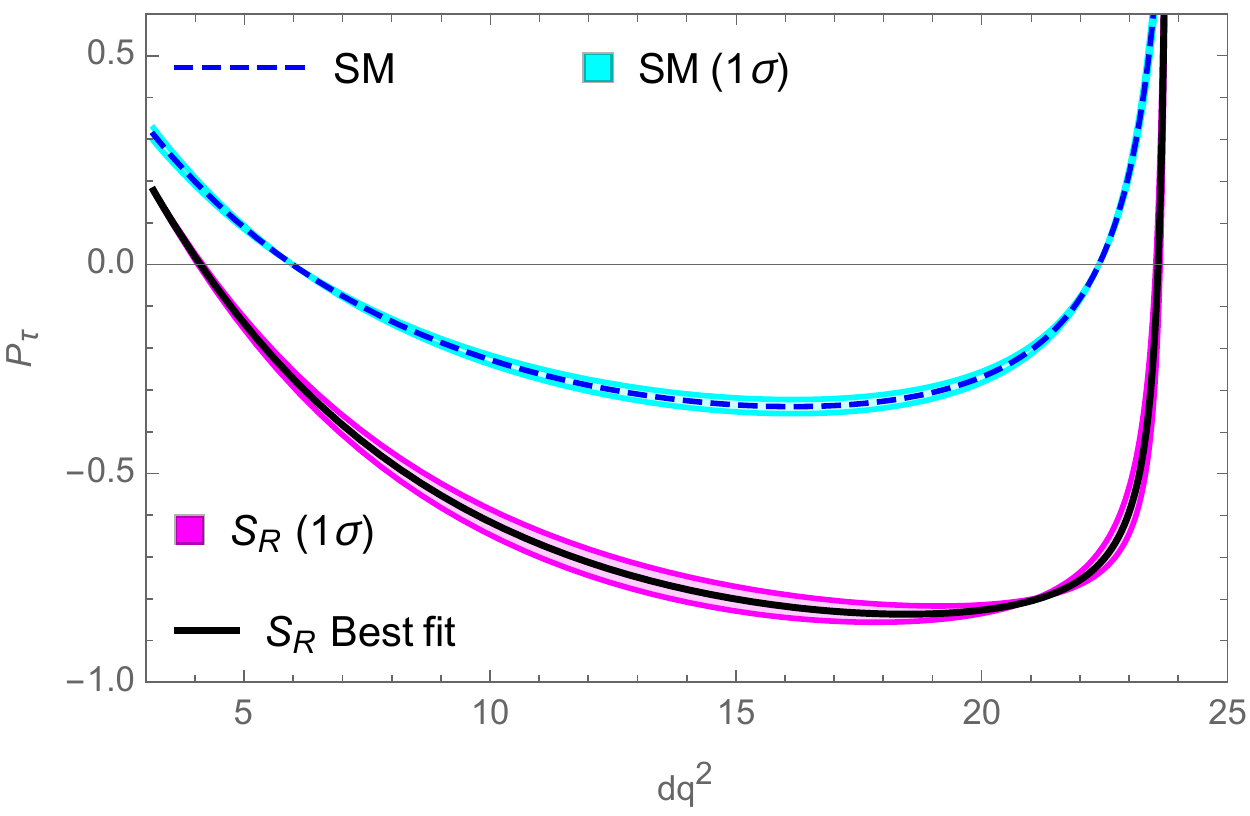}
\caption{The $q^2$ (in ${\rm GeV}^2$) variation of branching fraction, $R_K^{\tau/\ell}$, $A_{\rm FB}$, $P_\tau$ observables for $B_s \to K \tau  \bar \nu_\tau$ process in the presence of $V_L$, $V_R$ and $S_R$ NP scenarios. }
\label{Bs2K}
\end{figure}
%%%%%%%%%%%%%%%%%%%%%%%%%%%%%%%%%%%%%%
\begin{figure}
\includegraphics[scale=0.4]{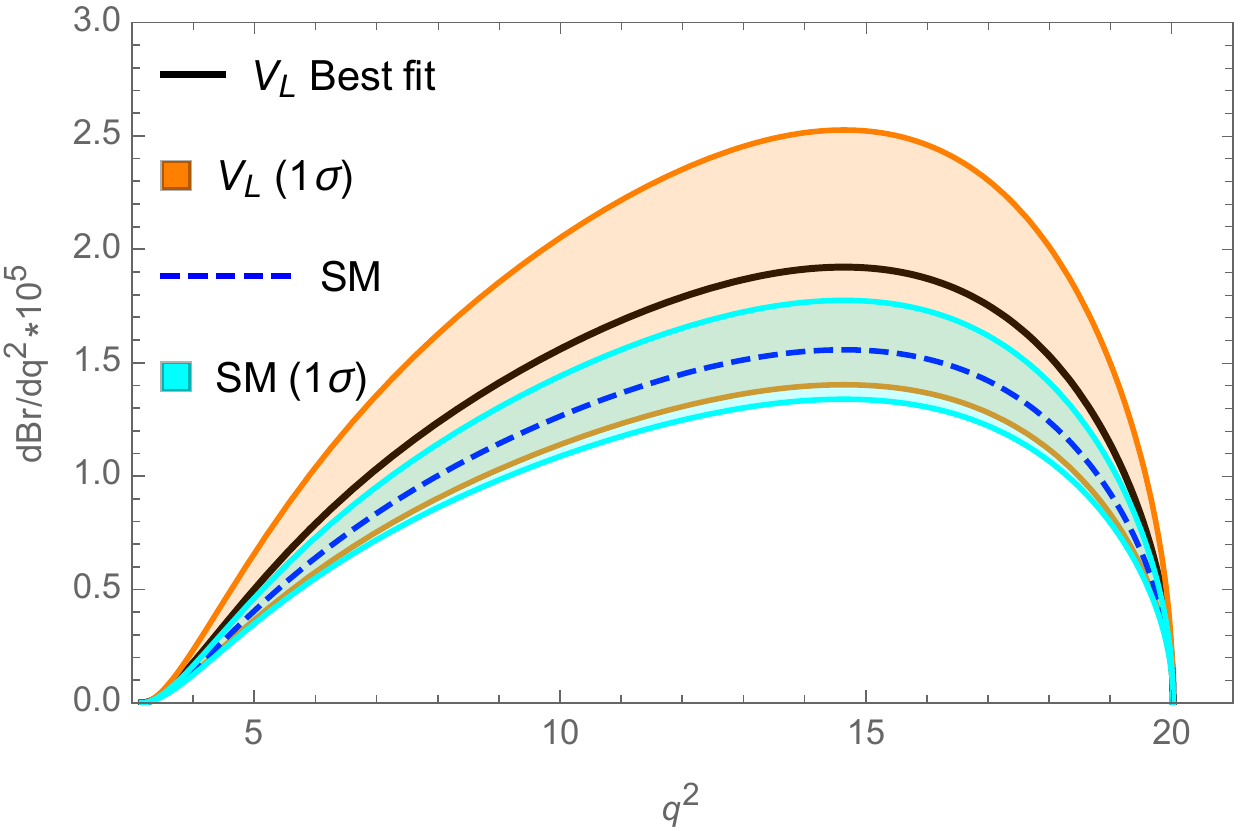}
\quad
\includegraphics[scale=0.4]{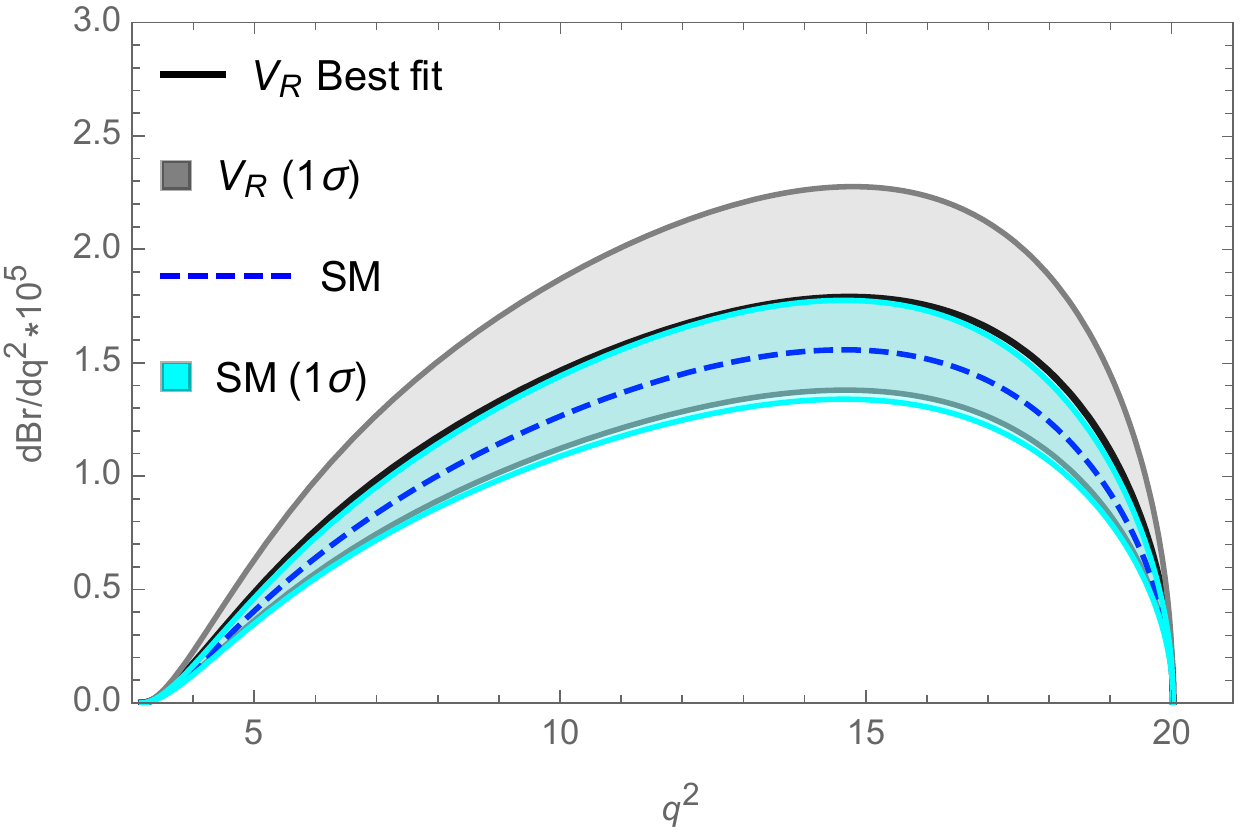}
\quad
\includegraphics[scale=0.4]{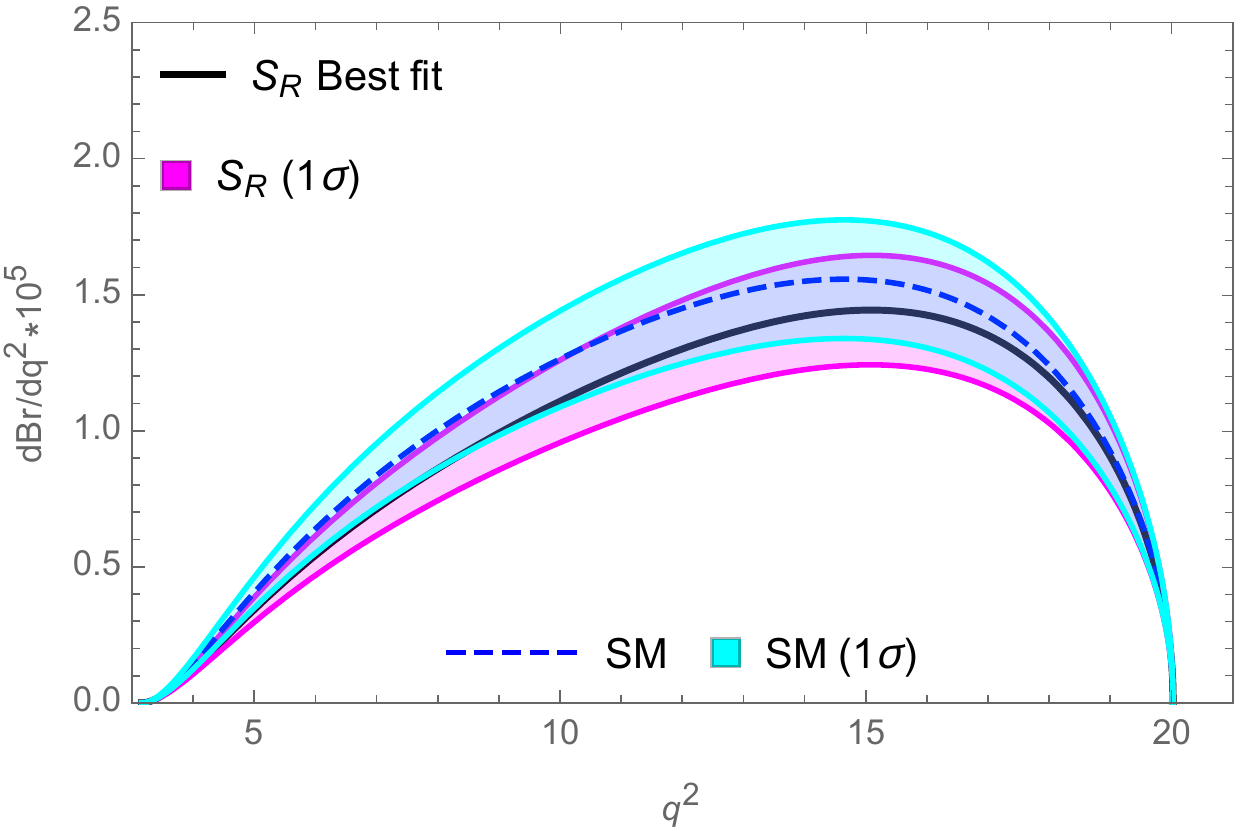}
\quad
\includegraphics[scale=0.4]{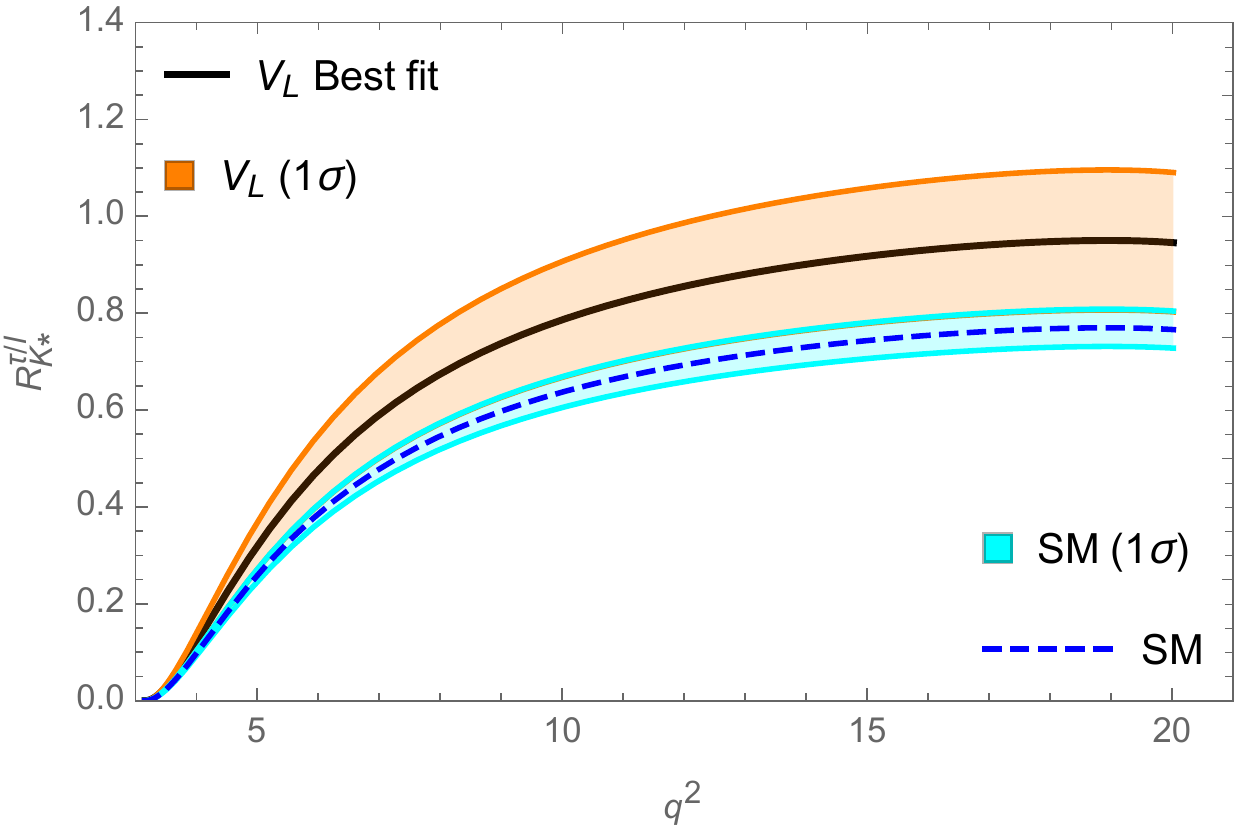}
\quad
\includegraphics[scale=0.4]{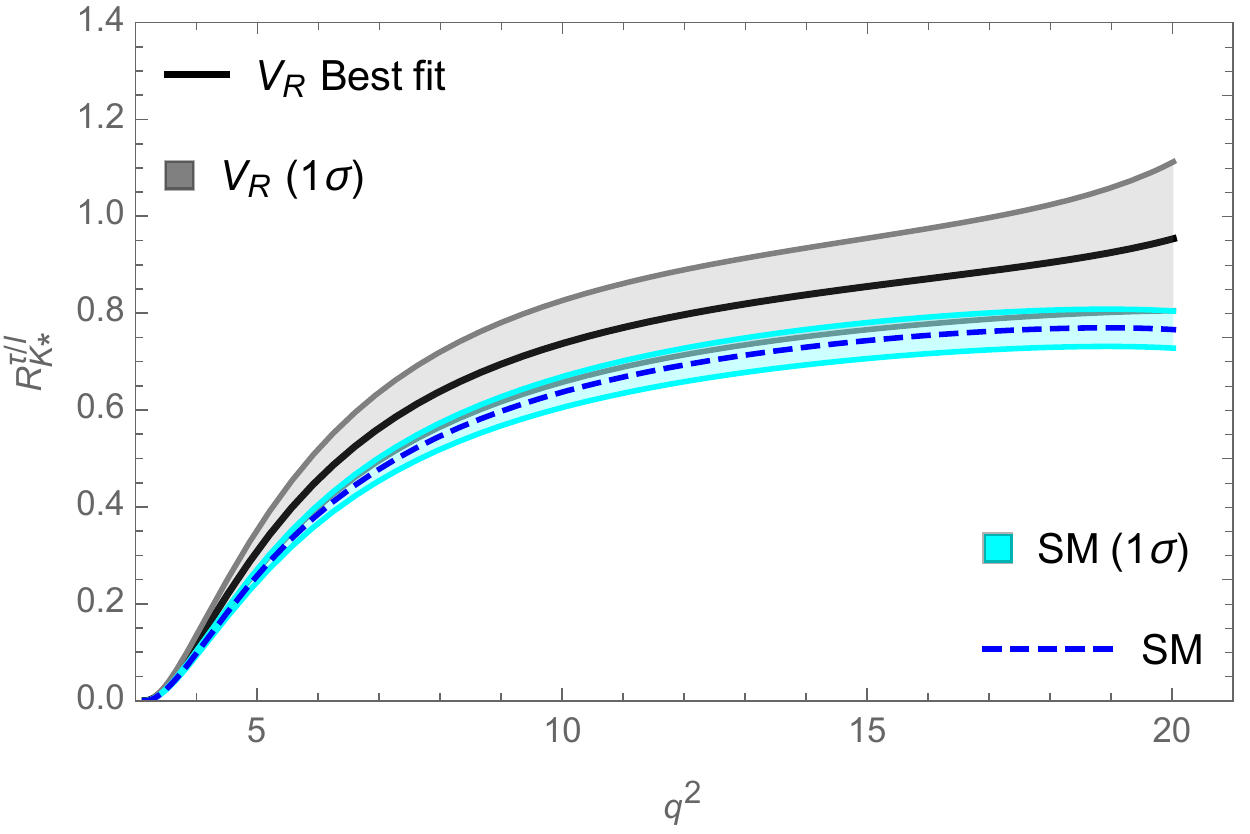}
\quad
\includegraphics[scale=0.4]{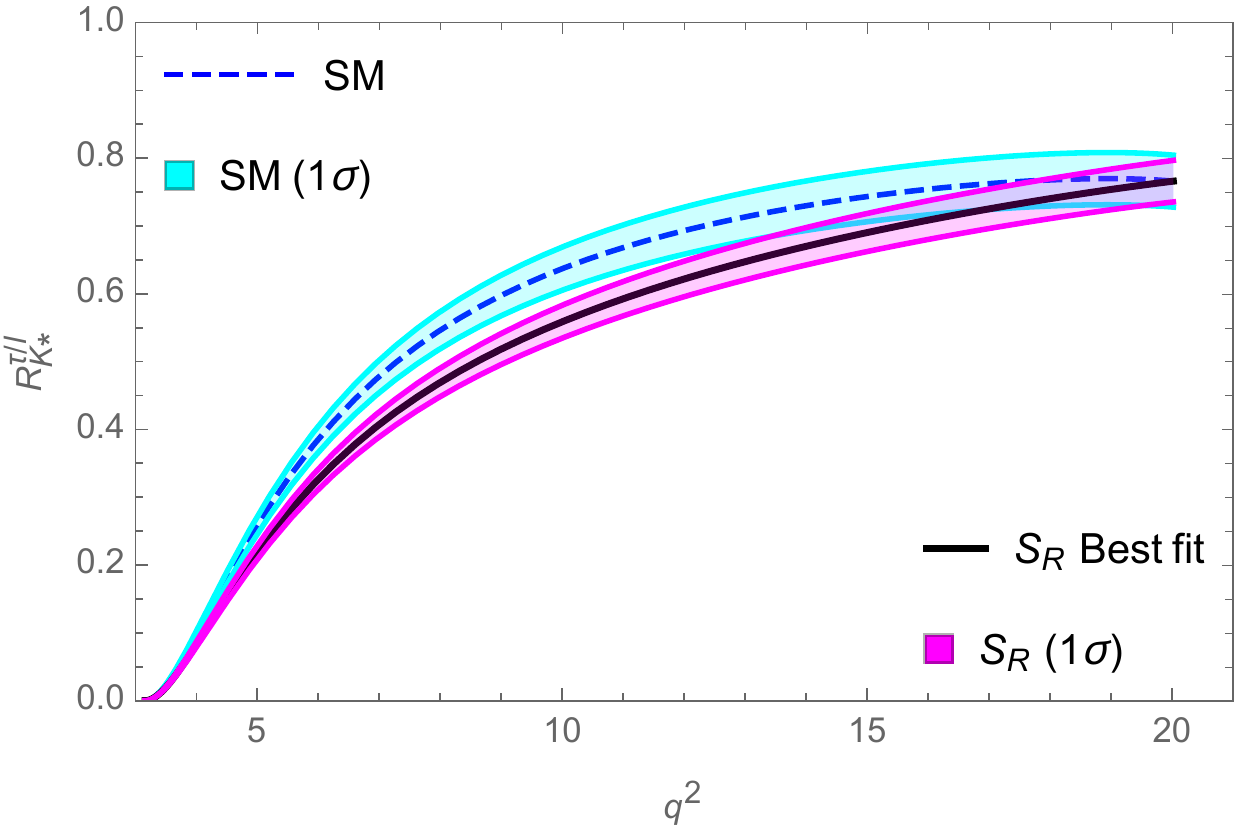}
\quad
\includegraphics[scale=0.4]{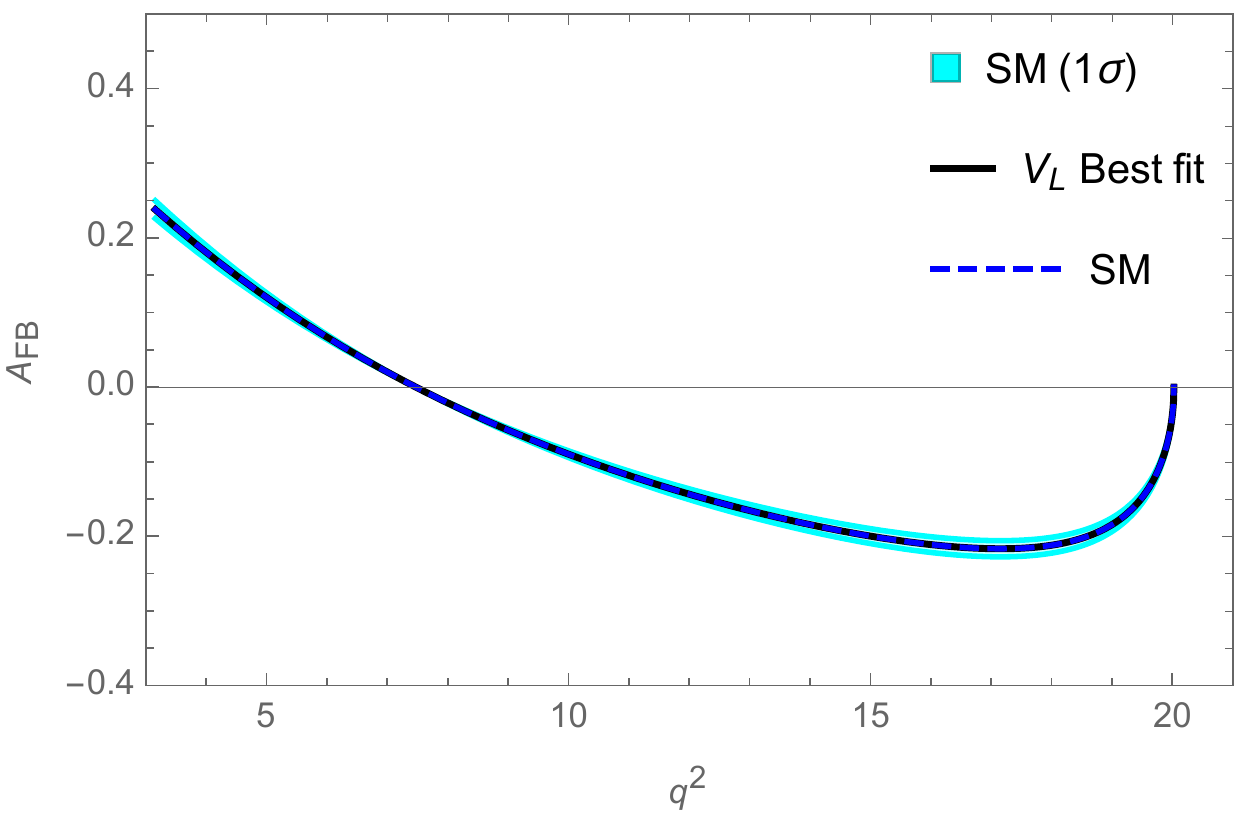}
\quad
\includegraphics[scale=0.4]{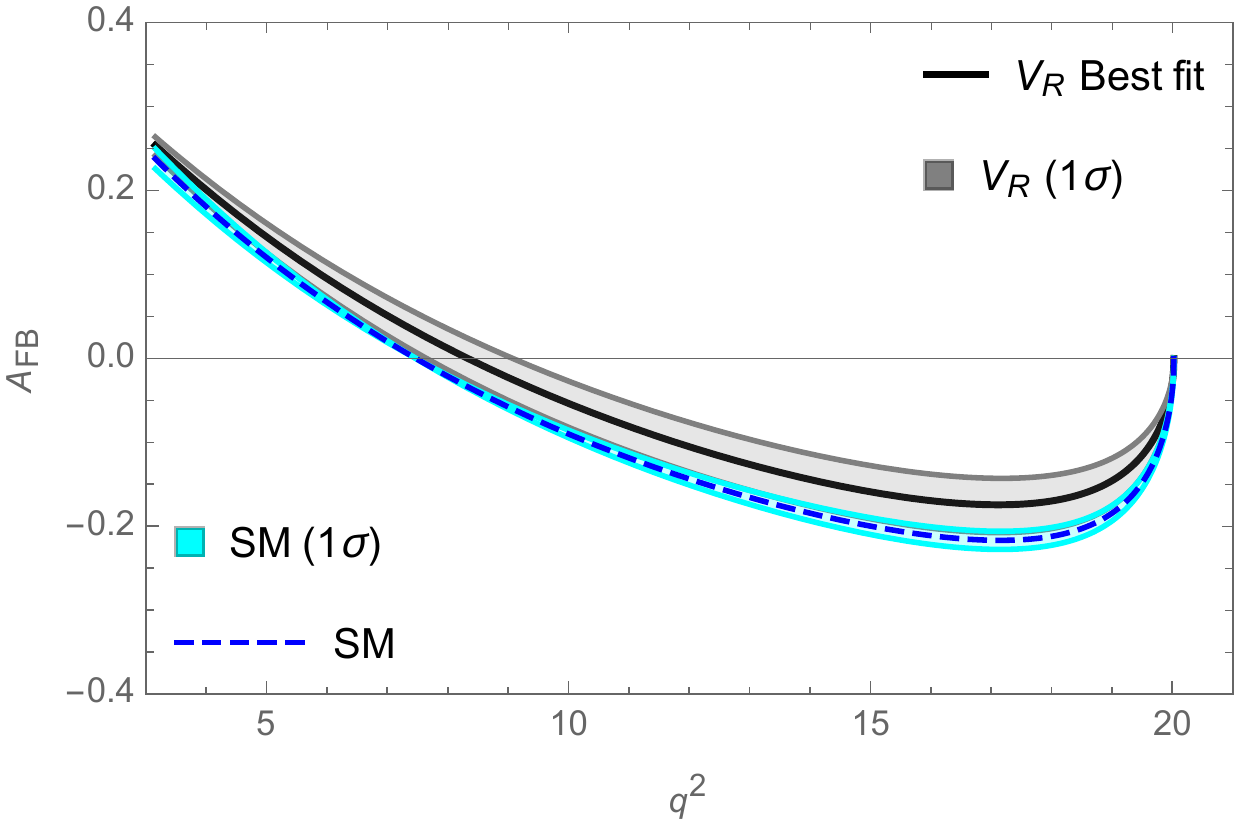}
\quad
\includegraphics[scale=0.4]{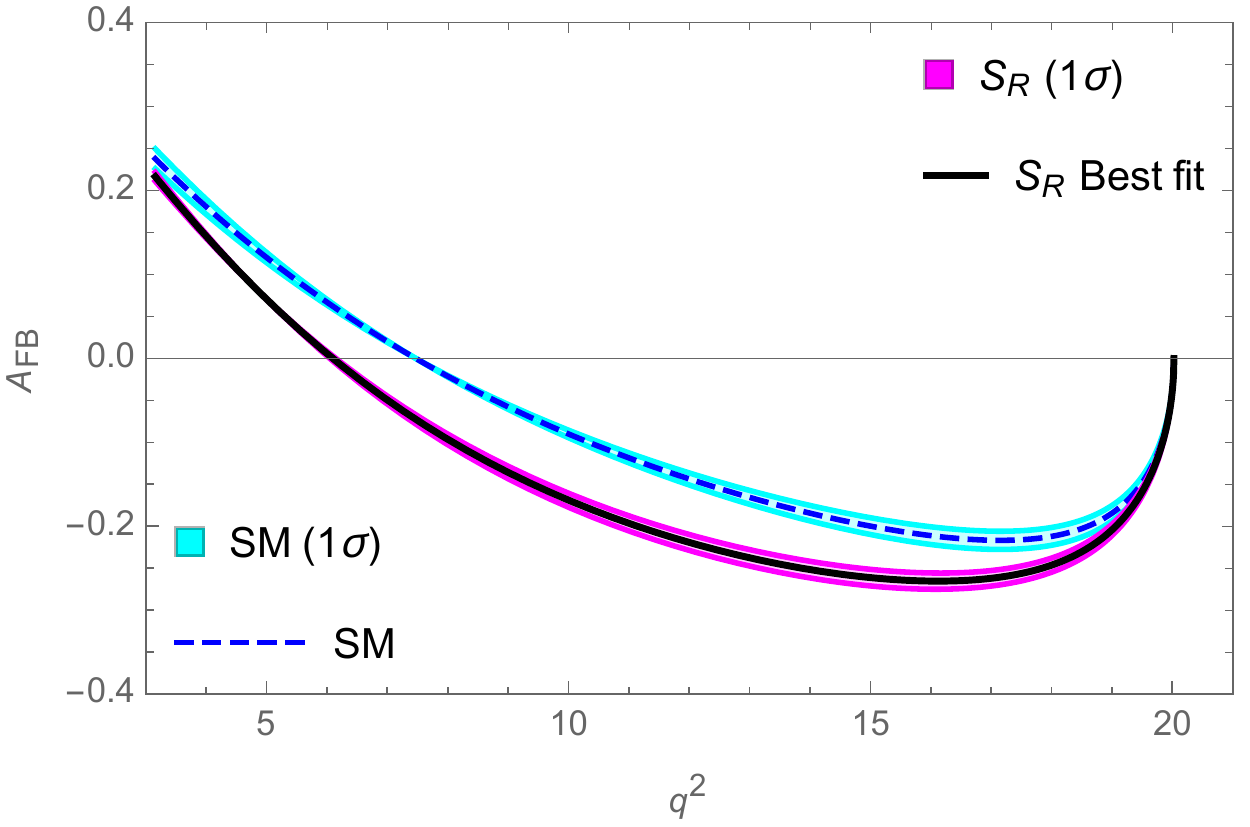}
\quad
\includegraphics[scale=0.4]{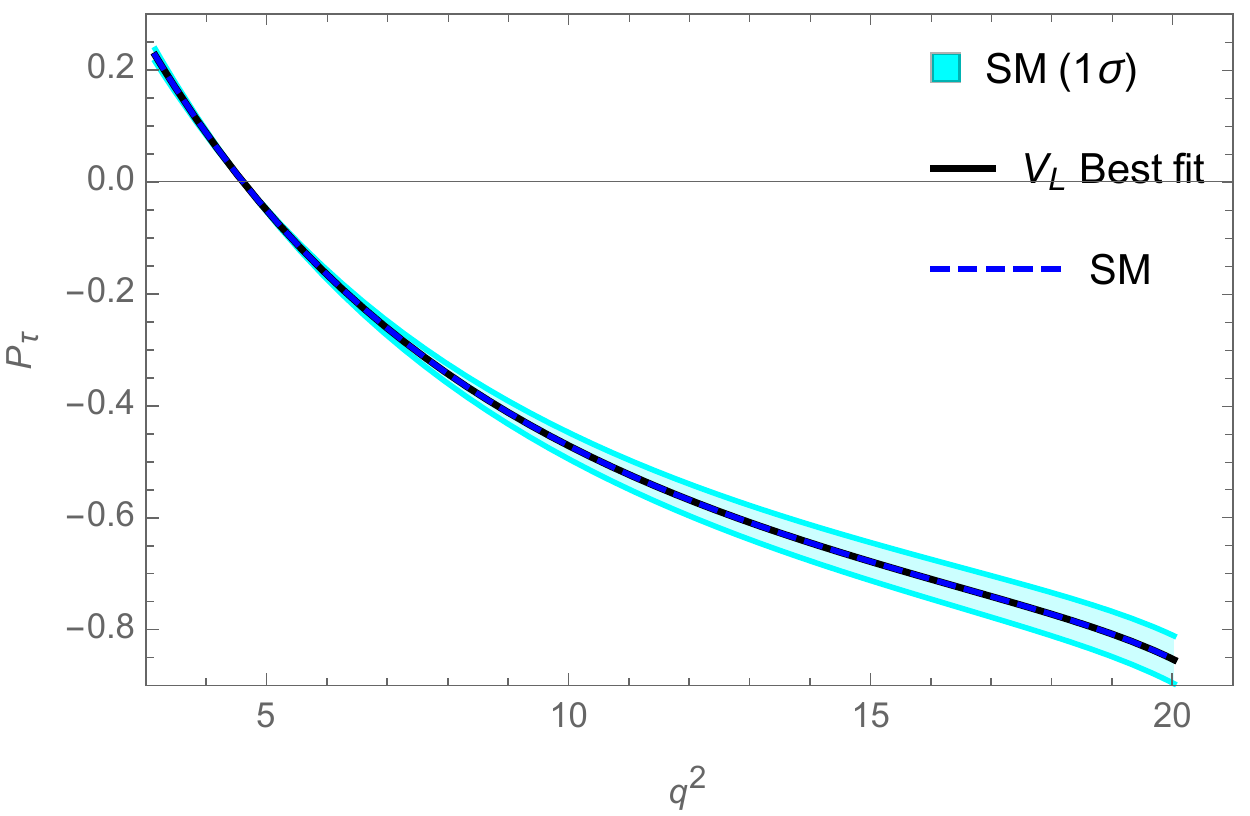}
\quad
\includegraphics[scale=0.4]{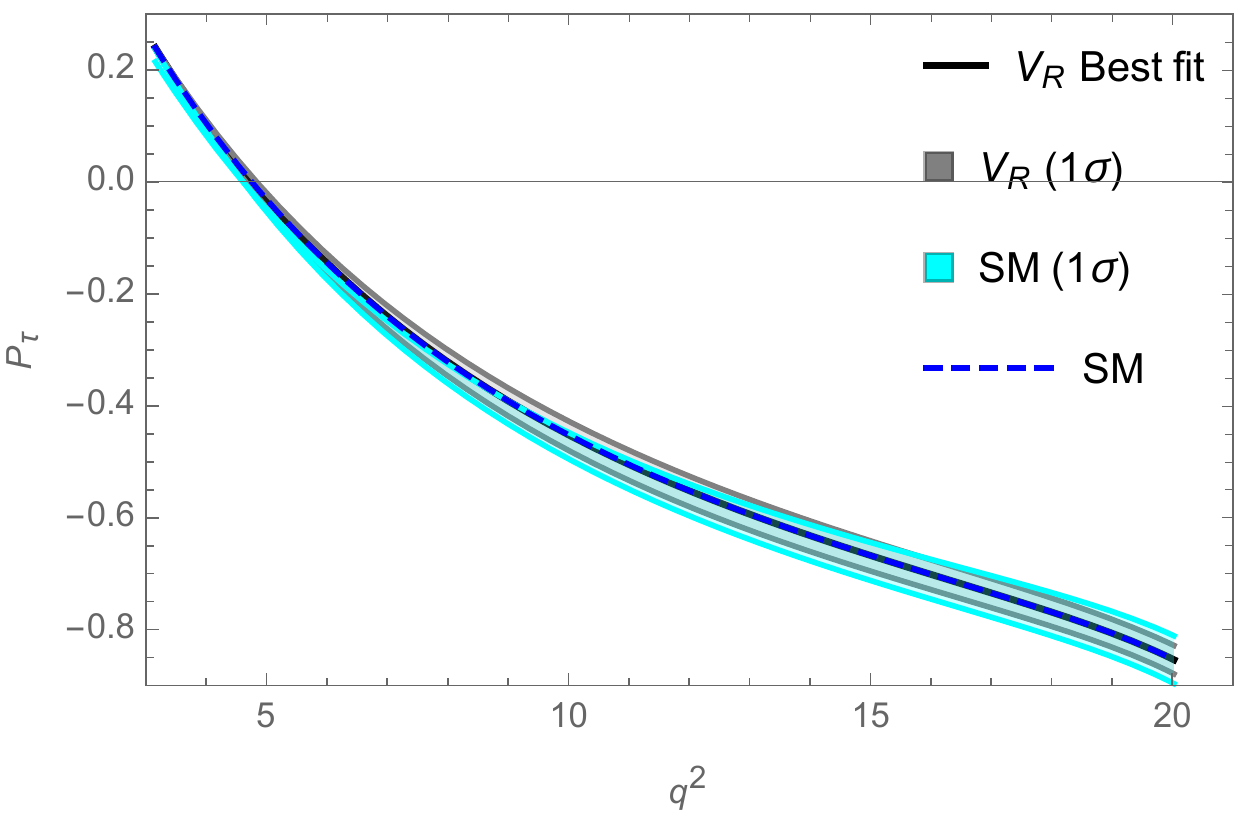}
\quad
\includegraphics[scale=0.4]{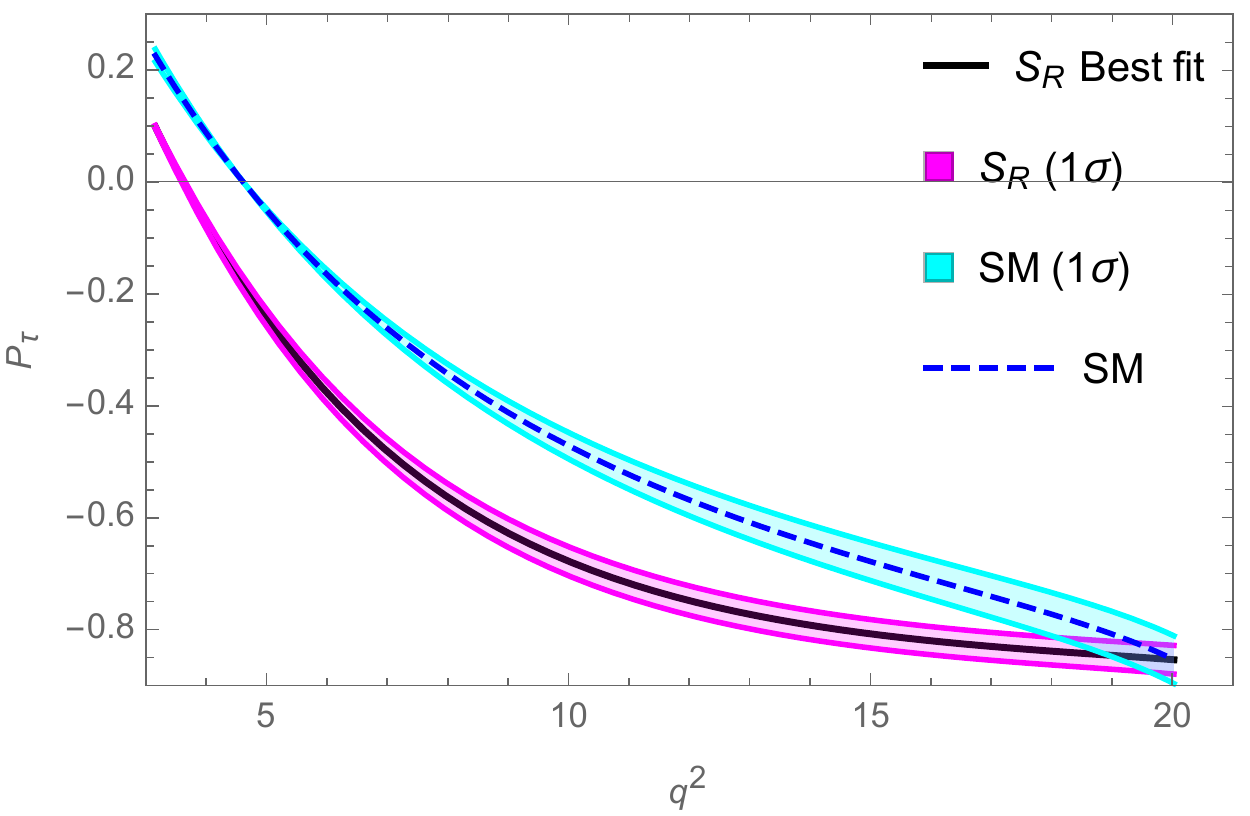}
\quad
\includegraphics[scale=0.4]{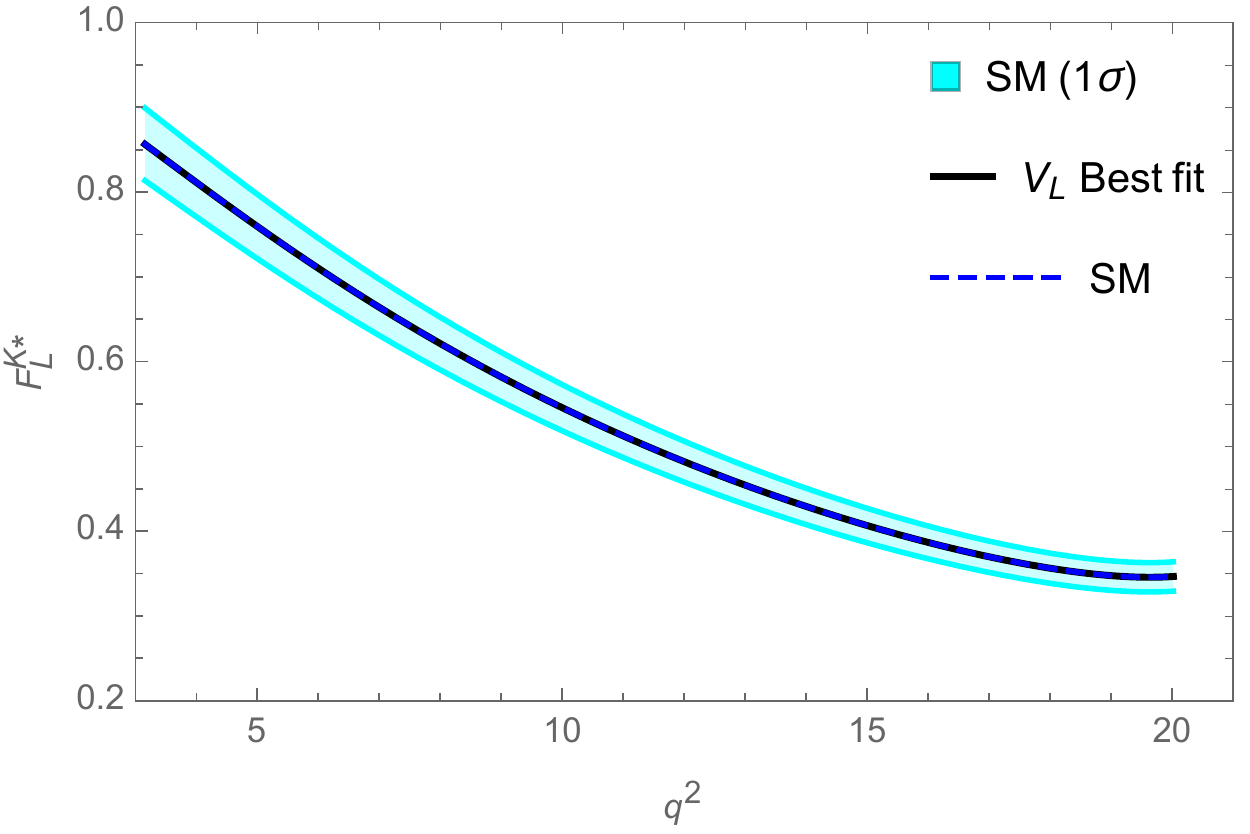}
\quad
\includegraphics[scale=0.4]{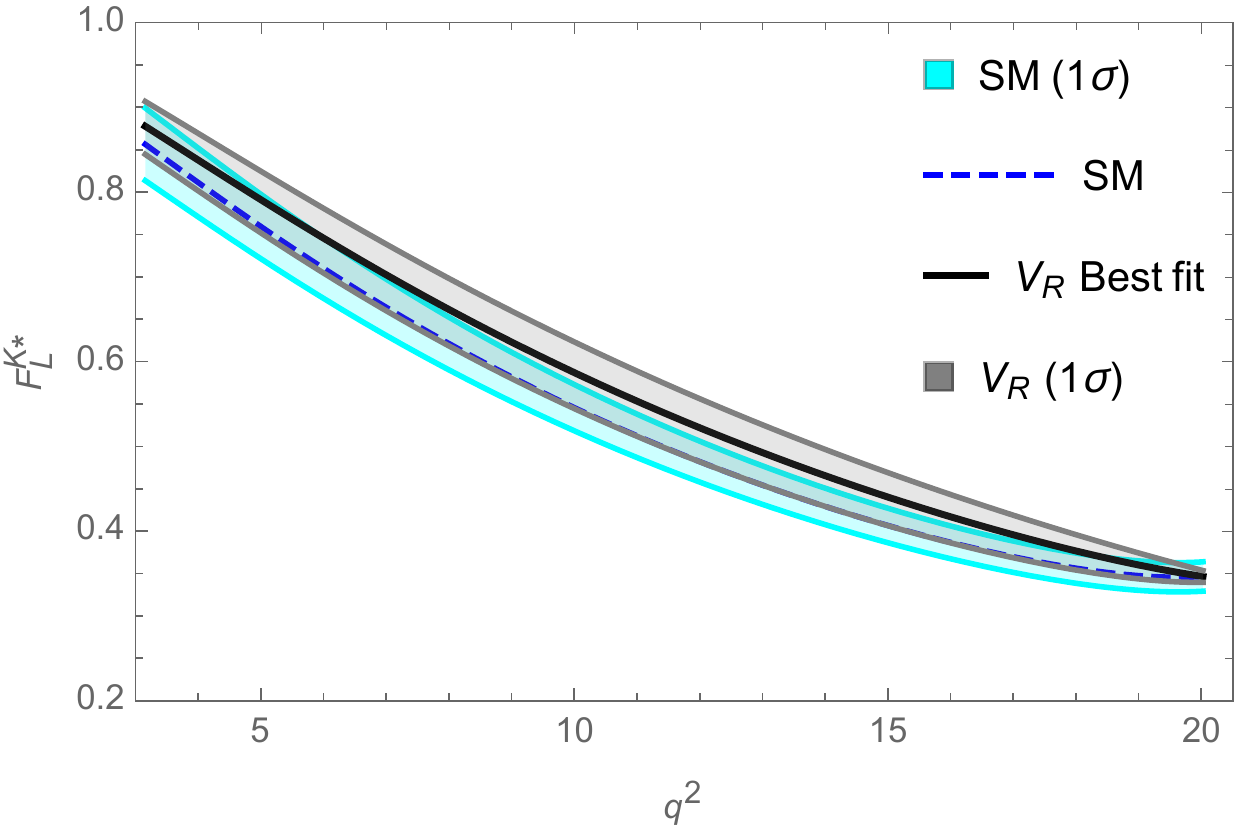}
\quad
\includegraphics[scale=0.4]{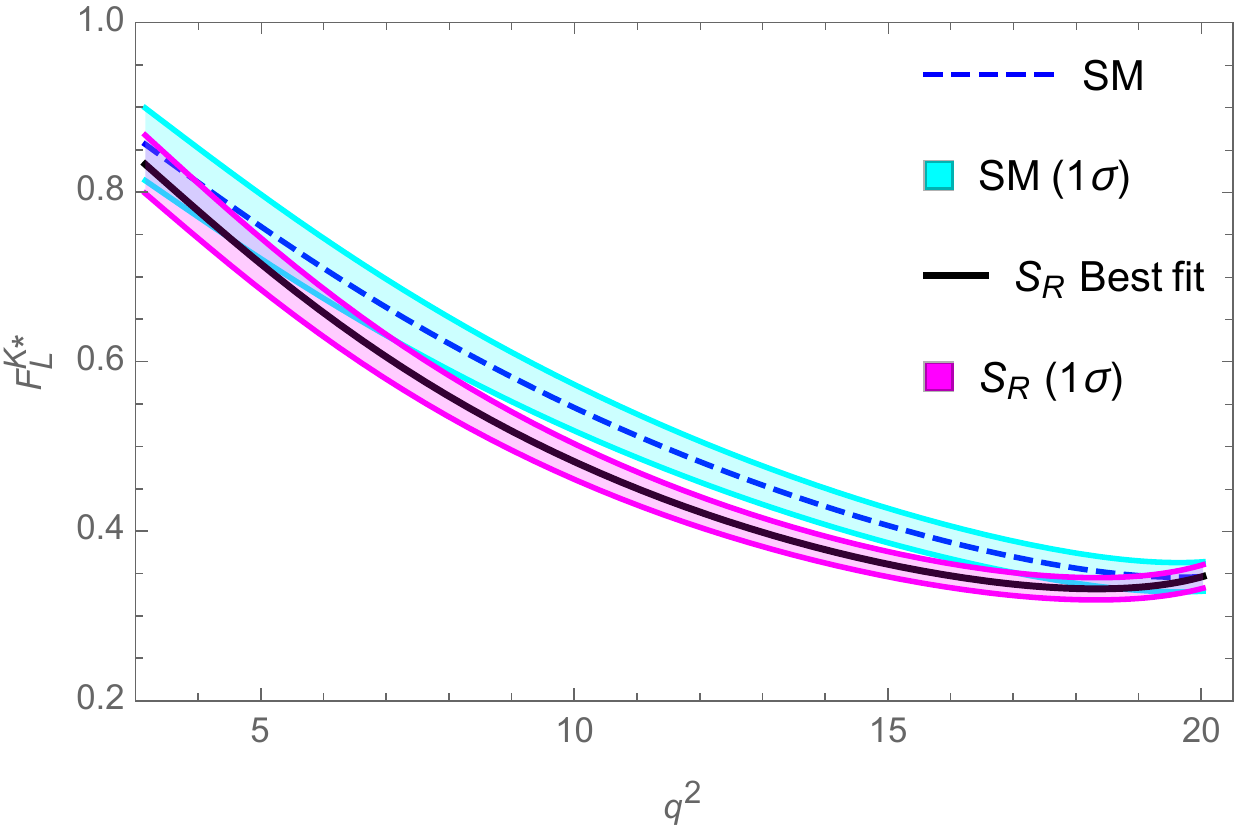}
\caption{Same as Figure \ref{B2rho}, for $B_s \to K^* \tau \bar \nu_\tau$ process.} 
\label{Bs2Ks}
\end{figure}
%%%%%%%%%%%%%%%%%%%%%%%%%%%%%%%%%%%%%%%
\begin{figure}
\includegraphics[height=70mm,width=80mm]{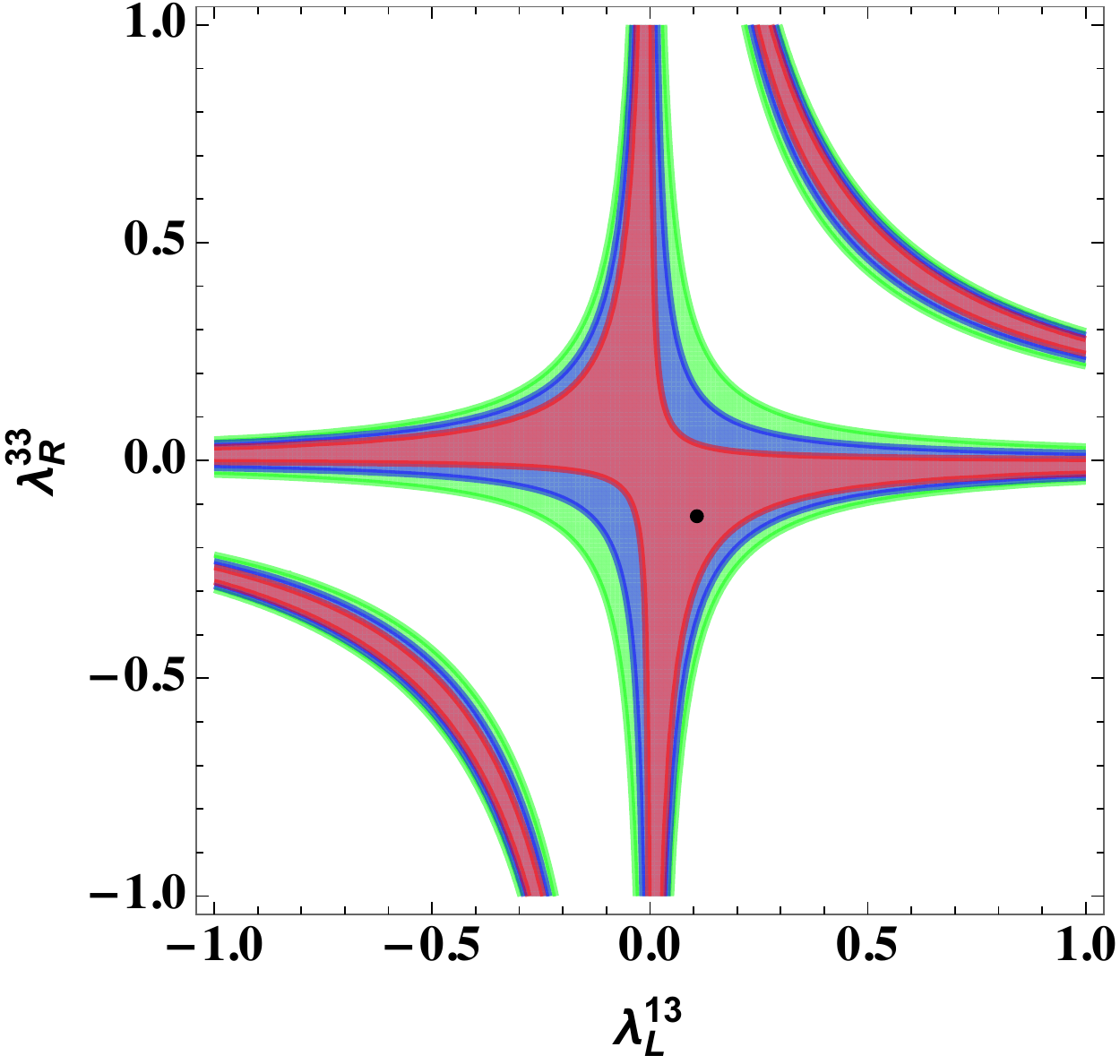}
\caption{Constraints on LQ parameter space from the current experimental data on ${\rm Br}(B_u \to \tau \bar \nu_\tau)$, ${\rm Br}(B \to \pi \tau \bar \nu_\tau)$ and  $R_\pi^\ell$. Different colors represent the $1 \sigma$, $2 \sigma$ and $3 \sigma$ contours and the black dot represents the best-fit value. }
\label{LQ-fig}
\end{figure}
%%%%%%%%%%%%%%%%%%%%%%%%%%%%%%%%%%%%

With these values of the form factors,  we show the $q^2$ variation of branching fraction, lepton non-universality parameter, forward-backward asymmetry, and tau polarization asymmetry in Fig. \ref{Bs2K}. From the figure,  it can be seen that the branching fraction and the LNU parameter deviate significantly from their SM values in the presence of $V_L$, $V_R$ and $S_R$ NP scenarios. However, due to the effect of $V_L$, the branching ratio is enhanced with respect to its SM value, whereas its value is found to be lower than the SM prediction in the presence of $V_R$ and $S_R$. Furthermore, though the forward-backward asymmetry remain unaffected due to the $V_L$  and $V_R$ contribution, the impact of $S_R$ is found to be quite substantial. Thus, the measurement of these observables will help to discriminate various kinds of NP scenarios. The numerical values of these observables are presented in Table \ref{Results}. 

For $B_s \to K^* \tau \nu$, the values of the form factors (\ref{FFV},\ref{FFV1}) are taken from \cite{Straub:2015ica}, and the corresponding expansion coefficients are provided in Table  \ref{Table:FF}. With these values, the  $q^2$ dependence of the various observables is shown in Fig. 
\ref{Bs2Ks}. In this case also, the branching fraction and the LNU parameters  have substantial deviations from their SM values  in the upward direction for $V_{L,R}$ and in downward direction for $S_R$. The $P_\tau$ and $F_L^{K^*}$ observables show only marginal deviation for $V_R$ and $S_R$ scenarios. The numerical values of these observables are presented in Table \ref{Results}.\\
\section{Leptoquark:  An example of New Physics scenario}
In this section, we will discuss the effect of leptoquarks on $b \to u \tau \nu$ transitions as a possible new physics scenario. We will consider two possible leptoquark models: the scalar leptoquark $R_2(3,2,7/6)$ and the vector leptoquark $U_1(3,1,2/3)$, which are found to be quite successful in addressing the recent flavour anomalies associated with  $b \to c \ell \bar\nu_\ell$ transition.
\subsection{Comment on effect of  scalar leptoquark $R_2(3,2,7/6)$}
Here we consider the example of scalar leptoquark (LQ) $R_2(3,2,7/6)$ as the NP scenario, where the quantum numbers in the parenthesis represent its values under the SM gauge group $SU(3)_C \times SU(2)_L \times U(1)_Y$ and briefly discuss  its implication on various  observables of $b \to u \tau \bar \nu$ transition.  The $SU(2)_L$  doublet scalar LQ can generate significant contribution to $b \to c \ell \bar \nu$  processes and can explain the observed experimental data quite well \cite{Sakaki:2013bfa,Iguro:2018vqb}. Additionally, it also safeguards the proton decay, as the diquark coupling is absent.  It couples to quark and lepton fields flavour dependently via Yukawa couplings and the  interaction Lagrangian involving $R_2$ can be expressed as
\begin{eqnarray}
{\cal L}_{\rm int}=\lambda_R^{ij} \bar Q_{Li} \ell_{Rj} R_2 - \lambda_L^{ij} \bar u_{Ri} R_2 i \tau_2 L_{Lj} +{\rm h.c.},\label{Int-L}
\end{eqnarray}
where $\lambda_{L,R}$ are the $3 \times 3$ complex matrices, $Q_L (L_L)$ is the left-handed quark (lepton) doublets, $u_R (\ell_R)$ is the right-handed up-type quark (charged lepton) singlet and $i,j$ are the generation indices. After expansion of the $SU(2)$ indices, the interaction Lagrangian (\ref{Int-L}) in the mass basis  can be expressed as
\begin{eqnarray}
{\cal L}_{\rm int}&=&(V_{\rm CKM}\lambda_R)^{ij} \bar u_{Li} \ell_{Rj} R_2^{(5/3)} +\lambda_R^{ij} \bar d_{Li} \ell_{Rj} R_2^{(2/3)}\nn\\
&+& \lambda_L^{ij} \bar u_{Ri} \nu_{Lj} R_2^{(2/3)}  -\lambda_L^{ij} \bar u_{Ri} \ell_{Lj} R_2^{(5/3)} +{\rm h.c.},\label{Int-L1}
\end{eqnarray}
where the superscripts in $R_2$ denote its electric charge and we consider the mass basis for quark doublet fields as $ ( (V_{\rm CKM}^\dagger u_L)^i, d_L^i )^T$ and lepton fields as $(\nu_L^i, \ell_L^i)^T$, ignoring the mixing in the lepton sector, i.e., the lepton mixing matrix is assumed to be unit matrix. Thus, it can be noted from (\ref{Int-L1})  that the exchange of $R_2^{(2/3)}$ can give rise to new contribution to $b \to u \tau \bar \nu_\tau$ transition at tree-level and generate the scalar and tensor operators at the LQ mass scale $(\mu_{\rm LQ})$ as: 
\bea
{S_L}(\mu_{\rm LQ})=4 {T_L} (\mu_{\rm LQ}) = \frac{1}{4 \sqrt 2 G_F V_{ub}} \frac{\lambda_L^{13} (\lambda_R^{33})^*}{m_{ \rm LQ}^2}\;,\label{SLQ}
\eea
where $m_{ \rm LQ}$ is the mass of the leptoquark, and we consider a typical representative value for LQ mass as 1 TeV, in this analysis. The new coefficients in (\ref{SLQ}) depend on the NP scale $(\mu(m_{\rm LQ}))$, and it is imperative to 
consider the renormalization-group (RG) equation to evolve their values from NP scale to effective Hamiltonian matching scale $\mu=m_b$, and are related as \cite{Gonzalez-Alonso:2017iyc, Blanke:2018yud}
\bea
\begin{pmatrix}
{S_L}(m_b)\\
{T_L}(m_b)
\end{pmatrix}=  \begin{pmatrix} 1.752 & -0.287\\
-0.004 & 0.842  \end{pmatrix}        
\begin{pmatrix}{S_L}(1~{\rm  TeV})\\
{T_L}(1~{\rm  TeV})
\end{pmatrix}. \label{SLQ1}
\eea
Performing a $chi$-square fit to the current experimental data on ${\rm Br}(B_u \to \tau \bar \nu_\tau)$, ${\rm Br}(B \to \pi \tau \bar \nu_\tau)$ and  $R_\pi^\ell$, and assuming the LQ couplings to be real,  the best-fit values for the  couplings are found to be $(\lambda_L^{13}, \lambda_R^{33})=(0.110,-0.129)$ and the corresponding allowed parameter space is shown in Fig. \ref{LQ-fig}, where different colors represent  the contours for $1\sigma$, $2\sigma$ and $3 \sigma$ allowed regions. Translating the obtained values of LQ coupling to the new scalar coupling through (\ref{SLQ}) and (\ref{SLQ1}), we obtain
\bea
 {S_L}=-0.033,
\eea
which is basically same order as the obtained value following model-independent approach. Therefore, one can conclude that the effect of the scalar LQ $R_2$ on various observables of $b \to u \tau \bar \nu$ is quite minimal and hence,  we do not provide their explicit values again  for this scenario. 
\subsection{Comment on effect of  scalar leptoquark $U_1(3,1,2/3)$}
\begin{figure}
\includegraphics[scale=0.6]{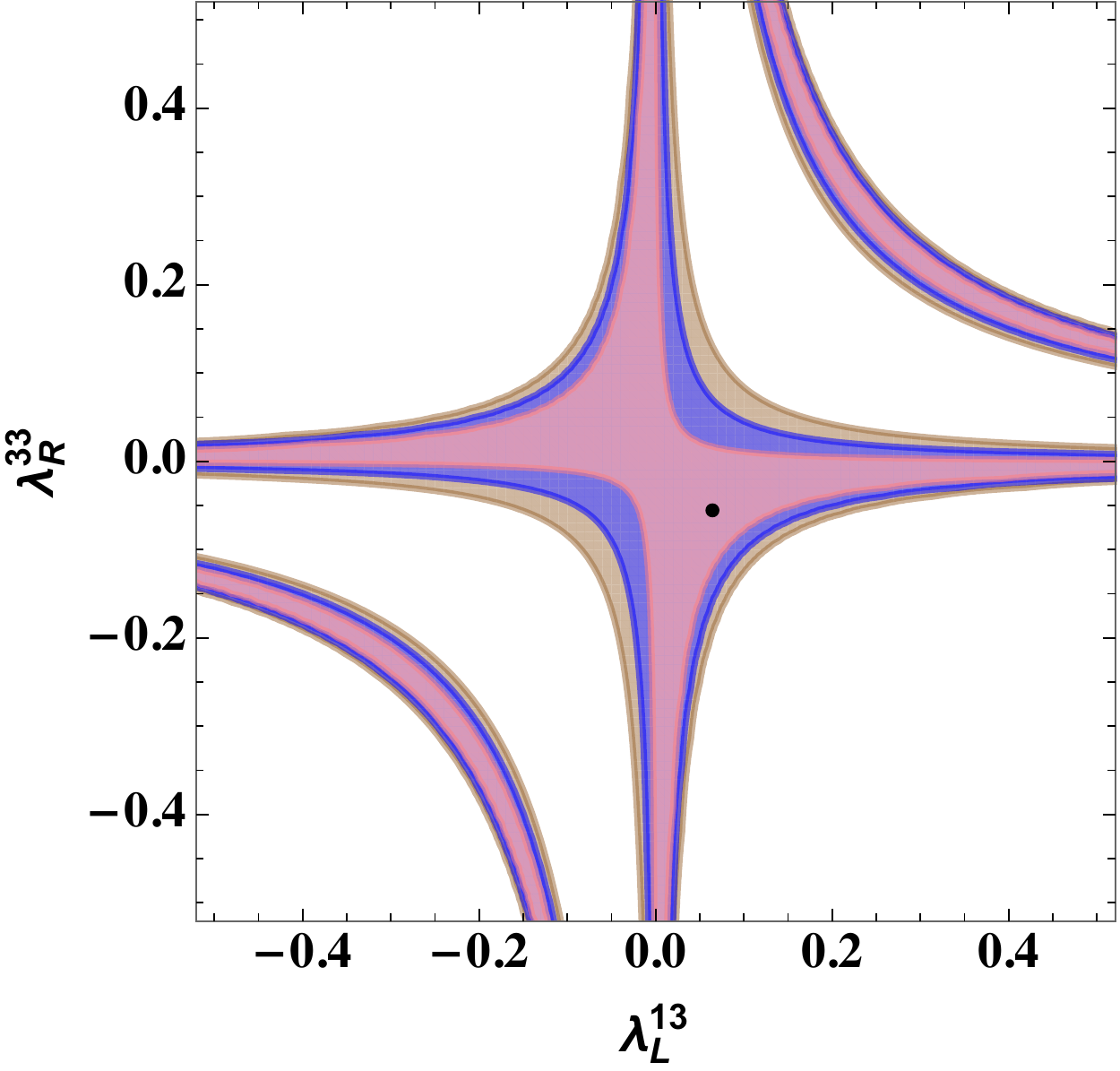}
\quad
\includegraphics[scale=0.6]{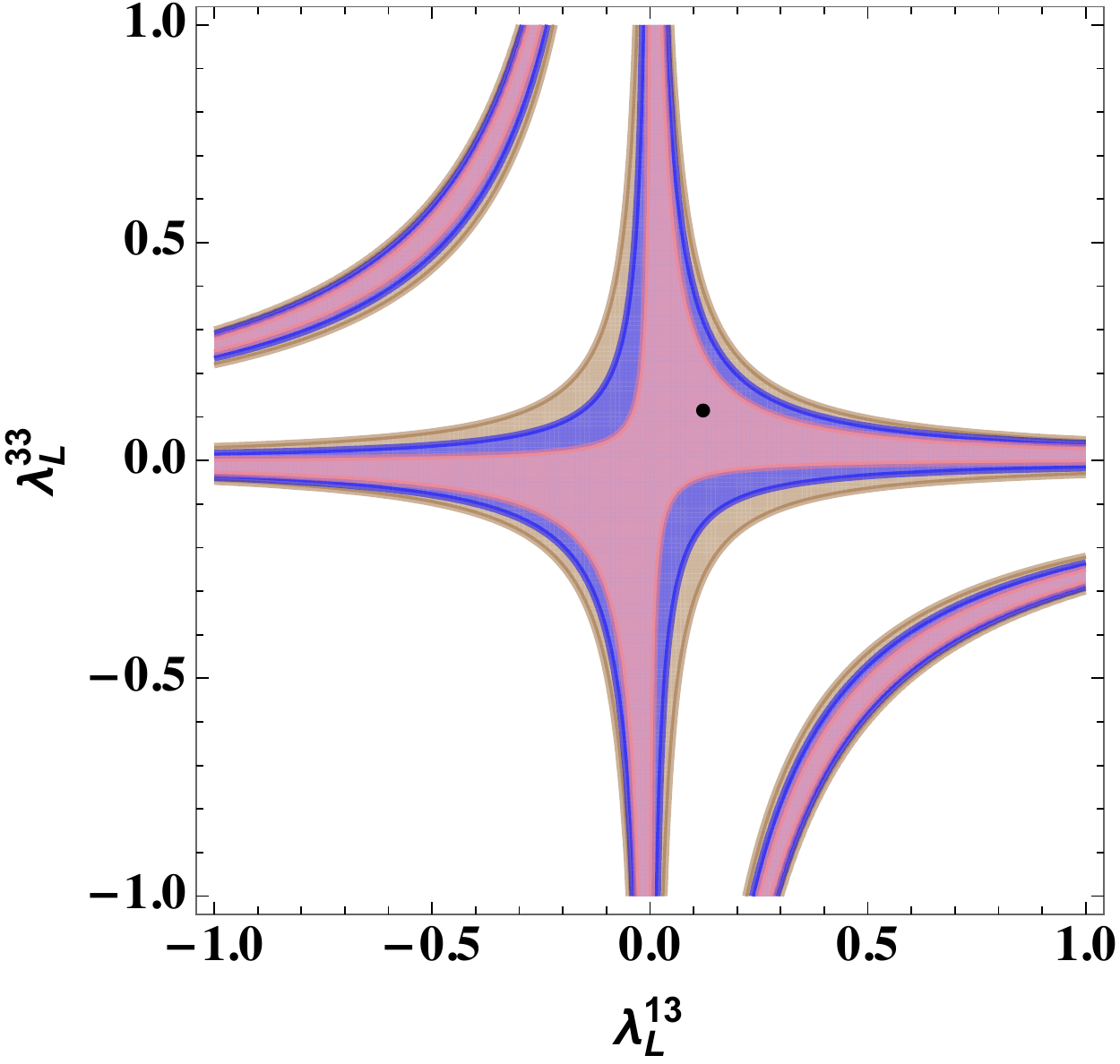}
\caption{Allowed parameter space for  $U_1$ Leptoquark couplings in $\lambda^{13}_L -\lambda_R^{33}$ plane (left panel) and $\lambda^{13}_L -\lambda_L^{33}$ plane (right-panel), where pink, blue and brown colors show the one, two and three-sigma allowed regions and the black points represent the best-fit values.} 
\label{U1-LQ}
\end{figure}

\begin{figure}
\includegraphics[scale=0.6]{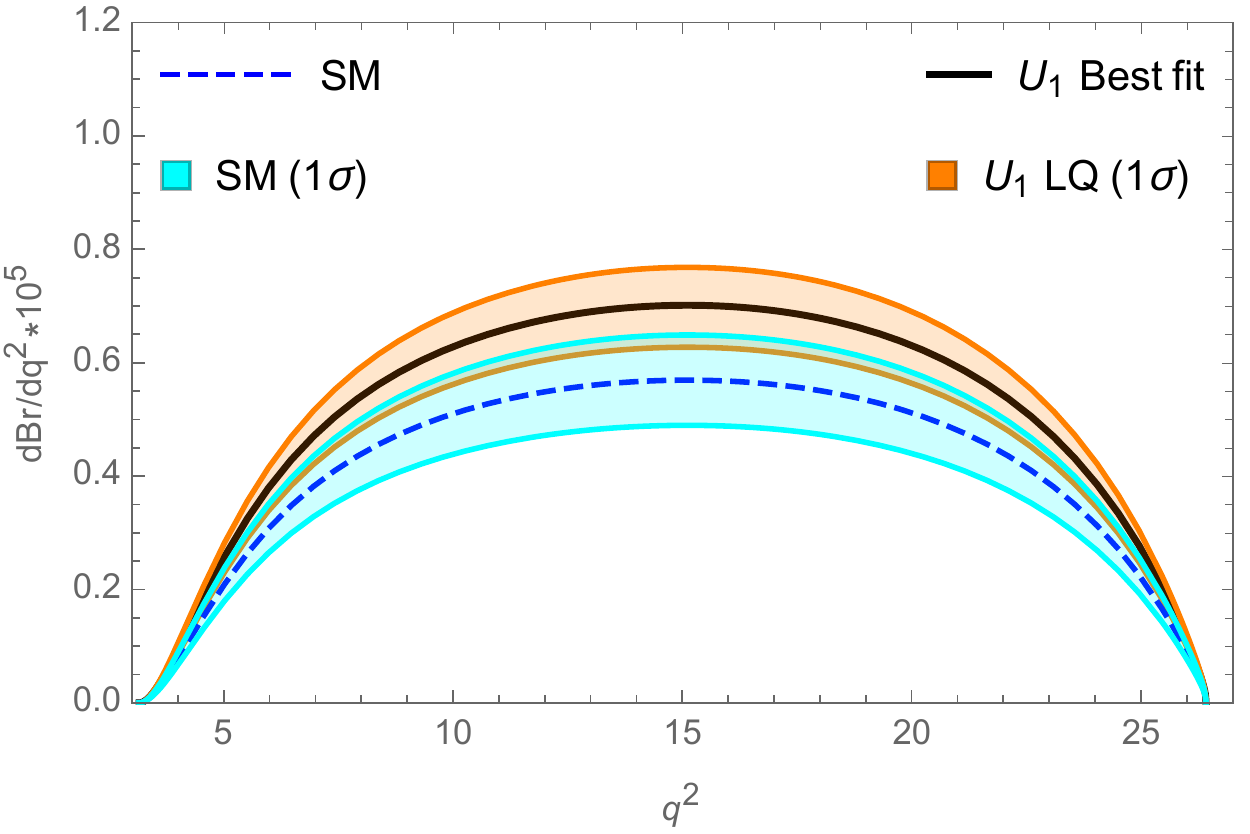}
\quad
\includegraphics[scale=0.6]{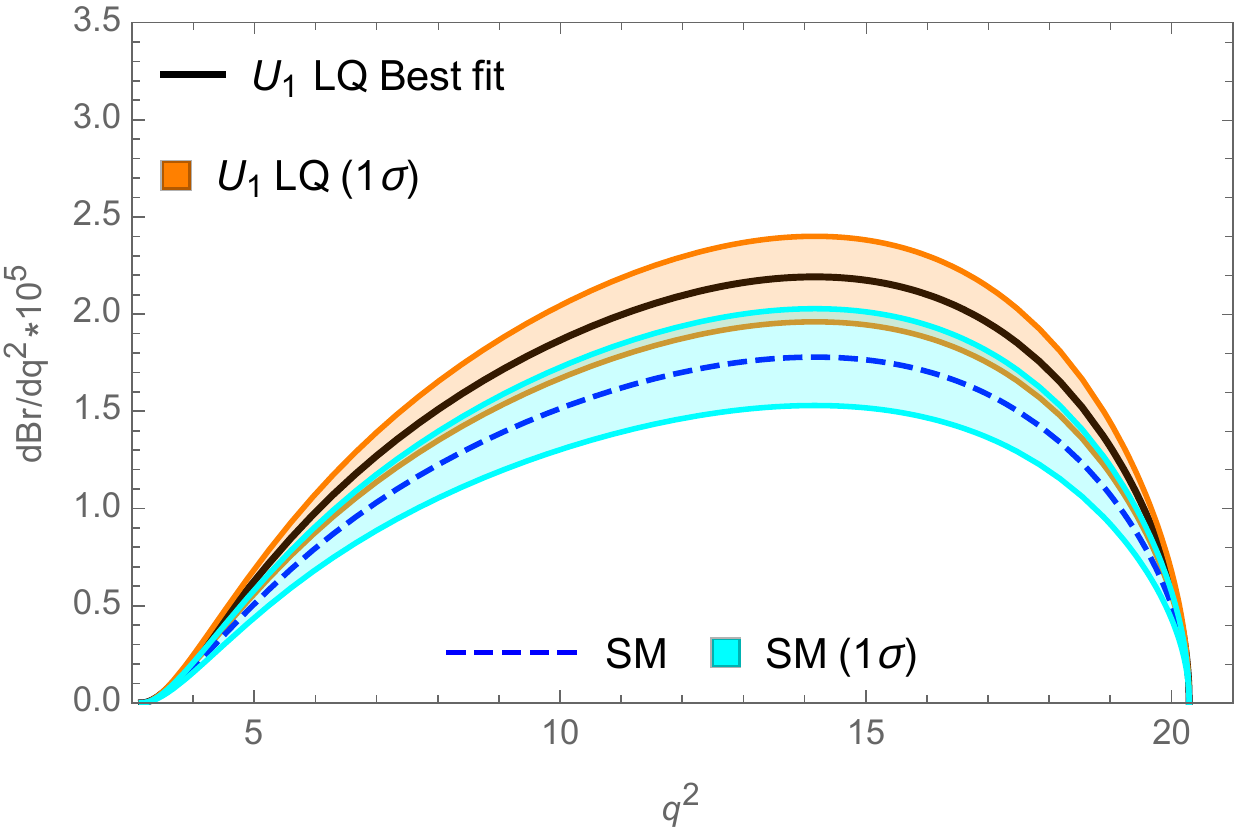}
\includegraphics[scale=0.6]{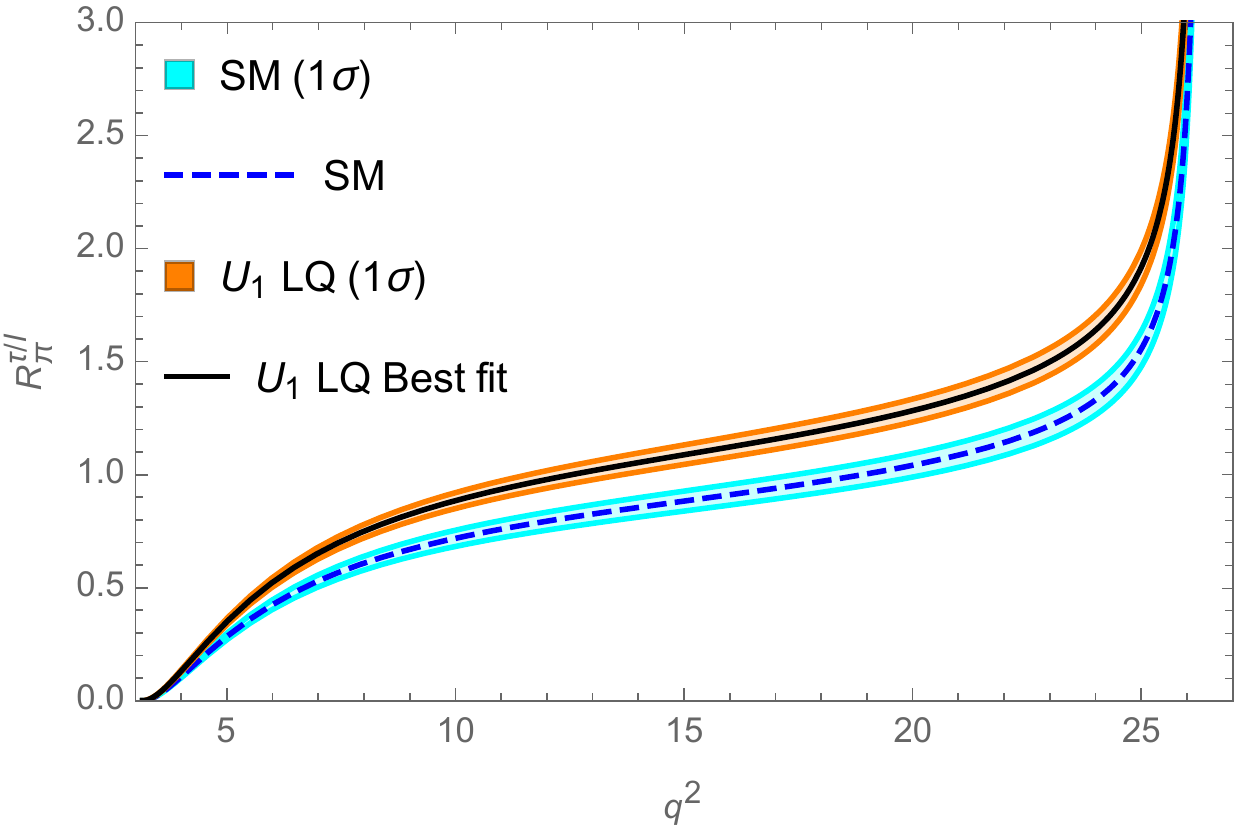}
\quad
\includegraphics[scale=0.6]{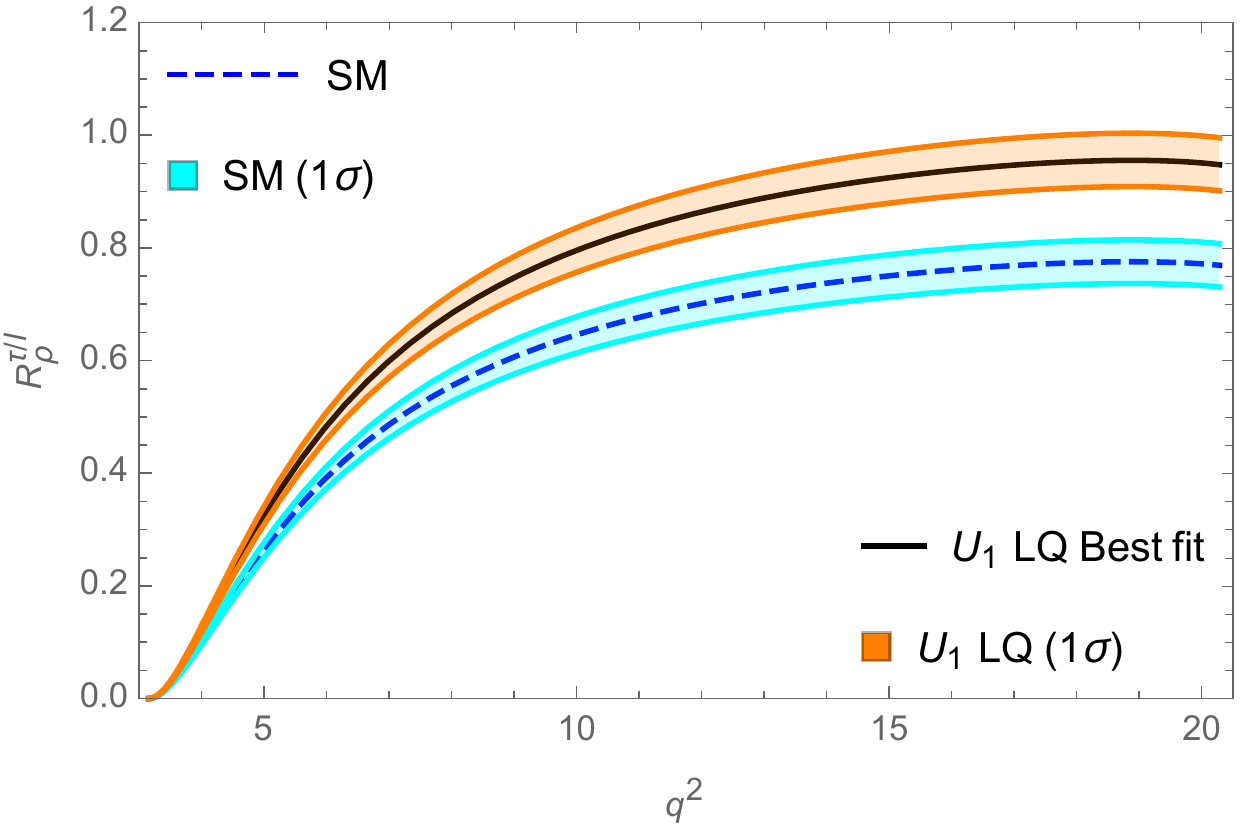}
\caption{The $q^2$ variation of branching fraction and the LNU observable for the $\bar B^0 \to \pi^+ \tau \bar \nu_\tau$ $(\bar B^0 \to \rho^+ \tau \bar \nu_\tau)$ process shown in the left (right) panel. } 
\label{U1-LQ3}
\end{figure}

The vector leptoquark $U_1(3,1,2/3)$ has received a lot of attention in recent times as it provides a simultaneous explanation to the observed flavour anomalies   
associated with $b \to c \ell \bar \nu_\ell$ and $b \to s \ell^+ \ell^-$ transitions.
The interaction Lagrangian describing the interaction between the $U_1$ LQ and the SM fermions can be represented as
\bea
{\cal L}= \lambda_L^{ij} \overline Q_i \gamma_\mu U_1^\mu L_j + \lambda_R^{ij} \overline d_{Ri} \gamma_\mu U_1^\mu \ell_{Rj}+{\rm h.c.},
\eea
where $\lambda_{L,R}^{ij}$ are the $3 \times 3$ complex matrices. After integrating out the heavy vector leptoquark $U_1$, the new Wilson coefficients contributing to $b \to u \tau \bar \nu_\tau$ are expressed as
\bea
&&{V_L}(\mu_{\rm LQ})= \frac{(V_{\rm CKM} \lambda_L)^{13} (\lambda_L^{33})^*}{2 \sqrt 2
G_F V_{ub}~ m_{U_1}^2}\nn\\
&&{S_R}(\mu_{\rm LQ})= -\frac{(V_{\rm CKM} \lambda_L)^{13} (\lambda_R^{33})^*}{ \sqrt 2
G_F V_{ub}~ m_{U_1}^2}.\label{U1_LQ}
\eea
For simplicity, we consider only the diagonal CKM matrix element $V_{11}$ to reduce the number of LQ couplings. We further assume  these couplings to be real. The values of these coefficients at the $m_b$ scale is obtained using the renormalization group equation \cite{Gonzalez-Alonso:2017iyc, Blanke:2018yud}
\bea
V_L(m_b)=V_L(1~{\rm TeV}),~~~~~S_R(m_b)=1.737S_R(1~{\rm TeV}).\label{U1-LQ1}
\eea
Since, there are three new couplings, i.e., $\lambda_L^{13}$, $\lambda_L^{33}$, $\lambda_R^{33}$,  it would be challenging to constrain them with three observables ${\rm Br}(B_u \to \tau \bar \nu_\tau)$, ${\rm Br}(B \to \pi \tau \bar \nu_\tau)$ and  $R_\pi^\ell$, we therefore assume that either ${V_L}$ or ${S_R}$ coupling will present at a given instant, (i.e., the presence of only two real couplings  at a given time).
Now considering the presence of $\lambda_L^{13}$ and $\lambda_R^{33}$,  the bounds on the LQ couplings are obtained by  performing a $\chi^2$ fit, with the best-fit values obtained as $(\lambda_L^{13}, \lambda_R^{33})=(0.064,-0.057)$ and the allowed parameter space in the $\lambda_L^{13}- \lambda_R^{33}$ plane is shown in the left panel of Fig. \ref{U1-LQ}, where different colors correspond to 1, 2, and 3$\sigma$ regions respectively and the black point represents the best-fit value. Similarly,  considering  $\lambda_L^{13}$ and $\lambda_L^{33}$ couplings, the best-fit values obtained are  $( \lambda_L^{13}, \lambda_L^{33})=(0.121,0.1155 )$ and the corresponding allowed parameter space is  shown  in the right panel of  Fig. \ref{U1-LQ}.  With Eqns (\ref{U1_LQ}) and  \ref{U1-LQ1}), these best-fit results give the values of the new couplings as
\bea
{S_R}=0.033,~~~~~{\rm and}~~~~~{V_L}=0.11.
\eea
As, the value of the scalar coupling ${S_R}$ is negligibly small, we  show the  effect of $U_1$ leptoquark with two non-zero real couplings, i.e.,  due to ${V_L}$, on the branching fraction and the LNU observable for the process  $B \to \pi \tau \bar \nu~(B \to \rho \tau \bar \nu)$  in he left (right) panel of Fig. \ref{U1-LQ3}.  
From the Figure, it can be seen that these observables deviate significantly from their SM predictions due the effect of $U_1$ leptoquark.
Other observables like forward-backward asymmetry and polarization asymmetries do not get affected by the new vector coupling and will remain consistent with their corresponding SM values. The integrated values of these observables for various processes are provided in Table \ref{U1-LQ4}.

\begin{table}[h!]
\caption{Predicted values of branching fractions (in units of $10^{-4}$) and LNU observables $R_{P,V}^{\tau/\ell}$  for $B\to (P,V)\tau\bar{\nu}_\tau $ processes in $U_1$ LQ model.}
\label{U1-LQ4}
\begin{tabular}{|c|c|c|}
\hline
~~Decay process~~ &~~ Branching fraction~~ &~~ $R^{\tau/\ell}$~~\\
\hline
$B \to \pi \tau \bar \nu$ &$(1.21\pm 0.12 )$ &~ $0.789 \pm0.136$~~\\
\hline
$B \to \rho \tau \bar \nu$ &$(2.64 \pm 0.26 )$ & $0.656 \pm 0.112$\\
\hline
$B \to \omega \tau \bar \nu$ &$(2.40\pm 0.24)$ &$0.659 \pm 0.113$ \\
\hline
$B_s \to K \tau \bar \nu$ & $(0.90\pm 0.09)$&$0.947 \pm 0.163$\\
\hline
$B_s \to K^* \tau \bar \nu$ & $(2.24\pm 0.22 )$ &$0.660 \pm 0.113$\\
\hline
\end{tabular}
\end{table}

\section{Conclusion}

It is well-known that the Standard Model  gauge interactions strictly respect lepton flavour universality and  any violation of it would point towards the possible role of new physics. The recent observation of several LFU violating signals in the charged current transitions $b \to c \ell \bar \nu_\ell$  in the form of $R_{D}$, $R_{D^*}$ and $R_{J/\psi}$ created huge excitement in the flavour physics community. To account for these discrepancies, it is generally assumed the attribution of new physics  to the semitauonic process $b \to c \tau \bar \nu_\tau$. Thus, if indeed new physics is  present in this decay process, its footprint can also be seen  in the allied charged-current process $b \to u \tau \bar \nu_\tau$, as these two processes have the same topologies, apart from the fact that the latter process is Cabibbo suppressed. Therefore,  in this article, we have performed a model independent analysis of semileptonic processes mediated through  $b \to u \tau \bar \nu$ transition in the presence of new physics.    In particular, we focus on the decay modes $B^0 \to \pi^+ \tau^- \bar \nu_\tau$, $B \to (\rho,\omega) \tau \bar \nu_\tau$ and $B_s \to (K,K^*) \tau \bar \nu_\tau$.  The new physics couplings are constrained by  using experimental data on the branching fractions of $B_u \to \tau \bar \nu_\tau$, $B \to \pi \tau \bar \nu_\tau$ and  $R_\pi^\ell$. Using the best-fit values and the corresponding $1\sigma$ ranges of NP couplings,  we show the $q^2$ variation of different observables and their  sensitivity towards new physics. In particular, we have estimated the values of branching fractions,  lepton non-universality parameters, forward-backward asymmetry, $\tau$ polarization asymmetry and the longitudinal polarization of the final vector meson in the presence of individual new coupling.  The differential branching fractions of all the processes showed a spectacular deviation from their SM predictions in the presence of $V_L, V_R$ and $S_R$ couplings whereas no deviation is found  in  the presence of $S_L$ coefficient. However, the nature of deviation in ${\rm Br}(B \to V \tau \bar \nu)$ transitions for $V_L$ type NP is opposite to those of $V_R$ and $S_R$ couplings. We  also noticed appreciable deviation in the LNU parameters in the presence of the $V_L$, $V_R$ and $S_R$  coefficients. Lepton spin asymmetry parameters almost consistent  with their SM values for $V_L,V_R$ and $S_L$ couplings, but in the presence of $S_R$ coupling they deviate  considerably from their SM values. $S_L$ coefficient  remains almost insensitive for all the observables. These observed features can help us to discriminate between different NP scenarios and to reveal the true nature of NP, if at all its presence is affirmed. We also investigated the leptoquark model as an example and considered two specific scenarios: the $R_2(3,2,7/6)$ scalar leptoquark and $U_1(3,1,2/3)$ vector leptoquark. Assuming the coupling between the leptoquark and the SM fermions  to be real, it has been found that the effect of $R_2(3,2,7/6)$ scalar leptoquark is negligible while the  vector leptoquark $U_1(3,1,2/3)$ can significantly enhance the values of branching fractions and LNU observables. 
Concerning the future prospects of these decay modes, they have great potential to be observed in the LHCb and Belle-II experiments and thus observation of these modes will definitely shed light on the interplay of new physics on $b \to u \tau \bar \nu_\tau$ transition. In addition, the search for lepton nonuniversality observables  $R_{P,V}^{\tau/\ell}$  is very promising as they also have significant deviation from their SM values for all these decay processes. Hence, observation of these observables can be used as an ideal probe to either confirm or rule out the presence of new physics.
To conclude, these decay processes  offer an  alternative probe to study  the implications of NP associated with the current $B$ anomalies in semileptonic transitions and could be accessible with the currently running LHCb and Belle II experiments.

\section*{Appendix: Helicity dependent differential decay rate}\label{helicity}

The $q^2$ distribution of the $B \to P \tau \bar \nu$  decay rates  for a given $\tau$ polarization are given as
\bea
\frac{d \Gamma(\lambda_\tau=1/2)}{dq^2}&=& \frac{G_F^2 |V_{ub}|^2}{192 \pi^3  m_B^3}q^2 \sqrt{\lambda_P(q^2)}\left (1- \frac{m_\tau^2}{q^2} \right )^2\Big\{\frac{1}{2}|1+V_L+V_R|^2 \frac{m_\tau^2}{q^2}(H_{V,0}^{s~2}+3 H_{V,t}^{s~2}\nn\\
&+& \frac{3}{2}|S_R+S_L|^2 H_S^{s~2}+8|T_L|^2 H_T^{s~2}-4{\rm Re}[(1+V_L+V_R)T_L^*] \frac{m_\tau}{\sqrt{q^2}}H_T^s H_{V,0}^s\nn\\
&+&3{\rm Re}[(1+V_L+V_R)(S_R^*+S_L^*)] \frac{m_\tau}{\sqrt{q^2}}H_S^s H_{V,t}^s
 \Big\},\\
 \frac{d \Gamma(\lambda_\tau=-1/2)}{dq^2}&=& \frac{G_F^2 |V_{ub}|^2}{192 \pi^3  m_B^3}q^2 \sqrt{\lambda_P(q^2)}\left (1- \frac{m_\tau^2}{q^2} \right )^2\Big\{|1+V_L+V_R|^2H_{V,0}^{s~2}+16 |T_L|^2 \frac{m_\tau^2}{q^2}  H_{T}^{s~2}\nn\\
 &-& 8 {\rm Re}[(1+V_L+V_R)T_L^*]\frac{m_\tau }{ \sqrt{q^2}}H_T^s H_{V,0}^s\Big \}.
\eea

Helicity dependent differential decay rate for $B \to V \tau \bar \nu$ process can be expressed as,
\beqa
\nn
\frac{d\Gamma(\lambda_\tau=\frac{1}{2})}{dq^2 }&=&\frac{G_F^2|V_{ub}|^2}{192 \pi^3 m_{\bar{B}}^3}q^2 \sqrt{\lambda_V(q^2)}\left(1-{m_\tau^2 \over q^2}\right)^2 \nn\\
&\times &\Big \{\frac{1}{2} (|1+V_L|^2 +|V_R|^2){m_\tau^2 \over q^2} (H_{V,+}^2+H_{V,-}^2+H_{V,0}^2+3H_{V,t}^2)\\ \nn
&& -{\rm Re}[(1+V_L)V_R^*]{m_\tau^2\over q^2}(H_{V,0}^2+2H_{V,+}H_{V,-}+3H_{V,t}^2)
 +{3\over 2} |S_R-S_L|^2 H_S^2\\ \nn 
 &&+8|T_L|^2(H_{T,+}^2+H_{T,-}^2+H_{T,0}^2)+3{\rm  Re}[(1+V_L-V_R)(S_R^*-S_L^*)]\frac{m_\tau^2} {\sqrt{q^2}}H_S H_{V,t}\\ \nn
 &&-4 {\rm Re} [(1+V_L)T_L^*]\frac{m_\tau}{\sqrt{q^2}}(H_{T,0}H_{V,0}+H_{T,+}H_{V,+}-H_{T,-}H_{V,-})\\
 &&+4 {\rm Re}[V_RT_L^*]\frac{m_\tau}{\sqrt{q^2}}(H_{T,0}H_{V,0}+H_{T,+}H_{V,-}-H_{T,-}H_{V,+})\Big \},\\
\nn
\frac{d\Gamma(\lambda_\tau=-\frac{1}{2})}{dq^2 }&=&\frac{G_F^2|V_{ub}|^2}{192 \pi^3 m_{\bar{B}}^3}q^2\sqrt{\lambda_{V}(q^2)}\left(1-\frac{m_\tau^2}{q^2}\right)^2 \Big\{  (|1+V_L|^2+|V_R|^2)(H_{V,+}^2+H_{V,-}^2+H_{V,0}^2)\\ \nn
&& -2{\rm Re}[(1+V_L)V_R^*](H_{V,0}^2+2H_{V,+}H_{V,-})+16|T_L|^2\frac{m_\tau^2}{q^2}(H_{T,+}^2+H_{T,-}^2+H_{T,0}^2)\\ \nn 
&&-8 {\rm Re} [(1+V_L)T_L^*]\frac{m_\tau}{\sqrt{q^2}}(H_{T,0}H_{V,0}+H_{T,+}H_{V,+}-H_{T,-}H_{V,-})\\
&&+8 {\rm Re}[V_RT_L^*]\frac{m_\tau}{\sqrt{q^2}}(H_{T,0}H_{V,0}+H_{T,+}H_{V,-}-H_{T,-}H_{V,+})
\Big \} .
\eeqa

The decay distribution for the longitudinal polarization of final $V$ meson is given as
\bea
\frac{d \Gamma(\lambda_V=0)}{d q^2}&=& \frac{G_F^2 |V_{ub}|^2}{192 \pi^3 m_B^3}q^2 \sqrt{\lambda_V(q^2)} \left (1-\frac{m_\tau^2}{q^2} \right )^2
\Big\{|1+V_L-V_R|^2\left [\left (1+ \frac{m_\tau^2}{2 q^2}\right )H_{V,0}^2 +\frac{3}{2} \frac{m_\tau^2}{q^2} H_{V,t}^2    \right ] \nn\\
&+& \frac{3}{2}|S_R-S_L|^2 H_S^2 +8|T_L|^2 \left (1+ \frac{2 m_\tau^2}{q^2}  \right ) H_{T,0}^2 -12 {\rm Re}[(1+V_L-V_R)T_L^*] \frac{m_\tau}{\sqrt{q^2}}H_{T,0} H_{V,0}\nn\\
&+& 3{\rm Re}[(1+V_L-V_R)(S_R^*-S_L^*)] \frac{m_\tau}{\sqrt{q^2}}H_S H_{V,t}
\Big\}.
\eea

 \acknowledgements
AB would like to acknowledge DST INSPIRE program for financial support.
RM and AR would like to thank Science and Engineering Research Board (SERB), Govt. of India for financial support through grant no. EMR/2017/001448. The computational work done at CMSD, University of Hyderabad is duly acknowledged.
\medskip
 \bibliography{BL}

\end{document}